\documentclass[aps,reprint,amsmath,amssymb,twocolumnm,nofootinbib]{revtex4-2}
\usepackage{aas_macros}
\usepackage{natbib}
\usepackage{color}
\usepackage{graphicx}% Include figure files
\usepackage{dcolumn}% Align table columns on decimal point
\usepackage{bm}% bold math
\usepackage{pifont}% http://ctan.org/pkg/pifont
\usepackage{comment}
\usepackage[colorlinks=true, linkcolor=blue, citecolor=blue, urlcolor=blue]{hyperref}

\begin{document}

\title{Long-term Monte Carlo-based neutrino-radiation hydrodynamics simulations for a black hole-torus system}

\author{Kyohei Kawaguchi}
\affiliation{Max Planck Institute for Gravitational Physics (Albert Einstein Institute), Am M\"{u}hlenberg 1, Potsdam-Golm, 14476, Germany}
\affiliation{Institute for Cosmic Ray Research, The University of Tokyo, 5-1-5 Kashiwanoha, Kashiwa, Chiba 277-8582, Japan}
\affiliation{Center of Gravitational Physics and Quantum Information,
 Yukawa Institute for Theoretical Physics, 
Kyoto University, Kyoto, 606-8502, Japan} 
\author{Sho Fujibayashi}
\affiliation{Frontier Research Institute for Interdisciplinary Sciences, Tohoku University, Sendai 980-8578, Japan}
\affiliation{Astronomical Institute, Graduate School of Science, Tohoku University, Sendai 980-8578, Japan}
\affiliation{Max Planck Institute for Gravitational Physics (Albert Einstein Institute), Am M\"{u}hlenberg 1, Potsdam-Golm, 14476, Germany}
\author{Masaru Shibata}
\affiliation{Max Planck Institute for Gravitational Physics (Albert Einstein Institute), Am M\"{u}hlenberg 1, Potsdam-Golm, 14476, Germany}
\affiliation{Center of Gravitational Physics and Quantum Information,
 Yukawa Institute for Theoretical Physics, 
Kyoto University, Kyoto, 606-8502, Japan} 

\newcommand{\angstrom}{\text{\normalfont\AA}}
\newcommand{\rednote}[1]{{\color{red} (#1)}}
\newcommand{\ms}[1]{{\color{blue} {\em #1}}}
\newcommand{\kk}[1]{{\color{magenta} #1}}
\newcommand{\sho}[1]{{\color{green} #1}}

\begin{abstract}
We present our new general relativistic Monte Carlo (MC)-based neutrino radiation hydrodynamics code designed to solve axisymmetric systems with several improvements. The main improvements are as follows: (i) the development of an extended version of the implicit MC method for multi-species radiation fields; (ii) modeling of neutrino pair process rates based on a new numerically efficient and asymptotically correct fitting function for the kernel function; (iii) the implementation of new numerical limiters on the radiation-matter interaction to ensure a stable and physically correct evolution of the system. We apply our code to a black hole (BH)-torus system with a BH mass of $3\,M_\odot$, BH dimmensionless spin of 0.8, and a torus mass of $0.1\,M_\odot$, which mimics a post-merger remnant of a binary neutron star merger in the case that the massive neutron star collapses to a BH within a short time scale ($\sim10\,{\rm ms}$). We follow the evolution of the BH-torus system up to more than $1\,{\rm s}$ with our MC-based radiation viscous-hydrodynamics code that dynamically takes into account non-thermal pair annihilation. We find that the system evolution and the various key quantities, such as neutrino luminosity, ejecta mass, torus $Y_e$, and pair annihilation luminosity, are broadly in agreement with the results of the previous studies. We also find that the $\nu_e{\bar \nu}_e$ pair annihilation can launch a relativistic outflow for a time scale of $\sim 0.1\,{\rm s}$, and it can be energetic enough to explain some of short-hard gamma-ray bursts and the precursors. Finally, we calculate the indicators of the fast flavor instability directly from the obtained neutrino distribution functions, which indicate that the instability can occur particularly near the equatorial region of the torus.
\end{abstract}

\keywords{radiative transfer}

\maketitle

\section{Introduction}\label{sec:intro}
Merger of neutron star (NS) binaries is one of the most interesting scientific targets in multi-messenger astrophysics. A NS binary gradually decreases its orbital separation and eventually merges by emitting gravitational waves (GWs). GWs emitted during the orbital evolution and at the time of the merger are the main targets of ground-based GW detectors~\cite{TheLIGOScientific:2014jea,TheVirgo:2014hva,Kuroda:2010zzb}. At the onset of the merger, a fraction of the neutron-rich matter is ejected by tidal disruption and collisional shock heating~\cite[e.g.,][]{Rosswog:1998hy,Ruffert:2001gf,Hotokezaka:2012ze}. After the binary merger, a massive NS or black hole (BH) surrounded by a strongly magnetized hot and dense accretion torus is formed~\citep{Price:2006fi,Kiuchi:2017zzg}, and the further outflows can be launched by magnetic pressure and tension, viscous heating due to magnetohydrodynamic turbulence, and neutrino irradiation~\cite[e.g., ][]{1977MNRAS.179..433B,BP82,Balbus:1998ja,Dessart:2008zd,2013MNRAS.435..502F,Metzger:2014ila,Perego:2014fma,Just:2014fka,Siegel:2017nub,Shibata:2017xdx,Fujibayashi:2017puw,Siegel:2017jug,Fernandez:2018kax,Christie:2019lim,Miller:2019dpt,Fujibayashi:2020qda,Fujibayashi:2020jfr,Fujibayashi:2020dvr,Ciolfi:2020wfx,Vsevolod:2020pak,Fernandez:2020oow,Mosta:2020hlh,Shibata:2021bbj,Shibata:2021xmo,Fujibayashi:2022ftg,Kiuchi:2022nin,Foucart:2022kon,Sun:2022vri,Just:2023wtj,Curtis:2023zfo,Sprouse:2023cdm}. These outflows will be the source of various high-energy electromagnetic transients, such as gamma-ray bursts~\cite{1991AcA....41..257P,Nakar:2007yr,Berger:2013jza}, kilonovae~\cite{Li:1998bw,Kulkarni:2005jw,Metzger:2010sy,Kasen:2013xka,Tanaka:2013ana}, and synchrotron flares~\cite{Nakar2011Natur,Hotokezaka:2015eja,Hotokezaka:2018gmo,Margalit2020MNRAS}. In the neutron-rich outflows, a suitable condition can be realized for the $r$-process nucleosynthesis of heavy elements to proceed~\cite{Lattimer:1974slx,Eichler:1989ve,Freiburghaus1999a,Cowan:2019pkx}. Hence, a merger of NSs is considered to be one of the important production sites for the about half of the elements heavier than iron in the universe~\cite{Lattimer:1974slx,Eichler:1989ve,Freiburghaus1999a,Cowan:2019pkx}. The simultaneous detection of GWs and EM signals from NS binaries, of which first detection is indeed achieved in GW170817~\cite{2017PhRvL.119p1101A,LIGOScientific:2017ync} and more detections are expected to be achieved in the next few years~\citep{2018LRR....21....3A,2020MNRAS.493.1633S,2020arXiv200802921K,2022ApJS..260...18A}, will surely give a great opportunity to understand these important astrophysical phenomena.

Weak interactions and neutrino radiative transfer play an important role in determining the post-merger dynamics. They determine the dynamics and thermodynamic properties of the merger remnants, the post-merger environment, and the abundance of elements synthesized in the ejecta.~\cite[e.g.,][]{Metzger:2010sy,Goriely:2010bm,Wanajo:2014wha,Just:2014fka,Sekiguchi:2015dma,Sekiguchi:2016bjd,Radice:2016dwd,2019PhRvD.100b3008M,Fujibayashi:2017puw,Fujibayashi:2020qda,Fujibayashi:2020jfr,Fujibayashi:2020dvr,Foucart:2020qjb,Just:2021cls,2021arXiv211104621H,2022arXiv220505557F,2022arXiv220710680F,Kiuchi:2022nin}. Neutrino-antineutrino pair annihilation could also be the important mechanism for the system to launch a jet powerful enough to explain short-hard gamma-ray bursts~\citep{1996A&A...305..839J,1999ApJ...518..356P}. To maximize the scientific return from the observed signals, a quantitative prediction of the merger outcome is crucial. Hence, accurately solving neutrino radiation is a task for this purpose. However, solving radiative transfer is for many cases computationally expensive due to its large dimensionality of the phase space dependence; seven dimensions which come from time, 3 real-space dimensions, and 3 momentum-space dimensions. Moreover, the physical time scale of the local radiation-matter coupling can often be much shorter than the dynamical time scale of the system. This fact also requires the implementation of some complicated schemes, such as implicit solvers, to solve the system numerically in a feasible computational time.

To overcome the computational difficulties, various approximation methods have been proposed. One of the most successful approximation methods among them is the moment scheme. In a moment scheme, two lowest moments of radiation in the momentum space are solved as dynamical variables with an approximate closure relation to higher moments~\citep{1980RvMP...52..299T,2011PThPh.125.1255S}. In the context of relativistic problems, many numerical codes are developed employing moment schemes sometimes with a combination of the leakage algorithm, and enabled to quantitatively understand the post-merger dynamics and outcomes consistently taking neutrino radiative transfer effects into account~\citep{Galeazzi:2013mia,Neilsen:2014hha,2013ApJ...772..127T,2014MNRAS.439..503S,2014MNRAS.441.3177M,2015MNRAS.447...49S,Sekiguchi:2015dma,2015PhRvD..91l4021F,Sekiguchi:2016bjd,Radice:2016dwd,Radice:2021jtw,2022arXiv220504487K,Sun:2022vri,Werneck:2022exo}. However, the moment schemes require an auxiliary closure relation for the higher moments to derive the system equations in a closed form. For the limited accuracy of the closure relation models~\cite{Murchikova:2017zsy}, the moment schemes do not necessarily provide a solution which converges to the correct solution of the full radiation-transfer equations (but see~\cite{Izquierdo:2023fub,Foucart:2017mbt} for the improved method for modeling the closure relation). It should also be noted that the moment schemes employed for NS merger simulations are often energy-integrated, and thus, the information of energy distribution is lost in these simulations (see Refs.~\cite{Just:2015fda,Kuroda:2015bta,Roberts:2016lzn,Cheong:2024buu} for multi-energy moment schemes). Therefore, it is not guaranteed that the results derived from moment schemes are always quantitatively accurate. %\ms{"M1 scheme" is not defined so changed to "moment scheme". }

The recent significant progress of computer resources and numerical techniques have made it possible to directly solve radiation-transfer equations by the full discretization of a radiation field ~\cite[e.g.,][]{2014ApJS..214...16N,2017ApJS..229...42N,2014ApJS..213....7J,2022arXiv220906240J,2016ApJ...818..162O,2020ApJ...901...96A}. However, the size and resolution of the problems that can be solved are still limited for such approaches. As an alternative approach for directly solving the radiation-transfer equation, recently, radiation hydrodynamics codes based on the Monte Carlo (MC) scheme are developed by several groups~\citep{2012ApJ...755..111A,2015ApJS..217....9R,2015ApJ...807...31R,2018MNRAS.475.4186F,2019PhRvD.100b3008M,2019ApJS..241...30M,Foucart:2020qjb,2021ApJ...920...82F,2022ApJ...933..226R} (see also~\cite{Richers:2015lma,Sumiyoshi:2020bdh}). The MC-based methods have the advantage that the solution obtained by the MC scheme manifestly converges to the solution of radiative transfer equation in the limit of large packet numbers. Moreover, the energy dependence and the complicated angular dependence as well as the relativistic effects in radiative transfer can be incorporated in a straightforward manner. While there are several drawbacks in the MC approach, such as the slow convergence of the statistical error of the MC packets (the ``MC shot noise''), the study based on the MC scheme and the comparison with the previous study will provide important insights for understanding the possible systematic errors in the results derived by the approximated radiative transfer schemes.

In fact, quantitative differences between the results from the MC and moment schemes are pointed out in Refs.~\cite{2018PhRvD..98f3007F,Foucart:2020qjb,Foucart:2024npn}. While the neutrino luminosity obtained by moment schemes agrees with that by a MC scheme within $\approx 10\%$-$30\%$, nearly 50\% disagreement in the angular dependence of the neutrino flux is present, and this could lead the errors in the neutrino-antineutrino pair annihilation rate in a NS merger simulation by a factor of 2--3. However, most of the studies based on MC schemes follow the evolution of the system only for a short time scale ($\sim$10\,ms) (but see~\cite{Sprouse:2023cdm} for a 1.2 s BH-torus simulation with a MC scheme), and the impact of the radiative transfer scheme on the dynamics and outcomes from the merger remnant, for which the evolution of the time scale of the system is $\agt 1\,{\rm s}$, is not comprehended yet.

The determination of the neutrino distribution functions by directly solving the radiative transfer equation also can contribute to understanding the possible flavor conversion of neutrinos. In particular, it is pointed out that the neutrino fast flavor instability (FFI) can take place ubiquitously in the post-merger system and significantly modify the resulting composition of the ejected matter~\cite{Wu:2017qpc,Just:2022flt}. Since the FFI is due to quantum effects,  solving the quantum kinematic equation is required for the detailed analysis, while directly solving the FFI is also challenging for its extremely short time scale ($\sim 1\,{\rm ns},$~\cite{Richers:2022dqa}; see also~\cite{Kato:2021cjf,Froustey:2023skf} for the method based on MC and moment schemes based on quantum kinematic equations). However, linear stability analysis has shown that solving neutrino distribution functions in the classical level still can be useful to indicate the possible location in which the FFI can occur (see also~\cite{Just:2022flt}).

In this paper, we present our new general relativistic MC-based neutrino radiation viscous-hydrodynamics code designed to solve axisymmetric systems. In particular, we present several improvements to the code from previous studies. We then apply our code to a BH-torus system with a BH mass of $3\,M_\odot$, BH dimmensionless spin of 0.8, and a torus mass of $0.1\,M_\odot$, which mimics a post-merger remnant of a binary NS merger in the case that the massive NS collapses to a BH in a short time scale ($\sim10\,{\rm ms}$). We follow the evolution of the BH torus system up to more than $1\,{\rm s}$. We note that this is the first study to perform radiation viscous-hydrodynamics simulations for BH-torus systems dynamically taking into account non-thermal pair annihilation by employing an MC-based scheme (see, e.g.,~\cite{Just:2015fda,Kuroda:2015bta,Fujibayashi:2017xsz} for the study based on moment schemes dynamically taking into account the non-thermal pair process). We find that the system evolution and the various key quantities, such as neutrino luminosity, ejecta mass, torus $Y_e$, and pair annihilation luminosity, are broadly in agreement with the results of the previous studies in which the leakage or moment schemes are employed~\cite[e.g.,][]{Fujibayashi:2020qda,Just:2021cls} (and also the works which employ MC schemes~\cite{Richers:2015lma,Miller:2019dpt,2020ApJ...902...66M,Foucart:2022kon,Sprouse:2023cdm,Mukhopadhyay:2024zzl}). We also find that $\nu_e{\bar \nu}_e$ pair annihilation can launch a relativistic outflow for a time scale of $\sim 0.1\,{\rm s}$, and it can be energetic enough to explain some of short gamma-ray bursts~\cite{Nakar:2007yr} and the precursors~\cite{2010ApJ...723.1711T,Xiao:2022quv}. Finally, we demonstrate that our code enables to directly calculate the indicators of FFI introduced in~\cite{Richers:2022dqa}. We show a strong indication that FFI can take place particularly around the equatorial region of the torus, which is broadly in agreement with the previous study~\cite{Mukhopadhyay:2024zzl}.
%\ms{Is it right? Isotorpic energy may be large enough.}

This paper is organized as follows: In Sec.~\ref{sec:method}, we describe the formulation and methods employed in our code. In Sec.~\ref{sec:model}, we present our model setup of a BH-torus system studied in this paper. In Sec.~\ref{sec:diag}, we describe the definitions of several key quantities used for presenting the results. Sec.~\ref{sec:res} presents the results of the simulations for a BH-torus system. Finally, Sec.~\ref{sec:sum} is devoted to a summary of this paper. Throughout this paper, $c$ and $G$ denote the speed of light and gravitational constant, respectively, and the units of $c=G=1$ are employed unless otherwise stated.

\section{Formulation and Methods}\label{sec:method}
In this section, we describe the method implemented in our MC-based neutrino radiation viscous hydrodynamics code. Before presenting the details of the formulation and methods, we summarize the important assumptions and simplifications imposed in our code. 

First, the axisymmetry and equatorial plane symmetry are imposed in our code to reduce the computational cost. This is motivated by the fact that, the merger remnant typically relaxes to a nearly axisymmetric structure within the dynamical time scale ($\sim 10\,{\rm ms}$). We note that, however, the non-axisymmetric structure in the remnant torus and fall back tail can remain even for longer time scale and may play an important role in the dynamics particularly for unequal-binaries (see also~\cite{Fernandez:2014bra,Fernandez:2016sbf} in the context of BH-NS binaries). Hence, we should note that such effects are not taken into account in our simulations. 

Second, a fixed space-time metric is employed for solving viscous-hydrodynamics and radiative transfer in this paper. This is a reasonable simplification for a BH-torus system studied in this paper, of which torus mass is much smaller ($0.1\,M_\odot$) than the BH mass ($3\,M_\odot$). This simplification dramatically reduces the computational cost for the system with a BH because it requires a large resolution for the stable dynamical evolution~\cite{Fujibayashi:2020qda,Fujibayashi:2020jfr}. However, we should note that the long-term evolution of the BH, such as the increase in the BH mass and spin, may also have certain quantitative influences on the results, which are not taken into account by the space-time metric to be fixed.

The third is the simplification in the microphysics. In this paper, as a first step toward developing our neutrino radiation hydrodynamics code, the effect of the finite electron mass is consistently neglected for the equation of state (EoS) and neutrino interaction rates. This allows us to describe the EoS and neutrino interaction rates with simple analytical expressions and to focus more on checking that the matter-radiation interaction is correctly solved. This simplification is also qualitatively reasonable in the BH-torus system studied in this paper since the typical matter temperature is above $1\,{\rm MeV}$, although this simplification may cause some quantitative differences in the results (see Sec.~\ref{sec:res}). We note that the mass difference between proton and neutron, and binding energy of $\alpha$ particles, which have the same order of magnitude with the electron mass, are taken into account because it is important to realize the similar configuration of the initial data for BH-torus simulations as in~Ref.~\cite{Fujibayashi:2020qda}.

\subsection{Viscous hydrodynamics}
In a strongly magnetized hot and dense accretion torus formed after NS mergers, angular momentum transport is likely to be induced effectively by a magnetohydrodynamics process~\cite{Balbus:1998ja}. In this work, we approximately describe this process by viscous hydrodynamics following the formulation of~\cite{Shibata:2017jyf}. The basic equations for the viscous hydrodynamics are formulated in the framework of the 3+1 decomposition of the space-time. In the 3+1 formulation, the metric tensor $g_{\mu\nu}$ is decomposed as
\begin{align}
	ds^2&=g_{\mu\nu}dx^\mu dx^\nu\nonumber\\
	&=-\alpha^2dt^2+\gamma_{ij}\left(dx^i+\beta^idt\right)\left(dx^j+\beta^jdt\right),\label{eq:metric}
\end{align}
where $\mu$ and $\nu$ denote the space-time indices, $i$ and $j$ the spatial indices, $\alpha$, $\beta^i$, and $\gamma_{ij}$ the lapse function, shift vector, and spatial metric, respectively. In this work we employ the fixed background metric of the rotating BH described in the Kerr-Schild coordinates.

Following~\cite{Shibata:2017jyf}, the energy-momentum tensor of a viscous fluid is written as
\begin{align}
	T^{\mu\nu}_{\rm fl}=\rho hu^\mu u^\mu+Pg^{\mu\nu}-\rho h \nu \tau^0_{\mu\nu},
\end{align}
where $\rho$, $h$, $u^\mu$, $P$, $\tau^0_{\rm \mu\nu}$, and $\nu$ denote the baryon mass density, specific enthalpy, four-velocity, pressure, stress tensor, and shear viscous coefficient, respectively. The equations of energy-momentum conservation and the continuity equation are given by
\begin{align}
	\gamma_{\nu i}\nabla_\mu T^{\mu\nu}_{\rm fl}&=\gamma_{\nu i} G^\nu\label{eq:eom}\\
	n_\nu \nabla_\mu T^{\mu\nu}_{\rm fl}&=n_\nu  G^\nu\label{eq:eoe}\\
	\nabla_\mu \left(\rho u^\mu\right)&=0,\label{eq:eoc}
\end{align}
with the covariant derivative, $\nabla_\mu$. Here, $n_\nu=-\alpha\nabla_\nu t$, $\gamma_{\mu\nu}=g_{\mu\nu}+n_\mu n_\nu$, and $G^\mu$ denotes the radiation four-force density. 

In presence of the time-like Killing vector, $t^\mu=\alpha n^\mu+\beta^\mu$, instead of solving the energy equation of Eq.~\eqref{eq:eoe}, we solve the conservation equation for the energy measured by the asymptotic observer given by
\begin{align}
    \nabla_\mu T^\mu_{{\rm fl},t}&= \frac{1}{\sqrt{-g}} \partial_\mu (\sqrt{-g} T^\mu_{{\rm fl},t})= G_t,\label{eq:eoecsv}
\end{align}
where $g$ is the determinant of $g_{\mu\nu}$. 

The time evolution of the viscous tensor is given by solving \cite{Shibata:2017jyf}
\begin{align}
    {\cal L}_u\tau_{\mu\nu}=-\zeta\tau^0_{\mu\nu},\label{eq:evtau}
\end{align}
where $h_{\mu\nu}=g_{\mu\nu}+u_\mu u_\nu$, $\tau_{\mu\nu}=\tau^0_{\mu\nu}-\zeta h_{\mu\nu}$, ${\cal L}_u$ denotes the Lie derivative with respect to $u^\mu$, and $\zeta$ denotes a non-zero constant which describes the time scale for the viscous tensor to relax to the local shear tensor.

During our calculation, we find that the component values of $\tau_{\mu\nu}$ sometimes accidentally become very large in the vicinity of the event horizon, and cause numerical instability in the simulation. To avoid the numerical instability caused by the unphysical increase of $\tau_{\mu\nu}$, we modify the evolution equation of the viscous tensor, Eq.~\eqref{eq:evtau}, by introducing a limiter term:
\begin{align}
    {\cal L}_u\tau_{\mu\nu}&=-\zeta\tau^0_{\mu\nu}-{\rm max}\left(\left|\tau^0\right|-\tau^0_{\rm max},0\right)\tau^0_{\mu\nu}.\label{eq:evtau2}
\end{align}
Here, $\left|\tau^0\right|=\sqrt{\tau^{0,\mu\nu}\tau^0_{\mu\nu}}$ and $\tau^0_{\rm max}$ denotes a constant parameter. After each step of the time evolution, we also normalize $\tau^0_{\mu\nu}$ so that $\left|\tau^0\right|$ is smaller than $\tau^0_{\rm max}$ (see~\cite{Fragile:2018xee} for the similar prescription). By these prescriptions, we find that the simulation becomes numerically stable by keeping the component values of $\tau^0_{\mu\nu}$ to be always smaller than $\tau^0_{\rm max}$, while the evolution of $\tau^0_{\mu\nu}$ with $\left|\tau^0\right|\leq\tau^0_{\rm max}$ kept the same. In the present BH-torus simulations in this work, we set $\tau^0_{\rm max}=3/M_{\rm BH}$, of which value is much larger than the physical value of the viscous tensor.

The electron fraction ($Y_e$) evolution is given by
\begin{align}
    \nabla_\mu \left(\rho Y_e u^\mu\right)&=\Lambda_e,\label{eq:ye1}
\end{align}
with $\Lambda_e$ being the electron number change rate density due to the matter-radiation interaction. By numerically solving Eq.~\eqref{eq:ye1} explicitly, the value of $Y_e$ sometimes goes out from the proper range ($\left[0,1\right]$) particularly in the optically thick region. To prevent this problem, we introduce a new variable $Y_e^{\rm res}$ and solve the following equations, which are equivalent to solving Eq.~(\ref{eq:ye1}) in the limit of infinitesimally small value of $\tau_{\rm res}$:
\begin{align}
    \nabla_\mu \left(\rho Y_e^{\rm res} u^\mu\right)&=\Lambda_e-\frac{\rho Y_e^{\rm res}}{\tau_{\rm res}},\label{eq:ye2res}\\
    \nabla_\mu \left(\rho Y_e u^\mu\right)&=\frac{\rho Y_e^{\rm res}}{\tau_{\rm res}}.\label{eq:ye2}
\end{align}
%\ms{Has this prescription developed by yourself? Or, do you have any reference to be cited? Anyway, it would be better to mention that $Y_e^{res}$ is usually zero; otherwise Eqs. (10)+(11) is not equal to (9).}\rednote{This has been developed by myself. I added the comment on the value for $Y_e^{\rm res}$ below.}
Here $\tau_{\rm res}={\rm max}\left(\left|Y_e^{\rm res}\right|/\Delta Y_e^{\rm tol},1\right)\Delta t$. We also modify the values of $Y_e$ and $Y_e^{\rm res}$ by
\begin{align}
    Y_e&\rightarrow {\rm min}\left[{\rm max}\left(Y_e,0\right),1\right],\nonumber\\
    Y_e^{\rm res}&\rightarrow Y_e^{\rm res}+Y_e-{\rm min}\left[{\rm max}\left(Y_e,0\right),1\right],
\end{align}
at the time that the $Y_e$ value leaves the range of $\left[0,1\right]$. By this prescription, the change in the value of $Y_e$ in each time step will be limited within $\sim \Delta Y_e^{\rm tol}$, and furthermore, the total lepton number of the system (including that in stored in $Y_e^{\rm res}$) is guaranteed to be conserved. In this study, we employ $\Delta Y_e^{\rm tol}=0.1$. We note that this prescription is justified in our simulations because the dynamical time scale of the system is much longer than the time interval of the numerical evolution. In fact, the mass averaged value of $Y_e^{\rm res}$ is found to be smaller than $10^{-4}$ for the present BH-torus simulations.

In our study, we impose the axisymmetry and equatorial symmetry to the system. Equations~\eqref{eq:eom},~\eqref{eq:eoc},~\eqref{eq:eoecsv},~\eqref{eq:evtau2},~\eqref{eq:ye2}, and~\eqref{eq:ye2res} are solved in the cylindrical coordinate system by employing a Kurganov-Tadmor scheme~\citep{2000JCoPh.160..241K} with a piecewise parabolic reconstruction for the quantities of cell interfaces and the minmod filter for the flux-limiter. The primitive recovery procedure is done by employing the method of Ref.~\cite{Kastaun:2020uxr} but with a small modification to take the viscous terms into account.

\subsection{Radiative transfer}
In this study, we numerically solve neutrino radiation fields by a MC scheme essentially in the same way as described in~\cite{Kawaguchi:2022tae}. In the MC scheme, neutrino radiation fields are described by the sets of particles (which we refer to as MC packets) of which each represents the set of neutrinos with certain energy and momentum. Each MC packet is created following the local deposition rate and propagates along the geodesic during the time evolution. At the same time, MC packets probabilistically experience the change in the energy-momentum and neutrino numbers or are removed from the system following the interaction cross-section to the matter field. The radiative-feedback to the matter field is determined by locally summing up the energy-momentum and neutrino number changes of the MC packets in each hydrodynamics cell. Our code employs a higher-order scheme introduced in~\cite{Kawaguchi:2022tae}, by which the 2nd order convergence both in time and space is realized in the limit of large MC packets.

The number of MC packets created in each time step in each hydrodynamics cell is determined in the same way as in the previous study: the number of the created MC packets is tuned so that, for each neutrino species, the fluid rest-frame radiation energy in the optically thick cell is resolved by a desired target number of MC packets, $N_{\rm trg}$, (for the case that the continuous absorption method is turned off~\cite{Kawaguchi:2022tae}) in thermal equilibrium. To reduce the MC shot-noise (statistical error), we also employ the continuous absorption method with the threshold parameter of $r_{\rm abs}$ (see~\cite{Kawaguchi:2022tae} for details). Note that, with the continuous absorption method, the fluid rest-frame radiation energy in the optically thick cell is typically resolved by $N_{\rm trg}/r_{\rm abs}$ MC packets. 

To suppress the number of MC packets which are energetically unimportant, we employ the numerical prescription of ``residual packets'' introduced in our previous paper~\cite{Kawaguchi:2022tae}. In this prescription, MC packets are created with a flag of "the residual packet" in the cell where only a small number of MC packets are created ($\leq 1$). During the evolution of a radiation field, the residual packets are evolved in the same way as for the normal packets, but at the end of the evolution of the radiation field, the residual packets are removed from the computation. The total laboratory-frame energy, momentum, and neutrino number of the removed MC packets are recorded for each cell in which the packets were located. At the beginning of the next radiation-field evolution, 2 residual packets are recreated in the center of each cell so that their total laboratory-frame energy, momentum, and neutrino number agree with those recorded in the last step. This allows us to avoid increasing too many low-energy MC packets, while at the same time to ensure total energy-momentum and lepton-number conservation, which can be important to follow the long-term evolution of the system quantitatively. This prescription helps the total MC packets to be reduced by a factor of $\approx 3$ in the present BH-torus simulations.

One drawback of this prescription is that the distribution functions may be slightly less accurately derived. This is because, while the energy-momentum and lepton-number are conserved, the information of the detailed energy and angular distribution are lost during the removal and recreation of the residual MC packets. In fact, the MC packets with the residual flag tend to create small bumps in the distribution functions at the average energy determined by the total energy and neutrino number recorded in the end of each evolution step (see Fig.~\ref{fig:nu_dist}). Nevertheless, the radiation energy of MC packets with residual-packet flag is always less than $\approx 10\%$ compared to the total radiation energy and MC packets with the residual flag are more located in optically thin region. We also checked that the results of a BH-torus simulation are essentially unchanged at least up to $\approx 0.1\,{\rm s}$ regardless of whether the prescription is used or not. For example, the differences in the mass averaged $Y_e$ value with/without this prescription is found to be smaller than $\sim 0.1\%$. Hence, we consider that the effect of this prescription to the dynamics of the system is minor.

To stably solve the region in which the emission/absorption time scales are much smaller than the dynamical time scale of the system, we employ the so-called the implicit MC technique~\cite{1971JCoPh...8..313F}. We employ essentially the same method as that introduced in~\cite{Kawaguchi:2022tae} but we generalize it for the multi-species radiation fields (see App.~\ref{app:imc-msp} for details). In the implicit MC method, the absorption rate, $\alpha_{{\nu_i},{\rm abs}}$, scattering rate, $\sigma_{{\nu_i},{\rm sct}}$, and emissivity, $j_{{\nu_i},{\rm ems}}$, of the neutrino species, $\nu_i=(\nu_e,{\bar \nu}_e,\nu_x)$, are modified as
\begin{align}
    \alpha_{{\nu_i},{\rm abs}}&\rightarrow \left<g\right>\alpha_{{\nu_i},{\rm abs}}\nonumber\\
    \sigma_{{\nu_i},{\rm sct}}&\rightarrow \sigma_{{\nu_i},{\rm sct}}+(1-\left<g\right>)\alpha_{{\nu_i},{\rm abs}}\nonumber\\
    j_{{\nu_i},\rm ems}&\rightarrow g_{\nu_i}j_{{\nu_i},\rm ems}.
\end{align}
Here, $g_{\nu_i}$ is the Fleck-Cummings factor for the neutrino species, $\nu_i$, and $\left<g\right>$ is the emissivity average of $g_{\nu_i}$ given by $\left(\sum_{\nu_i}g_{\nu_i}j_{{\nu_i},{\rm ems}}\right)/\left(\sum_{\nu_i}j_{{\nu_i},{\rm ems}}\right)$. In this study $g_{\nu_i}$ is given by
\begin{align}
    g_{\nu_i}={\rm min}\left[\frac{1}{\left<\alpha_{\rm abs}\right>_{\nu_i}(1+\beta)\Delta t'},1\right],
\end{align}
where $\Delta t'$ is the time interval of the evolution measured in the comoving frame, $\left<\alpha_{\rm abs}\right>_{\nu_i}$ and $\beta$ denote the Planck-mean absorption rate of the neutrino species, $\nu_i$, and $\beta=\partial u_{\rm th}/\partial u_{\rm fl}|_{\rho,Y_e}$, respectively, with $u_{\rm th}$ and $u_{\rm fl}$ being the total energy density of neutrino radiation fields in the thermal equilibrium and internal energy density of the matter field, respectively. 
By this prescription, the time scale of emission and absorption are artificially elongated so that the energy and momentum equations are solved numerically stably. 

Unfortunately, we have to confess that the implicit MC method currently employed is still not always sufficient to guarantee the numerically stability. The reason is that only the energy equation is implicitly solved in the implicit MC method. Hence, the time scale of the matter-radiation interaction in the electron fraction evolution sometimes becomes much smaller than the dynamical time scale even if the implicit MC method is applied. For this purpose we also employ a prescription to limit the radiative feedback, which is described in Sec.~\ref{sec:crr}.
%\ms{This sentence is not very clear because the implict method should not care the short time scale; implicit methods are used to neglect short time scales.}\rednote{I modified the sentence above to explain that this method is an implicit method only for the energy equation (actually it is implicit only with respect to the source term: not for the advection term).}

\subsection{Equation of state}
This subsection describes the EoS which we employ in this paper. Note that we recover $c$ in this subsection to clarify the physical dimension. 

Following~\cite{Shibata:2007gp}, we assume that the EoS is determined by the contributions of relativistic particles composed of electrons and positrons, non-relativistic particles composed of free protons, free neutrons, and $\alpha$-particles, and photons, respectively. Then, the pressure is written as
\begin{align}
    P=P_e+P_{\rm ion}+P_\gamma,\label{eq:eos}
\end{align}
where $P_e$, $P_{\rm ion}$, and $P_\gamma$ denote the pressure of relativistic particles composed of electrons and positrons, non-relativistic particles composed of free protons, free neutrons, and $\alpha$-particles, and photons, respectively. For simplicity, we consistently set the electron mass $m_e$ to be $0$ under the assumption that the correction from the finite electron mass is negligible for sufficiently high temperature ($k_{\rm B} T\gg m_e c^2$ where $k_\mathrm{B}$ is the Boltzmann constant and $T$ the matter temperature). We note that this assumption is not satisfied after $\gtrsim 0.5\,{\rm s}$ for the models studied in this work (see Figs.~\ref{fig:snap_a005} and~\ref{fig:snap_a015}). Hence, our treatment of the EoS would be less accurate after that, and we leave the improvement of the EoS as a future work.

Under the asumption of $m_e=0$, the pressure of relativistic electrons and positrons is written as 
\begin{align}
    P_e&=\frac{1}{12 \pi^2\left(\hbar c\right)^3}\left(k_{\rm B} T\right)^4\left(\eta_e^4+2\pi^2 \eta_e^2+\frac{7 \pi^4}{15}\right),
\end{align}
where $\hbar$ is the reduced Planck constant and $\eta_e$ is the degeneracy parameter of electron, which is determined by solving
\begin{align}
    \frac{\rho Y_e}{m_u}&=\frac{1}{3 \pi^2\left(\hbar c\right)^3}\left(k_{\rm B} T\right)^3\left(\eta_e^3+\pi^2 \eta_e\right),
\end{align}
for given baryon mass density, $\rho$, temperature, $T$, and electron fraction, $Y_e$. 

Following previous studies (e.g., \cite{Shibata:2007gp}) the pressure of non-relativistic ions is given by
\begin{align}
    P_{\rm ion}&=\frac{\rho k_{\rm B} T}{m_u}\frac{1+3X_{\rm nuc}}{4},
\end{align}
where $X_{\rm nuc}$ denotes the free nucleon fraction. $X_{\rm nuc}$ is determined by solving the Saha's equation for free protons, free neutrons, and $\alpha$-particles under the assumption of nuclear statistical equilibrium (NSE).

Finally, the pressure of photons is given by
\begin{align}
    P_\gamma&=\frac{\pi^2}{45\left(\hbar c\right)^3}\left(k_{\rm B} T\right)^4.
\end{align}

The total specific internal energy is given by
\begin{align}
    \epsilon &= \epsilon_e+\epsilon_{\rm ion}+\epsilon_\gamma,\nonumber\\
    \epsilon_e&=3\frac{P_e}{\rho},\,\epsilon_{\rm ion}=\frac{3}{2}\frac{P_{\rm ion}}{\rho},\,\epsilon_\gamma=3\frac{P_\gamma}{\rho},
\end{align}
with $\epsilon_e$, $\epsilon_{\rm ion}$, and $\epsilon_\gamma$ denote the contribution from the relativistic particles composed of electrons and positrons, non-relativistic particle composed of free protons, free neutrons, and $\alpha$-particles, and photons, respectively.

The total specific enthalpy is given by
\begin{align}
    h=c^2+\delta+\epsilon +\frac{P}{\rho},\label{eq:enth}
\end{align}
where $\delta$ denotes the averaged specific binding energy of non-relativistic particles. $\delta$ is given as a function of $X_{\rm nuc}$ and $Y_e$ by
\begin{align}
    \frac{\delta}{c^2}=&\frac{m_p n_p+m_n n_n+m_\alpha n_\alpha}{m_u n_{\rm b}}-1\nonumber\\
    &=\frac{1}{m_u}[(m_p+m_n-2m_u)\nonumber\\
    &+(m_p-m_n)(2 Y_e-1)\nonumber\\
    &+ (m_\alpha-m_u)(1-X_{\rm nuc})],\label{eq:mexc}
\end{align}
where $n_p$, $n_n$, and $n_\alpha$ denote the number densities, $m_u$, $m_p$, $m_n$, and $m_\alpha$ denote the masses of the atomic unit, free protons, free neutrons, and $\alpha$-particles, respectively.

\subsection{Viscous parameters}
In this paper, we model the viscous coefficient by the so-called $\alpha$-viscosity description~\cite{1973A&A....24..337S}. Specifically, we set the viscous coefficient to be $\nu=\alpha_{\rm vis}c_s H_{\rm vis}$ following~\cite{Fujibayashi:2020qda}. Here, $\alpha_{\rm vis}$, $H_{\rm vis}$, and $c_s$ denote the dimensionless $\alpha$-viscous parameter, a scale height, and the sound speed, respectively. We vary  $\alpha_{\rm vis}$ from $0.05$ to $0.15$, and $H_{\rm vis}$ and $\zeta$ are set to be $2 M_{\rm BH}$ and $\approx 1/M_{\rm BH}$, respectively, following the previous study~\cite{Fujibayashi:2020qda}.

\subsection{Neutrino processes}
In this subsection, we describe our setups of neutrino processes considered in this work. The expressions for the interaction rates are taken from~\cite{Bruenn:1985en,2004StellarCollapse,Horowitz:2001xf}. Note that we again recover $c$ in this subsection. In the present work, we consider electron/positron captures by free protons/neutrons and electron-positron pair annihilation for neutrino emission. For absorption processes, electron–type neutrino/antineutrino absorption by free protons/neutrons and neutrino/antineutrino pair annihilation are considered. For scattering process, elastic scattering by free protons and neutrons is considered. Consistently with the EoS, for simplicity, we set the electron mass $m_e$ to be $0$ under the assumption that the correction from the finite electron mass is negligible for sufficiently high temperature ($k_{\rm B} T\gg m_e c^2$). As we mentioned in the previous subsection, this assumption is not satisfied after $\gtrsim 0.5\,{\rm s}$ for the models studied in this work (see Figs.~\ref{fig:snap_a005} and~\ref{fig:snap_a015}). While we expect that the effect of neglecting the electron mass could not be significant since the weak interaction time scale is nevertheless much longer than the simulation time scale for the matter temperature less than $1\,{\rm MeV}$ (see~\cite{Fujibayashi:2020qda,Just:2021cls}), the effect may be important for the quantitative prediction, and we leave the improvement as a future task.

\subsubsection{Electron/positron captures}
The neutrino/antineutrino emissivity of electron/positron captures without the Fermi-blocking correction, $j_{\rm ec}$ and $j_{\rm pc}$, respectively, are given by~\cite{Bruenn:1985en}
\begin{align}
    j_{\rm ec}(\omega_{\nu_e})&=\frac{G_{\rm F}^2 c}{\pi(\hbar c)^4}(g_V^2+3g_A^2)n_p W_{\bar M}(\omega_{\nu_e})\nonumber\\
    &\times(\omega_{\nu_e}+Q)^2\frac{1}{h^3}F_e(\omega_{\nu_e}+Q),\\
    j_{\rm pc}(\omega_{\bar \nu_e})&=\frac{G_{\rm F}^2 c}{\pi(\hbar c)^4}(g_V^2+3g_A^2)n_n W_M(\omega_{\bar \nu_e})\nonumber\\
    &\times( \omega_{\bar \nu_e}-Q)^2\frac{1}{h^3}{\bar F}_{e}(\omega_{\bar \nu_e}-Q)\Theta(\omega_{\bar \nu_e}-Q),
\end{align}
% \begin{align}
%     j_{\rm ec}&=\frac{G_{\rm F}^2 c}{\pi(\hbar c)^4}(g_V^2+3g_A^2)n_p W_{\bar M}(\omega_{\nu_e}+Q)\nonumber\\
%     &\times(\omega_{\nu_e}+Q)^2\frac{1}{h^3}F_e(\omega_{\nu_e}+Q),\\
%     j_{\rm pc}&=\frac{G_{\rm F}^2 c}{\pi(\hbar c)^4}(g_V^2+3g_A^2)n_n W_M({\bar \omega}_{\nu_e}-Q)\nonumber\\
%     &\times( {\bar \omega}_{\nu_e}-Q)^2\frac{1}{h^3}{\bar F}_{e}({\bar \omega}_{\nu_e}-Q)\Theta({\bar \omega}_{\nu_e}-Q),
% \end{align}
where $G_{\rm F}$, $g_V$, $g_A$, and $Q$ are the Fermi coupling constant, vector coupling strength, axial vector coupling strength, and the rest-mass energy difference between a free neutron and proton, respectively, and $\omega_{\nu_e}$ and $\omega_{{\bar \nu}_e}$ are the electron-neutrino and electron-antineutrino energies in the rest-frame of the fluid motion, respectively. $\Theta$ denotes the Heaviside step function. $W_{\bar M}(\omega_{\nu_e})$ and $W_M(\omega_{\bar \nu_e})$ are the corrections for weak magnetism and recoil introduced in~\cite{Horowitz:2001xf}. $F_e$ and ${\bar F}_e$ are the normalized Dirac distribution functions given by
\begin{align}
F_e(\omega_{\nu_e})=\frac{1}{e^{\frac{\omega_{\nu_e}}{k_{\rm B}T}-\eta_e}+1},
{\bar F}_{e}({\bar \omega}_{\nu_e})=\frac{1}{e^{\frac{{\bar \omega}_{\nu_e}}{k_{\rm B}T}+\eta_e}+1},
\end{align}
where $\eta_e$ is the degeneracy parameter of electrons.

Absorption rates of electron neutrinos and electron antineutrinos by free protons and neutrons, $\alpha_{{\rm abs},\nu_e}$ and $\alpha_{{\rm abs},{\bar \nu_e}}$, respectively, are obtained from the neutrino/antineutrino emissivity introduced above by using the Kirchhoff's law. Employing the so-called ``stimulated absorption'' prescription~\cite{2004StellarCollapse} to take the Fermi-blocking effect into account, the effective absorption rates, $\alpha_{{\rm abs},\nu_e}^*$ and $\alpha_{{\rm abs},{\bar \nu_e}}^*$ are given by
\begin{align}
\alpha_{{\rm abs},\nu_e}^*=\frac{h^3 j_{\rm ec}}{F^\beta_{\nu_e}},
\alpha_{{\rm abs},{\bar \nu_e}}^*=\frac{h^3 j_{\rm pc}}{{\bar F}^\beta_{\nu_e}},
\end{align}
where $F^\beta_{\nu_e}$ and ${\bar F}^\beta_{\nu_e}$ are the normalized Dirac distribution functions of neutrinos/antineutrinos in the $\beta$-equilibrium state given by
\begin{align}
F_{\nu_e}^\beta(\omega)=\frac{1}{e^{\frac{\omega_{\nu_e}}{k_{\rm B}T}-\eta^\beta_{\nu_e}}+1},{\bar F}_{\nu_e}^\beta({\bar \omega}_{\nu_e})=\frac{1}{e^{\frac{{\bar \omega}_{\nu_e}}{k_{\rm B}T}+\eta^\beta_{\nu_e}}+1},
\end{align}
with $\eta^\beta_{\nu_e}=\eta_e+(\mu_p-\mu_n)/k_\mathrm{B}T$ the degeneracy parameter of electron-neutrinos in the $\beta$-equilibrium. Here, $\mu_p$ and $\mu_n$ are the chemical potentials of free proton and neutron, respectively.

\subsubsection{Neutrino/antineutrino pair process}
In the following, we describe how we handle neutrino-antineutrino pair process in our simulation. In particular, we consider neutrino-antineutrino pair annihilation with taking the effect of the non-thermal distributions into account. We explain the method by focusing on the interaction rates for neutrinos; those for antineutrinos can be obtained easily by swapping the roles of neutrinos and antineutrinos. The absorption rate of neutrino/antineutrino annihilation process for neutrinos is given by~\cite{Bruenn:1985en},
\begin{align}
    \alpha^{\rm pair}({\bf Q})&=c\int d^3 {\bar {\bf Q}} {\bar f}({\bar {\bf Q}}) R^{\rm ann}\left(\omega,{\bar \omega},\mu\right),
\end{align}
where ${\bf Q}$ and ${\bar {\bf Q}}$ represent the spatial momentum of the neutrino and antineutrino, respectively. $\omega$, ${\bar \omega}$, and $\mu$ are the neutrino energy, antineutrino energy, and cosine of the angle between the spatial momentum of the neutrino and antineutrino, and ${\bar f}$ is the distribution function of antineutrinos. Note that ${\bf Q}$, ${\bar {\bf Q}}$, $\omega$, ${\bar \omega}$, and $\mu$ are defined in the fluid rest-frame. $R^{\rm ann}$ denotes the interaction kernel of neutrino/antineutrino pair annihilation defined by
\begin{align}
    R^{\rm ann}&\left(\omega,{\bar \omega},\mu\right)=\frac{2 G_{\rm F}^2 c^6}{(2\pi)^2 (\hbar c)^4\omega {\bar \omega}} \int \frac{d^3p}{E}\int \frac{d^3{\bar p}}{{\bar E}}\nonumber\\
    &\times \delta^4(q+{\bar q}-p-{\bar p})\times\left[1-F_e(E)\right]\left[1-{\bar F}_e({\bar E})\right]\nonumber\\
    &\times\left[(C_V+C_A)^2 {\bar p}^\mu q_\mu p^\nu{\bar q}_\nu+(C_V-C_A)^2 p^\mu q_\mu {\bar p}^\nu{\bar q}_\nu\right],\label{eq:rabs}
\end{align}
where $q$, ${\bar q}$, $p$, and ${\bar p}$ are the four-momentum of the neutrino, antineutrino, electron, and positron, respectively. $E$ and ${\bar E}$ are the electron and positron energies in the fluid rest-frame, respectively, $C_V=1/2+{\rm sin}^2\theta_{\rm W}$ for $\nu_e$ and ${\bar \nu}_e$, $C_V=-1/2+{\rm sin}^2\theta_{\rm W}$ for heavy-lepton type neutrinos, and $C_A=1/2$ with ${\rm sin}^2\theta_{\rm W}\approx0.2319$.

%and $F_e(E)$ and ${\bar F}_e({\bar E})$ are the normalized Dirac thermal distribution function of electrons and positrons given, respectively, by
%\begin{align}
%F_e(E)=\frac{1}{e^{\frac{E}{k_{\rm B}T}-\eta_e}+1},{\bar F}_{e}({\bar E})=\frac{1}{e^{\frac{\bar E}{k_{\rm B}T}+\eta_e}+1}, 
%\end{align}

In our present work, we approximate $R^{\rm ann}$ by employing fitting functions $\phi_1$ and $\phi_2$ as
\begin{align}
    R^{\rm ann}&\left(\omega,{\bar \omega},\mu\right)\approx R^{\rm ann,fit}\left(\omega,{\bar \omega},\mu\right)\nonumber\\
    &=\frac{2G_{\rm F}^2c^4}{3\pi(\hbar c)^4\omega{\bar \omega}}(C_V^2+C_A^2)q_\mu q_\nu {\bar q}^\mu {\bar q}^\nu\nonumber\\
    &\times\left[\phi_1\left(x\right)\phi_1\left({\bar x}\right)-\phi_2\left(x\right)\phi_2\left({\bar x}\right)\right],\label{eq:rabs_fit}
\end{align}
where $x=\omega/k_{\rm B}T$ and ${\bar x}={\bar \omega}/k_{\rm B}T$. We note that  $\phi_1$ and $\phi_2$ are functions dependent also on $\eta_e$. The fitting functions $\phi_1$ and $\phi_2$ are optimized so that $R^{\rm ann,fit}$ reproduces the values of $R^{\rm ann}$ in the wide range of parameters (see App.~\ref{app:pair} for the motivation of the function form, optimization of $\phi_1$ and $\phi_2$, and the accuracy of the model).  

Employing $R^{\rm ann,fit}$, the neutrino absorption rate is simplified as 
\begin{align}
    \alpha^{\rm pair}&({\bf Q})=\nonumber\\
    &\frac{2G_{\rm F}^2c^3}{3\pi(\hbar c)^4} (C_V^2+C_A^2)\left[\phi_1 (x) {\bar \Phi}_1^{\mu\nu}-\phi_2 (x) {\bar \Phi}_2^{\mu\nu} \right]\frac{q_\mu q_\nu}{\omega}\label{eq:abs_pair}
\end{align}
where ${\bar \Phi}_i^{\mu\nu}$ is defined by 
\begin{align}
    {\bar \Phi}_i^{\mu\nu}=c^2\int d^3 {\bar {\bf Q}} {\bar f}({\bar {\bf Q}}) \phi_i\left({\bar x}\right)\frac{{\bar q}^\mu{\bar q}^\nu}{{\bar \omega}}.
\end{align}
We note that $\phi_1$ and $\phi_2$ are given so that $\phi_1{\bar \Phi}_1^{\mu\nu}-\phi_2{\bar \Phi}_2^{\mu\nu}$ reduces to the energy momentum tensor of the neutrino radiation field in the limit of non-degenerate electrons and high neutrino energies (see App.~\ref{app:pair}), and $\alpha^{\rm pair}({\bf Q})$ agrees with the exact expression for the neutrino absorption rate in such a limit~\cite{Fujibayashi:2017xsz,2021ApJ...920...82F}.

Since ${\bar \Phi}_i^{\mu\nu}$ is independent of the neutrino energy and momentum, practically in the simulations, ${\bar \Phi}_i^{\mu\nu}$ is calculated before the evolution of neutrino MC packets using the values obtained in the previous sub-step. The emissivity of pair process is also determined by employing the neutrino and antineutrino number densities obtained in the previous sub-step. This enables us to effectively treat pair annihilation as the one-body interaction with taking those non-thermal distribution effects into account, and hence, the neutrino MC packets can be evolved independently of the evolution of the antineutrino MC packets. However, as a drawback, the balance between neutrinos and antineutrinos that have been involved in pair process is not numerically guaranteed. This is problematic because the total numbers of neutrinos and antineutrinos which are emitted or absorbed through pair process should be the same. Such an imbalance can induce an unphysical $Y_e$ evolution in the matter field. To address this problem, a correction procedure described in Sec.~\ref{sec:crr} is applied at the end of each sub-step.

The neutrino emissivity by electron-positron pair annihilation is also computed employing $R^{\rm ann,fit}$ but with the simplification that the antineutrinos involved in the Fermi-blocking factor are in thermal equilibrium. While in principle we can employ the antineutrino distribution function directly from the calculated MC packets, we apply this simplification because the absolute value of the neutrino degeneracy parameter for $\nu_e$ and ${\bar \nu}_e$ is not high ($\leq 0.1$) for the problems studied in this paper (we note, however, this is not the case in the presence of a remnant massive NS).  This treatment significantly simplifies the numerical implementation by allowing the application of the Kirchhoff's law. The neutrino absorption rate of pair annihilation for the case that antineutrinos are in thermal equilibrium is given by%\ms{ [This sentence is something wrong.]}
\begin{align}
    \alpha^{\rm pair,th}&(\omega)=\nonumber\\
    &\frac{2G_{\rm F}^2c^3}{3\pi(\hbar c)^4} (C_V^2+C_A^2)\left[\phi_1 (x) {\bar \Psi}_1-\phi_2 (x) {\bar \Psi}_2 \right]\omega,
\end{align}
where $\Psi_i$ is define by
\begin{align}
    {\bar \Psi}_i=\frac{16\pi}{3c^3}\int d {\bar \omega} {\bar f}^{\rm th}({\bar \omega}) {\bar \omega}^3 \phi_i\left({\bar x}\right).
\end{align}
Here, ${\bar f}^{\rm th}({\bar \omega})$ is the thermal distribution function of antineutrinos which is given by
\begin{align}
{\bar f}^{\rm th}({\bar \omega})&=\frac{1}{h^3}{\bar F}^{\rm th}({\bar \omega})\nonumber\\
&=\frac{1}{h^3}\frac{1}{e^{\frac{{\bar \omega}}{k_{\rm B}T}+\eta_\nu}+1},
\end{align}
where $\eta_\nu$ is the degeneracy parameter of neutrinos obtained by the local neutrino/antineutrino number density and temperature. We note that ${\bar F}^{\rm th}$ is the thermal distribution realized under the fixed local neutrino/antineutrino number difference (and hence ${\bar F}^{\rm th}\neq{\bar F}^{\beta}$) because  the changes in the local neutrino/antineutrino numbers are balanced for pair process. Then, the neutrino emissivity by electron-positron pair annihilation, $j_{{\rm pair}}$, can be calculated with the ``stimulated absorption'' correction by 
\begin{align}
j_{{\rm pair}}(\omega)=\alpha^{\rm pair,th}&(\omega)\frac{1}{h^3} \frac{F^{\rm th}(\omega)}{1-F^{\rm th}(\omega)},
\end{align}
where
\begin{align}
F^{\rm th}(\omega)=\frac{1}{e^{\frac{\omega}{k_{\rm B}T}-\eta_\nu}+1}.
\end{align}
Finally, employing the expression for $j_{{\rm pair}}$, the effective absorption rates of neutrinos by pair annihilation including the stimulated absorption correction, $\alpha^{*,\rm pair}({\bf Q})$, is given by
\begin{align}
\alpha^{*,\rm pair}({\bf Q})=\alpha^{\rm pair}({\bf Q})+h^3j_{{\rm pair}}(\omega).
\end{align}

\subsubsection{Elastic scattering by free protons and neutrons}
The differential cross-section of elastic scattering by free protons and neutrons is given by~\cite{2004StellarCollapse}
\begin{align}
    \frac{d\sigma_s}{d\mu}=\sigma_s^0\left(1+\delta_s\mu\right)~(s=p,n),
\end{align}
where $\mu={\rm cos}\,\theta_{\rm sct}$ with $\theta_{\rm sct}$ being the angle between the spatial momenta in the fluid rest-frame before and after scattering. $\sigma^0_s$ denotes the total cross-section of the scattering process, which is given by
\begin{align}
    \sigma^0_p=\frac{G_F^2}{\pi(\hbar c)^4}\left(\frac{g_V^2+3 g_A^2}{4}+4{\rm sin}^4\theta_W-2{\rm sin}^2\theta_W\right)\omega^2
\end{align}
for scattering by protons and
\begin{align}
    \sigma^0_n=\frac{G_F^2}{\pi(\hbar c)^4}\frac{g_V^2+3 g_A^2}{4}\omega^2
\end{align}
by neutrons. $\delta_s$ is $\approx-0.2$ and $\approx-0.1$ for scattering by protons and neutrons, respectively~\cite{2004StellarCollapse}.

\subsubsection{Limiter to the radiative feedback}\label{sec:crr}
By employing the implicit MC method, the emissivity and absorption rates of neutrinos in each hydrodynamics cell are controlled so that the energy-momentum equations are solved numerically stably. However, even under this prescription, the source term in the electron fraction evolution equation sometimes becomes very large because the time scales of the energy-momentum change and electron number change are not necessarily the same. In such a situation, the correct thermal equilibrium is not realized or the calculation even becomes numerically unstable. 

Another serious problem can arise for the case that $\nu_e{\bar \nu}_e$ pair process is considered in the calculation. The net electron number change has to be zero for each $\nu_e{\bar \nu}_e$ pair process. However, neutrino and antineutrino are evolved effectively separately in our code, and hence, the balance between the radiative feedback on the electron number from neutrinos and antineutrinos is not numerically guaranteed. This induces an unphysical change in the matter electron fraction, which avoids the system to reach a correct thermal equilibrium state.

To overcome these problems, in each sub-step of the time evolution, we employ a limiter to the radiative feedback so that a) the net electron number change induced by $\nu_e{\bar \nu}_e$ pair process is guaranteed to be zero and b) the absolute net change in the electron fraction is limited to be within a given value ($\Delta Y_{e,{\rm tol}}$).

First, we describe our prescription to achieve the condition a). After solving the radiation fields in each sub-step of time evolution, the radiative feedbacks from electron neutrinos and electron antineutrinos to the matter energy-momentum equations, $G_\mu^{\nu_e}$ and $G_\mu^{{\bar\nu}_e}$, respectively, and electron number equation, $\Lambda^{\nu_e}$ and $\Lambda^{{\bar\nu}_e}$, respectively, are recorded for each hydrodynamics cell. We also separately record the fractions of these radiative feedbacks due to pair process: $G_\mu^{\nu_e,{\rm pair}}$ and $G_\mu^{{\bar\nu}_e,{\rm pair}}$, respectively, for the matter energy-momentum equations, and $\Lambda^{\nu_e,{\rm pair}}$ and $\Lambda^{{\bar\nu}_e,{\rm pair}}$, respectively, for the electron number equation. 

The condition of a) can be expressed by $\Lambda_{e,\nu_e}^{\rm pair}+\Lambda_{e,{\bar \nu}_e}^{\rm pair}=0$. To achieve this condition for every hydrodynamics cells, we modify the weight of the MC packets for electron neutrinos and electron antineutrinos at the end of the evolution. The number densities of electron neutrinos and electron antineutrinos in a given hydrodynamics cell, $n_{\nu_e}$ and $n_{{\bar \nu}_e}$, respectively, are given by
\begin{align}
n_{\nu_e}=\frac{1}{\sqrt{-g}\Delta^3 x}\sum_{k} w_k^{\nu_e},~n_{{\bar \nu}_e}=\frac{1}{\sqrt{-g}\Delta^3 x}\sum_{k} w_k^{{\bar \nu}_e},
\end{align}
where $w_k^{\nu_e}$ and $w_k^{{\bar \nu}_e}$ are the weights of electron neutrino and electron antineutrino MC packets, respectively, and $\Delta^3 x$ denotes the coordinate spatial volume of the hydrodynamics cell. The summations are taken for the MC packets located in the given cell. In our prescription, we modify $\Lambda^{\nu_e}$ and $\Lambda^{{\bar\nu}_e}$ with correction factors, $f_{\rm crr}^{\nu_e}$ and $f_{\rm crr}^{{\bar \nu}_e}$ by
\begin{align}
\Lambda^{\nu_e}&\rightarrow\Lambda^{\nu_e}+f_{\rm crr}^{\nu_e}\frac{m_u n_{\nu_e}}{\rho\Delta t},\nonumber\\
\Lambda^{{\bar \nu}_e}&\rightarrow\Lambda^{{\bar \nu}_e}-f_{\rm crr}^{{\bar \nu}_e}\frac{m_u n_{{\bar \nu}_e}}{\rho\Delta t},
\end{align}
with also modifying the weights of the MC packets by
\begin{align}
w_k^{\nu_e}\rightarrow (1-f_{\rm crr}^{\nu_e})w_k^{\nu_e},~w_k^{{\bar \nu}_e}\rightarrow (1-f_{\rm crr}^{{\bar \nu}_e})w_k^{{\bar \nu}_e},
\end{align}
so that the total lepton number of the cell is conserved. We note that $f_{\rm crr}^{\nu_e}$ and $f_{\rm crr}^{{\bar \nu}_e}$ should manifestly be less than unity. We further restrict the values to be within $[-0.9,0.9]$ so that to suppress the artifact due to this prescription. At the same time, we also modify $G_\mu^{\nu_e}$ and $G_\mu^{{\bar\nu}_e}$ by
\begin{align}
G_\mu^{\nu_e}&\rightarrow G_\mu^{\nu_e}+f_{\rm crr}^{\nu_e}\frac{T_{{\rm rad},\mu}^{\nu_e,t}}{\Delta t},\nonumber\\
G_\mu^{{\bar\nu}_e}&\rightarrow G_\mu^{{\bar\nu}_e}+f_{\rm crr}^{{\bar \nu}_e}\frac{T_{{\rm rad},\mu}^{{\bar\nu}_e,t}}{\Delta t},
\end{align}
so that the total energy and momentum of the cell are conserved. Here, $T_{{\rm rad},\mu\nu}^{\nu_e}$ and $T_{{\rm rad},\mu\nu}^{{\bar\nu}_e}$ denote the energy momentum tensors of electron neutrinos and electron antineutrinos, respectively. The values of the correction factors, $f_{\rm crr}^{\nu_e}$ and $f_{\rm crr}^{{\bar \nu}_e}$, are chosen so that to satisfy
\begin{align}
\Lambda_{e,\nu_e}^{\rm pair}&+\Lambda_{e,{\bar \nu}_e}^{\rm pair}+f_{\rm crr}^{\nu_e}\frac{m_u n_{\nu_e}}{\rho\Delta t}-f_{\rm crr}^{{\bar \nu}_e}\frac{m_u n_{{\bar \nu}_e}}{\rho\Delta t}=0\label{eq:crrcond1}
\end{align}
while minimizing the change in the source term of the Laboratory-frame energy
\begin{align}
\left|\Delta G_{\rm crr}^t\right|=\left|f_{\rm crr}^{\nu_e}T_{\rm rad}^{\nu_e,tt}+f_{\rm crr}^{{\bar \nu}_e}T_{\rm rad}^{{\bar\nu}_e,tt}\right|.
\end{align}

Unfortunately, the solutions for $f_{\rm crr}^{\nu_e}$ and $f_{\rm crr}^{{\bar \nu}_e}$ do not always exist (for instance, for the case that $n_{\nu_e}=n_{{\bar \nu}_e}=0$ with $\Lambda_{e,\nu_e}^{\rm pair}+\Lambda_{e,{\bar \nu}_e}^{\rm pair}\ne0$). For such a case, we create new electron neutrino or electron antineutrino MC packets in the cell to balance $\Lambda^{\nu_e,{\rm pair}}$ and $\Lambda^{{\bar\nu}_e,{\rm pair}}$ by the radiative feedbacks induced by the MC packet creation. For this prescription, the neutrino energy of the newly created packets is basically chosen from the thermal distribution of the local temperature and neutrino/antineutrino number densities.

The prescription to achieve the condition b) is essentially the same as that for a), expect for that $f_{\rm crr}^{\nu_e}$ and $f_{\rm crr}^{{\bar \nu}_e}$ are chosen to satisfy
\begin{align}
\left|\Lambda_{e,\nu_e}+\Lambda_{e,{\bar \nu}_e}+f_{\rm crr}^{\nu_e}\frac{m_u n_{\nu_e}}{\rho\Delta t}-f_{\rm crr}^{{\bar \nu}_e}\frac{m_u n_{{\bar \nu}_e}}{\rho\Delta t}\right|\leq \frac{\Delta Y_{e,{\rm tol}}}{\Delta t}\label{eq:crrcond2}
\end{align}
under the condition minimizing the change in the source term of the Laboratory-frame energy $\left|\Delta G_{\rm crr}^t\right|$. In the case that no solution exist within the given ranges of $f_{\rm crr}^{\nu_e}$ and $f_{\rm crr}^{{\bar \nu}_e}$, we simply gave up to exactly satisfy Eq.~\eqref{eq:crrcond2} but employ the values of $f_{\rm crr}^{\nu_e}$ and $f_{\rm crr}^{{\bar \nu}_e}$ which the left-hand side of Eq.~\eqref{eq:crrcond2} is the smallest.

This prescription obviously induces artificial emission and absorption of neutrinos. However, we find that the values of $f_{\rm crr}^{\nu_e}$ and $f_{\rm crr}^{{\bar \nu}_e}$ are at most as large as $10^{-2}$ and typically much smaller than $10^{-3}$. Hence, we consider that the artifact due to this prescription is always minor.

\subsection{Parallelization}
Our current MC radiative hydrodynamics code is parallelized using a hybrid of MPI and OpenMP programming. The computation for hydrodynamics evolution is parallelized by dividing the computational domain equally among MPI nodes, and the computation of the domain assigned for each node is further parallelized by OpenMP. On the other hand, the computation for evolving the radiation field is parallelized packet-wise by assigning the MC packets to each OpenMP thread in each MPI node regardless of the MC packet positions. In order to evolve the MC packets located outside the hydrodynamics domain assigned to the MPI node, the thermodynamic quantities such as the baryon mass density, $\rho$, temperature, $T$, and electron fraction, $Y_e$, as well as the four-velocity information, $u^\mu$, in the entire computational domain are collected and shared among all MPI nodes after each sub-step of the hydrodynamics evolution. After all the evolution of the MC packets is finished, the radiation feedback to the matter field as well as the statistical information of the radiation field are summed up for each hydrodynamics cell, and they are collected and shared among the MPI nodes. Note that in this way the full information of the MC packets is not needed to be communicated among different MPI nodes. 

\subsection{Validation}

Our code has been validated from various aspects. Since the hydrodynamics solver and basic infrastructure of the radiative transfer solver are tested in our previous studies~\cite{Kawaguchi:2020vbf,Kawaguchi:2022tae}, in this paper, we focus on examining  the microphysics implementation and matter-radiation interaction in the presence of multiple neutrino species. We validate our implementation of the EoS and neutrino interaction rates by reproducing the emission equilibrium value of $Y_e$ discussed in the previous study~\cite{Just:2021cls}. We confirm by comparison with Fig.~1 in \cite{Just:2021cls} that the emission equilibrium value of $Y_e$ with the weak magnetism and recoil correction is reproduced by our code for the temperature above $2\,{\rm MeV}$. For the temperature below $2\,{\rm MeV}$ we find some deviation from their results due to the neglect of the electron mass effect and the limited NSE ensembles in our microphysics implementation. Nevertheless, the discrepancy in the equilibrium $Y_e$ value at $1\,{\rm MeV}$ is still within $\sim10\%$. We also note that the discrepancy below $1\,{\rm MeV}$ does not affect the present results, because the time scale of weak interaction is very long ($\gtrsim10\,{\rm s}$) compared to the simulation time (i.e., evolution time scale of the system). We then validate that our code correctly solves the thermalization of matter and radiation fields in one-zone tests in various situations (see App.~\ref{app:th-test}).  

\section{Model}\label{sec:model}
\subsection{Grid setups and simulation parameters}

In this paper, we solve a BH-torus system under the assumption of axisymmetry and equatorial-plane symmetry. The cylindrical coordinate system is employed for solving the viscous-hydrodynamics, and $x$ and $z$ are assigned to the cylindrical radius and the vertical coordinate, respectively. The grid is set to be non-uniform in both $x$ and $z$ directions, of which grid-spacing increasing outward with a constant rate of 1.0125. The innermost (and hence the finest) grid-spacing is determined so that for each coordinate the grid covers from 0 to $2500\,M_\odot \approx 3750\,{\rm km}$ with $320$ grid-cells, that is, $\Delta x_0\approx 0.6\,M_\odot\approx900\,{\rm m}$. (Note that the BH mass is $M_\mathrm{BH}=3M_\odot$ and thus $\Delta x_0 \approx 0.2 M_\mathrm{BH}$.) The time interval of the simulation, $\Delta t$, is determined by the Courant–Friedrichs–Lewy condition with respect to the finest grid-spacing, and in this work we set the Courant number to be 0.5.

For a numerically stable simulation, an atmosphere of which baryon mass density, temperature, and $Y_e$ are $10\,{\rm g/cm^3}$, $0.036\,{\rm MeV}$, and 0.5, respectively, is artificially added outside the torus in the initial data following~\cite{Fujibayashi:2020qda}. We also set the floor values for the baryon mass density and temperature, of which values are given by $1\,{\rm g/cm^2}\times {\rm min}\left[(r/300\,M_\odot)^{-1},1\right]$ and $300\,{\rm eV}$, respectively. We note that the maximum baryon mass density and temperature of the torus are larger than $10^{11}\,\mathrm{g/cm^3}$ and $6$\,MeV, and hence, the values of the atmosphere are much lower than them (cf. Fig.~1). 

For solving the radiation fields, we set $N_{\rm trg}=120$ and $r_{\rm abs}=0.1$. With this setup, $\approx 10^7$--$10^8$ MC packets are solved in each time step. We confirm that the results are approximately unchanged even for $N_{\rm trg}=32$ (see App.~\ref{app:comp}). To reduce the computational cost, we limit the region of solving the radiation fields within $r\leq r_{\rm ext} = 300 M_\odot\approx 450\,{\rm km}$ in this work with $r=\sqrt{x^2+z^2}$ being the coordinate spherical radius. We justify this treatment by checking that the absorption and emission time scales are always more than an order of magnitude longer than the simulation time.

To check the dependence of the results on the viscous parameter, the numerical simulations are performed for $\alpha_{\rm vis}=0.05$ and $0.15$ employing the initial condition of an equilibrium torus around a rotating BH. The simulations are followed up to $t\approx 1.6\,{\rm s}$ and $\approx 1.2\,{\rm s}$ for the models with $\alpha_{\rm vis}=0.05$ and $0.15$, respectively. The simulations are run on the Sakura and Momiji clusters at Max Planck Computing and Data Facility with each employing 32 nodes and 1280 cores. The total computational cost is $\approx1$ million CPU hours for each simulation, and hence, $\approx1$ kilo CPU hours per 1\,ms simulation time.%\ms{How many nodes?}

\subsection{Initial conditions}
In this paper, we prepare an axisymmetric equilibrium torus around a rotating BH as the initial condition of our numerical simulation. This is motivated by the fact that, although the merger remnant has a non-axisymmetric structure in an early phase after the merger, it gradually relaxes to a nearly axisymmetric quasi-stationary state within the dynamical time scale ($\sim 10\,{\rm ms}$). We first compute the initial condition in the absence of neutrino radiation fields following the method of~\cite{Fujibayashi:2020qda}. Then, to minimize the artifact due to the relaxation of the neutrino radiation fields, we solve neutrino radiative transfer on the hydrodynamics solution freezing its profile except for $Y_e$. Finally, we employ the obtained matter profile and neutrino radiation fields as the initial condition of the dynamical simulation after they have settled into the approximately stationary configuration.

The line element is given in the Kerr-Schild coordinates in the 3+1 form of Eq.~\eqref{eq:metric}. Then the non-zero components of the metric in the spherical polar coordinates are $\alpha$, $\beta^r$, the diagonal component of $\gamma_{ij}$, and $\gamma_{r\varphi}=\gamma_{\varphi r}$ where $r$ and $\varphi$ denote the radial and toroidal angle coordinates of the spherical polar coordinates. We set the BH mass and dimensionless spin parameter to be $3\,M_\odot$ and $\chi=0.8$, respectively. We assume that the fluid is isentropic and the coordinate velocity given by $v^i=u^i/u^t$ satisfies $v^\varphi=\Omega$ and $v^x=0=v^z$ (i.e., $v^r=v^\theta=0$). Under stationary and axisymmetric conditions, the first integral of the Euler's equation for an isentropic fluid gives \cite{Shibata:2007zzb}
\begin{align}
    \frac{h}{u^t}+\int h u_\varphi d\Omega = C,\label{eq:EE1st}
\end{align}
where $C$ is a constant parameter. Here, $u_\varphi$ can be obtained by
\begin{align}
    u_\varphi&=u^t \gamma_{\varphi i}\left(v^i+\beta^i\right)\nonumber\\
    &=u^t\left[\gamma_{\varphi\varphi}(\Omega+\beta^\varphi)+\gamma_{\varphi x}\beta^x+\gamma_{\varphi z}\beta^z\right].\label{eq:ulphi}
\end{align}
From the normalize condition of the four velocity $g_{\mu\nu}u^\mu u^\nu=-1$, we have
\begin{align}
    u^t=\left[\alpha^2-\gamma_{ij}\left(v^i+\beta^i\right)\left(v^j+\beta^j\right)\right]^{-1/2}.
\end{align}

By assuming the specific angular momentum of the fluid, $hu_\varphi$, in the form of
\begin{align}
    j=h u_\varphi=A\Omega^{-n},
\end{align}
where $n$ is a constant, 
Eqs.~\eqref{eq:ulphi} and~\eqref{eq:EE1st} can be rewritten as
\begin{align}
    h u^t\left[\gamma_{\varphi\varphi}(\Omega+\beta^\varphi)+\gamma_{\varphi x}\beta^x+\gamma_{\varphi z}\beta^z\right]\Omega^n = A,\label{eq:aeq}
\end{align}
and
\begin{align}
    h\left(\frac{1}{u^t}-\frac{A\Omega^{-n+1}}{n-1}\right)=h\left(\frac{1}{u^t}-\frac{u_\varphi\Omega}{n-1}\right) = C,\label{eq:ceq}
\end{align}
respectively. The value of $h$ is identical at the inner and outer radii of the torus in the equatorial plane, $x_{\rm in}$ and  $x_{\rm out}$, respectively. This gives the conditions that
\begin{align}
    &\left\{u^t\left[\gamma_{\varphi\varphi}(\Omega+\beta^\varphi)+\gamma_{\varphi x}\beta^x+\gamma_{\varphi z}\beta^z\right]\Omega^n\right\}_{\rm in}\nonumber\\
    &=\left\{ u^t\left[\gamma_{\varphi\varphi}(\Omega+\beta^\varphi)+\gamma_{\varphi x}\beta^x+\gamma_{\varphi z}\beta^z\right]\Omega^n\right\}_{\rm out},\label{eq:omeeq1}
\end{align}
and
\begin{align}
    \left(\frac{1}{u^t}-\frac{u_\varphi\Omega}{n-1}\right)_{\rm in}=\left(\frac{1}{u^t}-\frac{u_\varphi\Omega}{n-1}\right)_{\rm out}\label{eq:omeeq2}.
\end{align}
$\Omega_{\rm in}=\Omega(x_{\rm in})$ and  $\Omega_{\rm out}=\Omega(x_{\rm out})$ can be determined by solving the algebraic Eqs.~\eqref{eq:omeeq1} and~\eqref{eq:omeeq2}. Then, constants $A$ and $C$ are determined from Eqs.~\eqref{eq:aeq} and~\eqref{eq:ceq} under the condition that $h=h_{\rm min}$ where $h_{\rm min}$ denotes the minimum specific enthalpy. Here,  $h=h_{\rm min}$ is given by Eqs.~\eqref{eq:enth} and~\eqref{eq:mexc} for a given value of $Y_e$ and for the floor values of the baryon mass density and temperature (by which the minimum enthalpy value among the baryon mass density and temperature above our setups of the floor values is indeed obtained).

In a binary NS merger, matter is also present in the vicinity of a BH just after the collapse of the remnant massive NS. Hence, it is desirable to have the inner edge of the torus to be located as close as possible to a BH. For this purpose, we impose a condition that the inner edge of the torus is cusp-like, and we determine $x_{\rm in}$ by the condition of
\footnote{
In this work, we found a convergence problem in the determination of $x_{\rm in}$. As a result, the initial condition of the simulations was determined as a quasi-stationary state employing $x_{\rm in}$ with a value $\approx 5\%$ larger than that of $x_{\rm in}$ determined by Eq.~\eqref{eq:cusp}. Nevertheless, we confirm that this does not affect the results in Sec.~\ref{sec:res} by more than $\approx 5\%$.
}
\begin{align}
    \left.\frac{\partial h}{\partial x}\right|_{x=x_{\rm in},~z=0}=0.\label{eq:cusp}
\end{align}

$h$ is obtained as a function of coordinates from Eq.~\eqref{eq:aeq} or Eq.~\eqref{eq:ceq} for a given value of $x_{\rm out}$ under the condition of Eq.~\eqref{eq:cusp}. Then, for a given value of $h$, the baryon mass density, $\rho$, is determined from a given EoS under a fixed value of specific total entropy, $s$, and the condition on $Y_e$. In this paper, we set the outer radius to be $x_{\rm out}=40\,M_{\rm BH}\approx180\,{\rm km}$, specific total entropy to be $6\,k_{\rm B}$ per baryon, and $n=1/7$, and assume the empirical relation between $Y_e$ and $\rho$ given by 
\begin{align}
    Y_e\left(\rho\right)=\left\{
    \begin{array}{ll}
         0.5    &  \rho\leq\rho_1\\
         0.07+0.43\displaystyle \frac{{\rm log}_{10}\left(\rho/\rho_2\right)}{{\rm log}_{10}\left(\rho_1/\rho_2\right)}&         \rho_1\leq\rho\leq\rho_2\\
         0.07   &  \rho_2\leq\rho
    \end{array}
    \right.
\end{align}
with $\rho_1=10^{11}\,{\rm g/cm^3}$ and $\rho_2=1.2\times 10^7\,{\rm g/cm^3}$ following~\citep{Fujibayashi:2020qda}. We then compute the initial configuration of the matter without neutrino radiation fields by employing the EoS described in Eq.~\eqref{eq:eos}.

To obtain the initial configuration with the neutrino radiation fields, we solve neutrino radiative transfer on the hydrodynamics solution freezing its profile but still allowing the evolution of the local $Y_e$. After neutrino radiation fields have relaxed, we recompute the matter configuration in the equilibrium state in the same manner as described above, but with the newly obtained $Y_e$ profile. We also consider the contribution of neutrinos in the EoS in addition to the contribution from electron/positron, nuclei, and photons described in Eq.~\eqref{eq:eos}, for which case the EoS can be written as
\begin{align}
    P=P_e+P_{\rm ion}+P_\gamma+P_\nu
\end{align}
and
\begin{align}
    \epsilon=\epsilon_e+\epsilon_{\rm ion}+\epsilon_\gamma+\epsilon_\nu
\end{align}
with
\begin{align}
    \epsilon_{\nu}=3\frac{P_\nu}{\rho}.
\end{align}
Here, $P_\nu$ consists of the contributions from electron neutrino, electron antineutrino, and heavy-lepton type neutrinos described by
\begin{align}
    P_{\nu}=c_{\nu_e}P_{\nu_e}+c_{{\bar \nu}_e}P_{{\bar \nu}_e}+c_{\nu_x}P_{\nu_x},
\end{align}
where
\begin{align}
    P_{\nu_e}&=\frac{1}{2\pi^2\left(\hbar c\right)^3}\left(k_{\rm B}T\right)^4F_3\left(\eta_e\right)\\
    P_{{\bar \nu}_e}&=\frac{1}{2\pi^2\left(\hbar c\right)^3}\left(k_{\rm B}T\right)^4F_3\left(-\eta_e\right)\\
    P_{\nu_x}&=\frac{2}{\pi^2\left(\hbar c\right)^3}\left(k_{\rm B}T\right)^4F_3\left(0\right),
\end{align}
with $F_3$ being the third-order Fermi-Dirac integral. $c_{\nu_e}$,  $c_{{\bar \nu}_e}$, and $c_{\nu_x}$ are factors introduced to describe how strongly neutrinos are coupled to matter, and these values become 0 and 1 for the optically thin and thick limits, respectively. We assume the following form of expression for $c_{\nu_i}$ $\left({\nu_i}={\nu_e},{{\bar \nu}_e},{\nu_x}\right)$:
\begin{align}
    c_{\nu_i}=\left(1-e^{-\alpha_{\nu_i}^{\rm abs}t_{\rm dyn}}\right){\rm min}\left(\frac{u_{\nu_i}}{u^\beta_{{\nu_i}, {\rm th}}},1\right).
\end{align}
Here, $\alpha_{\nu_i}^{\rm abs}$, $u_{\nu_i}$, and $u^\beta_{\nu, {\rm th}}$ denote the Planck-mean absorption rate, comoving energy density, and comoving energy density in the $\beta$-equilibrium of the neutrino species ${\nu_i}$, and $t_{\rm dyn}$ denotes the local dynamical time scale, which we approximate it with $t_{\rm dyn}\approx 2\pi\left(r^3/M_{\rm BH}\right)^{1/2}$.

To recompute the matter configuration in the equilibrium state, we note that, while we keep using the same fixed value for the outer radius ($40 M_{\rm BH}\approx180\,{\rm km}$) and specific total entropy ($6\,k_{\rm B}$ per baryon), we take into account the contribution of neutrinos to the total entropy.

\begin{figure}
 	 \includegraphics[width=1.\linewidth]{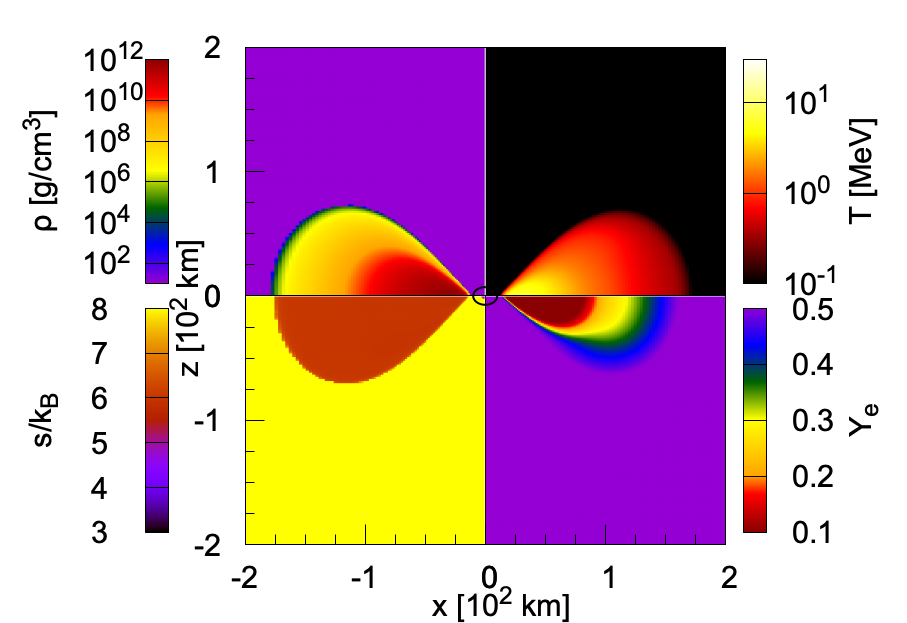}
 	 \caption{The baryon mass density, temperature, entropy per baryon, and $Y_e$ profiles of the initial condition.}
   %\ms{Outer radius seems to be much larger than $40M_\odot=60$\,km. (i) Should $40M_\odot$ be replaced by $40M_\mathrm{BH} \approx 180$\,km; (ii) the units of $x$ and $z$ are likely to be 100~km.}
	 \label{fig:snap_ini}
\end{figure}

Figure~\ref{fig:snap_ini} shows the baryon mass density, temperature, entropy per baryon, and $Y_e$ profiles obtained by the above procedure. The total mass, maximum baryon mass density, and maximum temperature of the torus are $\approx0.10\,M_\odot$, $7.6\times10^{11}\,{\rm g/cm^3}$, and $6.3\,{\rm MeV}$, which are broadly consistent with the initial condition calculated in~\cite{Fujibayashi:2020qda}.

\section{Diagnostics}\label{sec:diag}
Here, we briefly summarize various quantities used in this work for the analysis of the simulation results. The mass accretion rate on the BH, ${\dot M}_{\rm fall}$, is determined by integrating the mass flux on the event horizon as
\begin{align}
    {\dot M}_{\rm fall}&=\int_{{\rm EH}} dS_i\, \rho_* v^i,
\end{align}
where $dS_i$ is the surface element. The total mass accreted on the BH, $M_{\rm fall}$ is obtained by the time integral of ${\dot M}_{\rm fall}$ as
\begin{align}
    M_{\rm fall}&=\int dt{\dot M}_{\rm fall}.
\end{align}

In this work, the ejecta matter is determined by employing the so-called Bernoulli criterion: the matter in the hydrodynamics cell at which $-h u_t$ is larger than $h_{\rm min}$ is considered to be gravitationally unbound and becomes ejecta. Here, $u_t$ denotes the lower time component of the 4-fluid velocity. 

During the evolution of the system, the mater escapes from the computational domain. To determine the total ejecta mass taking into account the escaped components, we define the ``bulk'' region of the computational domain as the region where the cylindrical radius (i.e., the $x$-coordinate) and the half height (i.e., the $z$-coordinate) are smaller than $L_{\rm ext}$. Then, we compute the contributions from the ``bulk'' region and that escaped from the surface of the ``bulk'' region as
\begin{align}
    M_{\rm eje}&=M_{\rm eje, bulk}+\int dt{\dot M}_{\rm eje,esc},\\ 
    M_{\rm eje, bulk}&=\int_{{\rm bulk}} d^3x \rho_*\Theta\left(-h u_t-h_{\rm min}\right),\\
    {\dot M}_{\rm eje, esc}&=\int_{\partial{\rm bulk}} dS_i \rho_* v^i\Theta\left(-h u_t-h_{\rm min}\right).\label{eq:def-ejem}
\end{align}
In our work, we take $L_{\rm ext}=2000\,{\rm km}$, while the results are approximately unchanged by increasing $L_{\rm ext}$ even up to $3500\,{\rm km}$.

The asymptotic kinetic energy of the ejecta, $E_{\rm kin,eje}$, is measured by the same way as in Eq.~\eqref{eq:def-ejem}, but by substituting $\rho_*$ in the integrand with $\rho_*(-h/h_{\rm min}u_t-1)$. The mass-averaged $Y_e$ value of the ejecta is also measured from a quantity defined by Eq.~\eqref{eq:def-ejem} but with $\rho_*\rightarrow \rho_* Y_e$ divided by the total ejecta mass.

Neutrino luminosity and number emission rate for $\nu_i$ neutrino species ($L_{\nu_i}$ and ${\dot N}_{\nu_i}$, respectively) are obtained from the MC packets escaped at $r=r_{\rm ext}$:
\begin{align}
    L_{\nu_i}&=-\frac{1}{\Delta t}\sum_{k,{\rm escape}} w_k^{\nu_i} p_{(k),t},\\
    {\dot N}_{\nu_i}&=\frac{1}{\Delta t}\sum_{k,{\rm escape}} w_k^{\nu_i}.
\end{align}
Here, $p_{(k),t}$ denotes the lower time components of the neutrino 4-momentum for $k${\it -th} escaped MC packets. The average energy of emitted neutrinos, $\left<\omega_{\nu_i}\right>_{\rm ems}$, is calculated by $\left<\omega_{\nu_i}\right>_{\rm ems}=L_{\nu_i}/{\dot N}_{\nu_i}$.

The total pair annihilation energy deposition rate is calculated by
\begin{align}
    L_{\rm pair}=\sum_{{\nu_i}}\int d^3x\sqrt{\gamma}\,{\rm max}\left(-G_{{\nu_i},t}^{{\rm pair}},0\right),
\end{align}
where $\gamma$ and $G_{{\nu_i},t}^{{\rm pair}}$ denote the determinant of the spatial metric and the lower time component of the radiation 4-force density contributed by neutrino pair process, respectively.

We note that there is a time delay until emitted neutrinos are observed and reflected in the luminosities due to the propagation to the extraction radius. Hence, for direct comparison with other instantaneously determined quantities, we shift the time origin of the neutrino luminosities by the propagation time scale, i.e., $t\rightarrow t-r_{\rm ext}/c$. On the other hand, the mass accretion rate and pair annihilation energy deposition rate are measured instantaneously, and hence, the time shift is not applied to these quantities.

\section{Results}\label{sec:res}
In this section, we present the results of the simulations for a BH-torus system. We also compare our results with the models in the previous study~\cite{Fujibayashi:2020qda} with similar setups (K8 and K8s). The main differences between the previous simulation setups and the present work are as follows: In the previous work, 1) the spacetime is dynamically evolved, 2) neutrino radiative transport is solved by a gray moment scheme in combination with a leakage method~\cite{2010PThPh.124..331S,2011PThPh.125.1255S,Fujibayashi:2017xsz}, 3) neutrino interactions with heavy nuclei, nucleon-nucleon bremsstrahlung, and plasmon decay are considered in addition to those considered in the present work while $\nu_e{\bar \nu}_e$ pair process is neglected, and 4) the DD2 EOS~\cite{banik2014oct} extended to low-density and low-temperature ranges by an EOS of~\cite{2000ApJS..126..501T} is employed with taking larger NSE ensembles of heavy nuclei into account. In both previous and present studies, the initial condition is prepared employing the same BH and torus masses and BH spin with similar setups for the inner and outer radii of the torus. On the other hand, the self-gravity of the torus is taken into account for constructing the initial condition, and the initial radiation fields are set to be zero in the previous study.

\subsection{Evolution process}

\begin{figure*}
 	 \includegraphics[width=0.48\linewidth]{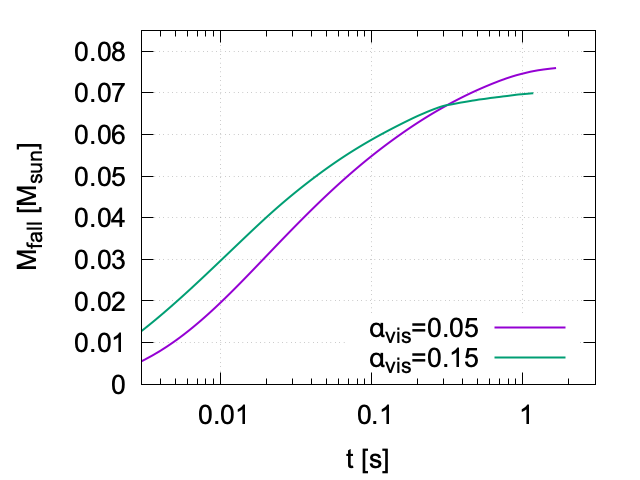}
 	 \includegraphics[width=0.48\linewidth]{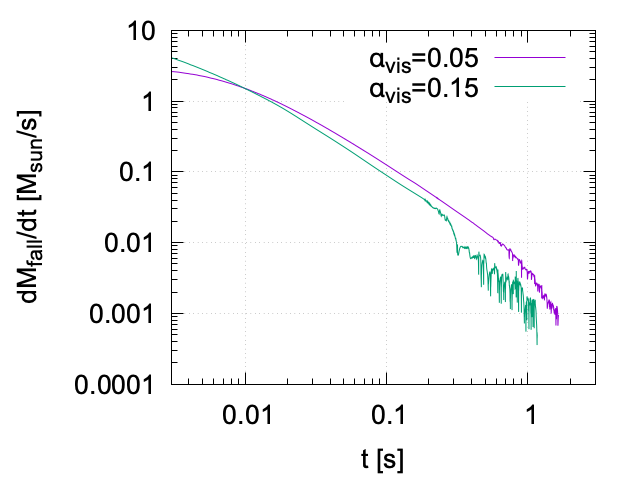}
 	 \caption{Total accreted mass (left) and mass accretion rate (right) onto the BH as functions of time.}
	 \label{fig:t-acc}
\end{figure*}

Figure~\ref{fig:t-acc} shows the total accreted mass and mass accretion rate on to the BH as functions of time. Based on the $\alpha$-thin or Shakura-Sunyaev disk model~\cite{1973A&A....24..337S}, the viscous time scale can be estimated by
\begin{align}
    \tau_{\rm vis}=\frac{R^2}{h\nu}\sim 0.7\,{\rm s}&\left(\frac{\alpha_{\rm vis}}{0.05}\right)^{-1}\left(\frac{h c_s}{0.075\,c^3}\right)^{-1}\nonumber\\
    &\times\left(\frac{H}{9\,{\rm km}}\right)^{-1}\left(\frac{R}{85\,{\rm km}}\right)^{2}.\label{eq:tauvis}
\end{align}
Here, we employed the typical mass-averaged values for the torus radius, $R$, and specific enthalpy weighted sound speed, $h c_s$ measured at $t=20\,{\rm ms}$ when the system settled into a quasi-stationary evolution phase. Figure~\ref{fig:t-acc} shows that more than $\approx 70\%$ of the initial torus mass is accreted onto the BH within the viscous time scale. The accretion is more rapid for the model with a large value of $\alpha_{\rm vis}$ due to more efficient angular momentum transport. The accretion rate decreases with time due to the decrease in the torus mass and outward expansion of the torus. In particular, the accretion rate is strongly suppressed after the onset of the outflow ($\approx1\,{\rm s}$ and $\approx0.3\,{\rm s}$ for $\alpha_{\rm vis}=0.05$ and 0.15, respectively, see Fig.~\ref{fig:eje}), at which the neutrino cooling time scale becomes longer than the viscous time scale (see below for more details). The evolution of the total mass of the accreted matter agrees within $\approx5\%$ with the similar models in the previous study~\cite{Fujibayashi:2020qda} (K8 and K8s).

\begin{figure*}
 	 \includegraphics[width=0.48\linewidth]{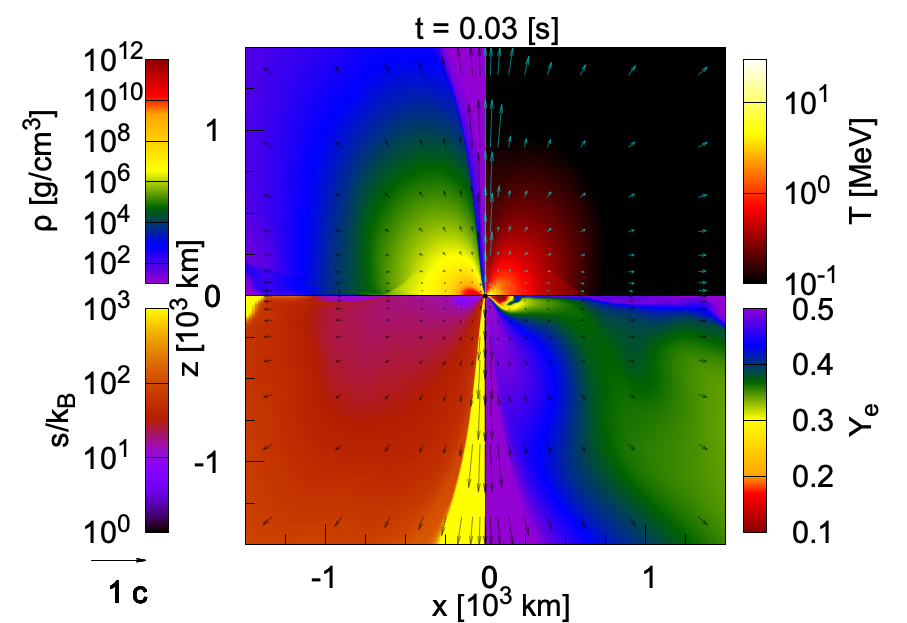}
 	 \includegraphics[width=0.48\linewidth]{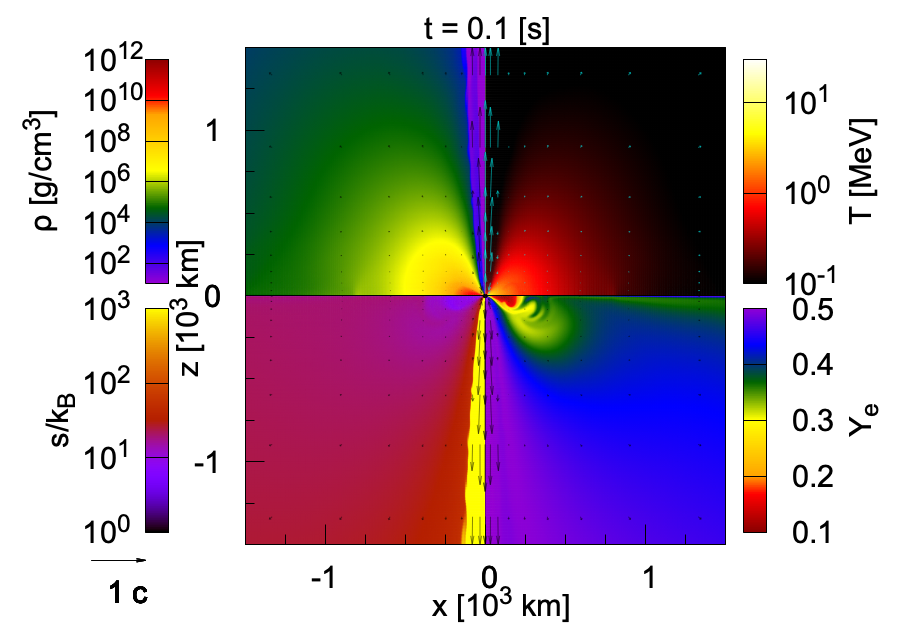}\\
 	 \includegraphics[width=0.48\linewidth]{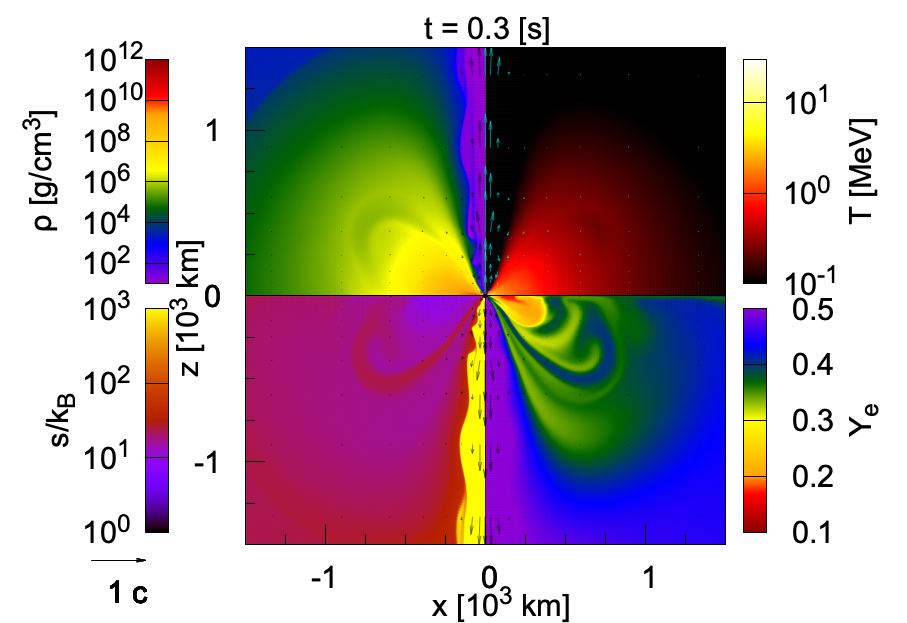}
 	 \includegraphics[width=0.48\linewidth]{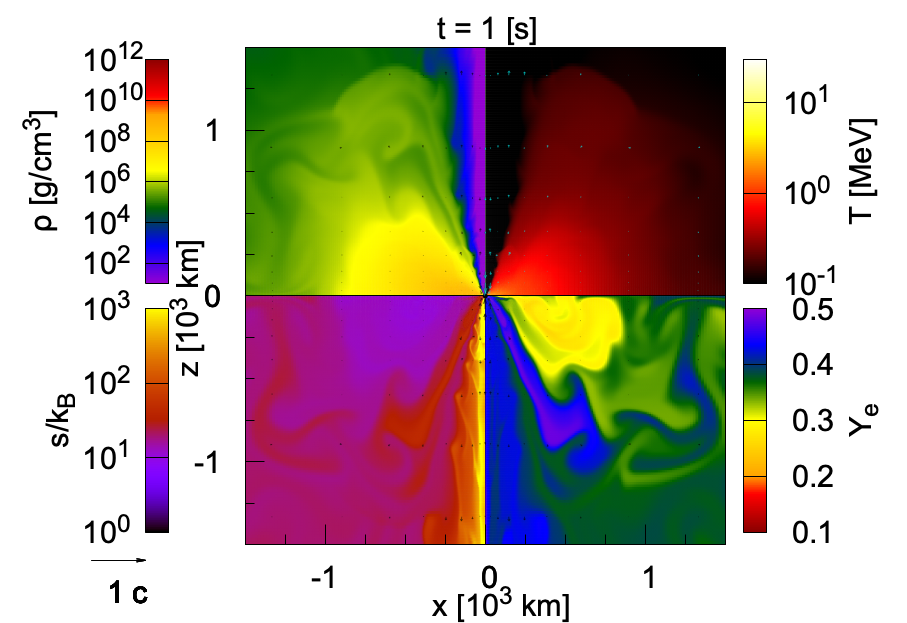}
 	 \caption{Snapshots for the baryon mass density, temperature, entropy per baryon, and $Y_e$ profiles for the model with $\alpha_{\rm vis}=0.05$.}
	 \label{fig:snap_a005}
\end{figure*}

Figure~\ref{fig:snap_a005} displays the evolution of the baryon mass density, temperature, entropy per baryon, and $Y_e$ profiles for the model with $\alpha_{\rm vis}=0.05$. The system evolution is broadly in agreement with that found in the previous study~\cite{Fujibayashi:2020qda}. In the first few tens of ms, the outer layer of the torus blows up due to the relaxation from the initial condition. After that, the system gradually settles into a quasi-stationary evolution phase, in which the torus gradually expands due to the angular momentum transport by the viscosity. The $Y_e$ values also increase in time due to weak interaction following the decrease in the baryon mass density~\cite{Fujibayashi:2020qda}. During this phase, the viscosity contentiously heats-up the torus, but the outflow is still suppressed because neutrino cooling is efficient. This quasi-stationary expansion phase lasts up to $\approx 1\,{\rm s}$ for the model with $\alpha_{\rm vis}=0.05$. But as the torus expands and the baryon mass density and temperature of the torus decrease, neutrino cooling becomes inefficient. Then after $\approx 1\,{\rm s}$, eventually convective motions in the torus are activated and the outflow starts to form.

\begin{figure*}
 	 \includegraphics[width=0.48\linewidth]{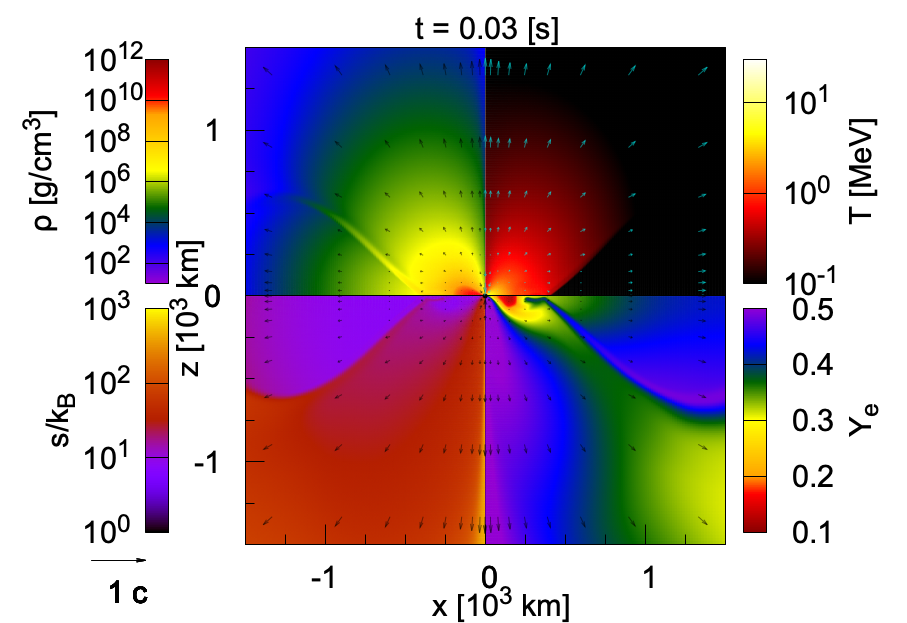}
 	 \includegraphics[width=0.48\linewidth]{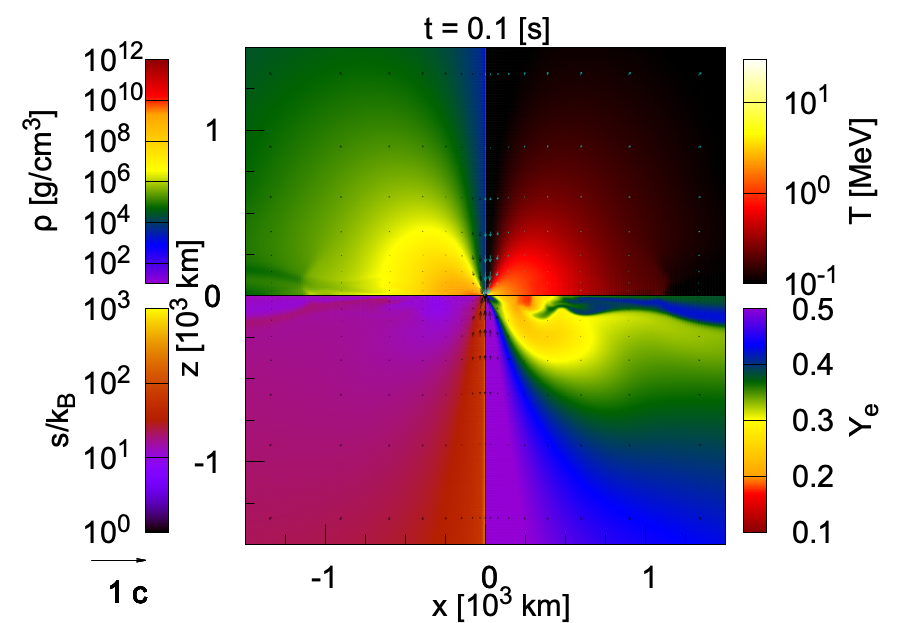}\\
 	 \includegraphics[width=0.48\linewidth]{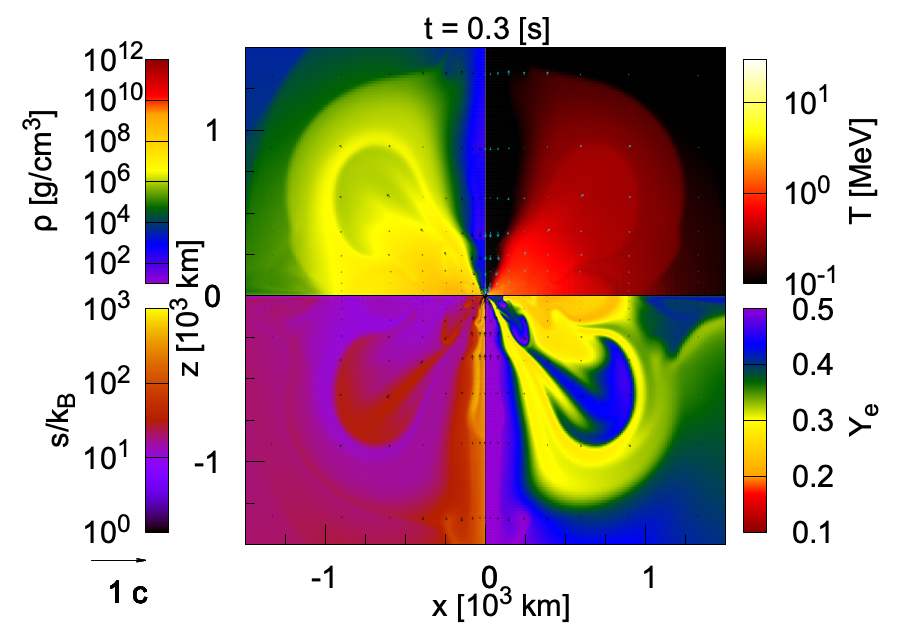}
 	 \includegraphics[width=0.48\linewidth]{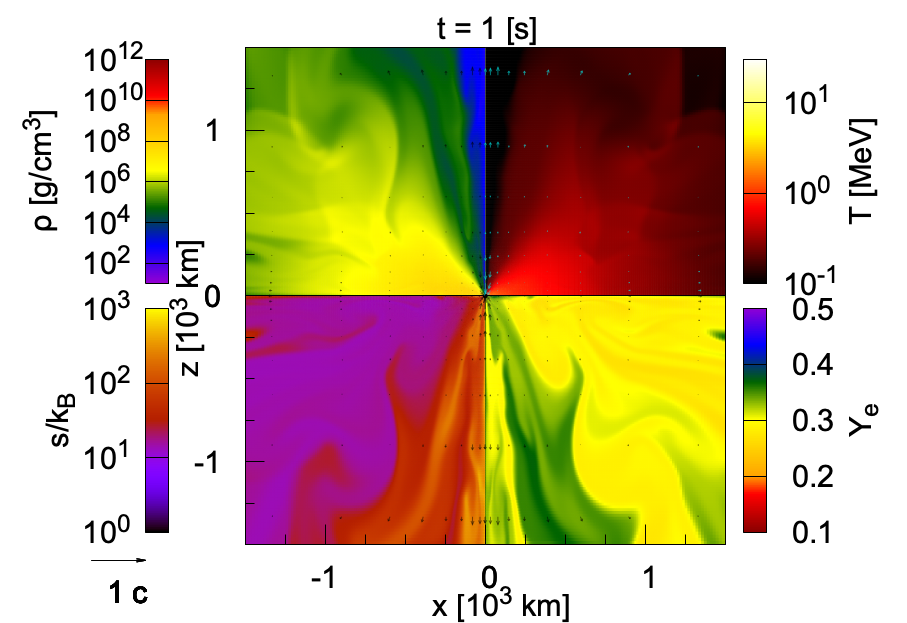}
 	 \caption{The same as Fig.~\ref{fig:snap_a005} but for the model with $\alpha_{\rm vis}=0.15$.}
	 \label{fig:snap_a015}
\end{figure*}

Figure~\ref{fig:snap_a015} displays the same as in Fig.~\ref{fig:snap_a005} but for the model with $\alpha_{\rm vis}=0.15$. The evolution process is broadly the same as for the model with $\alpha_{\rm vis}=0.05$, although the transient behavior during the relaxation from the initial condition is more violent. The evolution time scale for this model is shorter than with $\alpha_{\rm vis}=0.05$ due to more efficient angular momentum transport. In the quasi-stationary evolution phase after a few tens of ms, the torus expands more rapidly than the model with $\alpha_{\rm vis}=0.05$, and the outflow is launched at $\approx 0.3\,{\rm s}$. Since the mass ejection sets in earlier than for the model with $\alpha_{\rm vis}=0.05$, the material ejected from the system typically has a lower value of $Y_e$ for the model with $\alpha_{\rm vis}=0.15$.%\ms{For a given time, this is the case but for the final ejecta profile, I doubt this statement.}

\begin{figure*}
 	 \includegraphics[width=0.48\linewidth]{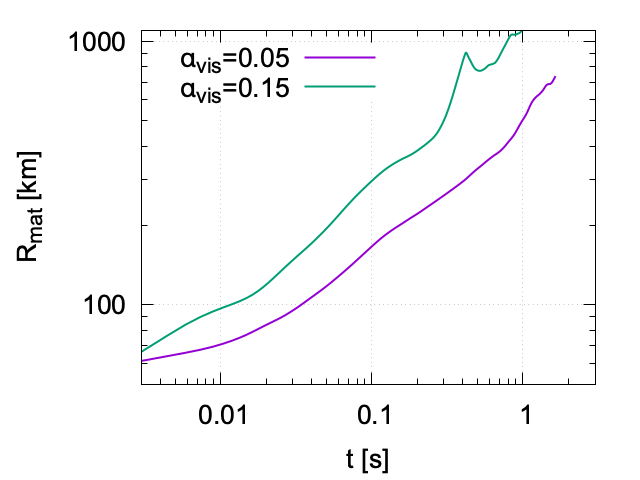}
 	 \includegraphics[width=0.48\linewidth]{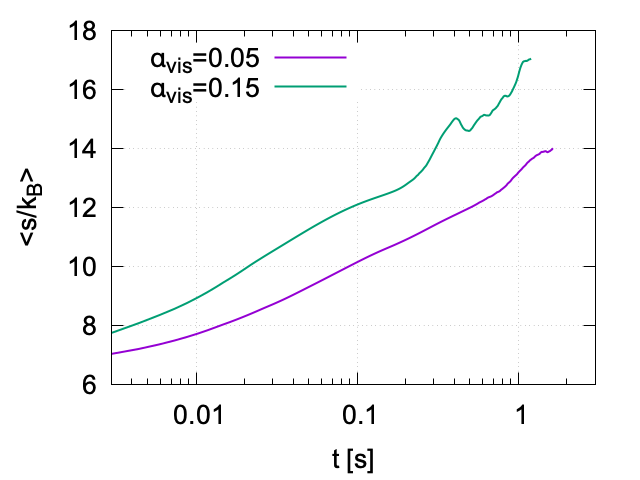}\\
 	 \includegraphics[width=0.48\linewidth]{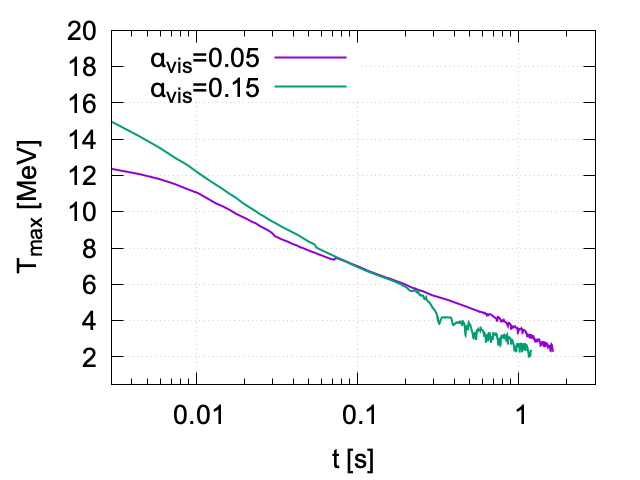}
 	 \includegraphics[width=0.48\linewidth]{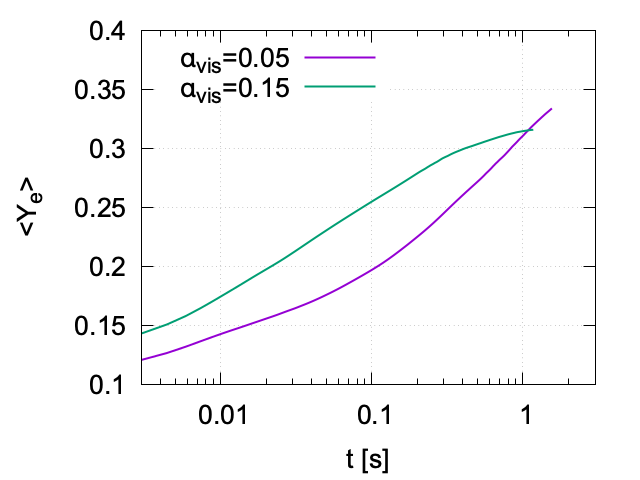}
 	 \caption{Time evolution of various quantities for the matter outside the event horizon: the mass-averaged cylindrical radius (top left), mass-averaged entropy per baryon (top right), maximum temperature (bottom left), and mass-averaged $Y_e$ (bottom right). We note that the sudden decreases in the average radius and entropy per baryon at $0.4\,{\rm s}$ found in the results of the $\alpha_{\rm vis}=0.15$ model are caused by the escape of the matter from the outer boundary. On the other hand, the contribution from the escaped matter is also taken into account for the average $Y_e$ value, and hence, it shows a smooth evolution.}
	 \label{fig:disk}
\end{figure*}

To take a close look at the evolution of the system more quantitatively, we show in Figure~\ref{fig:disk} various quantities defined for the matter outside the event horizon. The top left panel of Fig.~\ref{fig:disk} shows that the mass-averaged cylindrical radius gradually increases with time reflecting the expansion of the torus due to the angular momentum transport. The increase rate sharply rises after $\approx1\,{\rm s}$ and $\approx0.3\,{\rm s}$ for $\alpha_{\rm vis}=0.05$ and $0.15$, respectively, reflecting the onset of the mass ejection. The average entropy per baryon also increases with time due to entropy generation by viscous heating while neutrino cooling plays a role in reducing the entropy increase. The rate of increase becomes higher after the onset of mass ejection reflecting the fact that neutrino cooling becomes inefficient after this point. The average $Y_e$ value also increases with time. This reflects the fact that the $Y_e$ value at the emission equilibrium~\cite[e.g.,][]{Fujibayashi:2020qda, Just:2021cls} increases as the baryon mass density and temperature decrease. This change in the $Y_e$ value continues until the onset of the mass ejection, but the increase in the $Y_e$ value slows down thereafter because the weak interaction time scale becomes longer than the viscous evolution time scale. %neutrino emission becomes inefficient. 

The time evolution of the averaged radius is faster by a factor of $\approx3$ for the model with $\alpha_{\rm vis}=0.15$ than with $0.05$. This reflects the dependence of the viscous time scale on the $\alpha_{\rm vis}$ parameter, and indeed, the factor of the difference approximately matches to the difference in the value of $\alpha_{\rm vis}$. This is also the case for the evolution of average entropy per baryon and $Y_e$. Independent of the value of $\alpha_{\rm vis}$, the average radius, entropy per baryon, maximum temperature, and average $Y_e$ value are $400$--$500\,{\rm km}$, 12--13$\,k_{\rm B}$, $3$--$4\,{\rm MeV}$, and $\approx0.3$, respectively, at the onset time of the mass ejection. However, the model with the different value of $\alpha_{\rm vis}$ shows the different subsequent evolution. In particular, the $Y_e$ value shows different evolution after the onset of the mass ejection: the increase of the $Y_e$ value slows down after the onset of the mass ejection. This slowing down is more significant for the model with a larger $\alpha_{\rm vis}$ parameter.

The average $Y_e$ values for the present simulations are typically larger by $\approx0.05$ than those at the similar epochs for the models with the similar setups in the previous study~\cite{Fujibayashi:2020qda} (model K8 and K8s). This may be due to the fact that the electron mass is neglected for neutrino/antineutrino absorption by neutrons/protons in the present work. By neglecting the electron mass, the neutrino/antineutrino absorption rates for low energy neutrinos/antineutrinos can be enhanced, and hence, the $Y_e$ evolution in the torus can be accelerated. In fact, reference~\cite{Fujibayashi:2020qda} shows that the absorption of streaming neutrinos/antineutrinos plays a role in enhancing the $Y_e$ values (see Fig.~7 of~\cite{Fujibayashi:2020qda}). 

We also find quantitative differences in the onset time of mass ejection. While mass ejection sets in at $\approx 1\,{\rm s}\,(0.3\,{\rm s})$ for the present model, it happens at $\approx 0.5\,{\rm s}\,(0.1\,{\rm s})$ for the model with $\alpha_{\rm vis}=0.05\,(0.15)$ in the previous study, and hence, there is a factor of $\approx2$ delay in the present result. One possible reason for this difference may be due to the difference in the employed EoS. In our study, we only consider protons, neutrons, and $\alpha$ particles for the NSE ensemble elements. On the other hand, in the previous study, the NSE ensembles of nuclei also heavier than $\alpha$ particle are taken into account in the EoS~\cite{banik2014oct}. This difference can lead to a non-negligible difference in the nuclear binding energy (1--2$\,{\rm MeV}$ per baryon). Since the NSE ensemble setup in this paper gives less binding energy release in the recombination of the nuclei when the baryon mass density drops, the present setup tends to work to suppress the ejecta formation.

To clarify the reason for these quantitative discrepancies, a detailed comparison varying the microphysical setup under the same initial condition is needed. We should also note that not only the difference in the microphysical setups but also the difference in the radiation transfer solver may be responsible for the quantitative difference between the present and previous results. For example, taking into account the effect of the energy dependent absorption rates under the non-thermalized distribution function may cause some difference in the lepton number transfer in the torus. Thus, also to evaluate the impact of the difference in the radiative transfer solver, it is crucial to make a comparison with the same microphysical setup, and we leave this as a future work.

\begin{figure*}
 	 \includegraphics[width=0.32\linewidth]{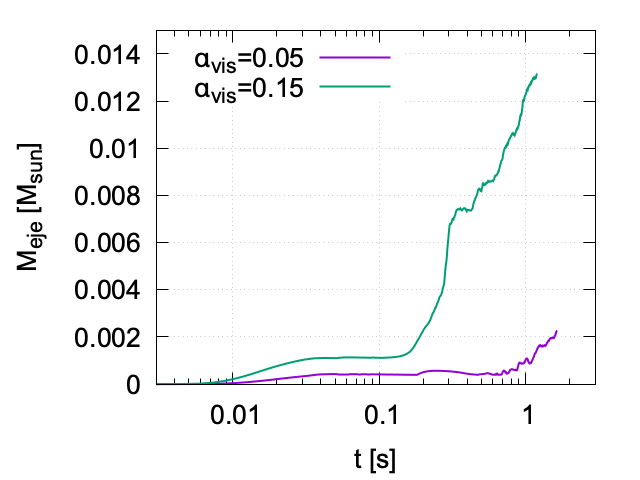}
 	 \includegraphics[width=0.32\linewidth]{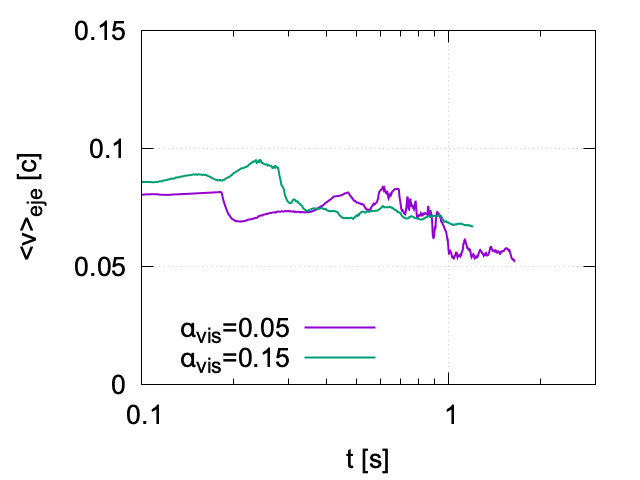}
 	 \includegraphics[width=0.32\linewidth]{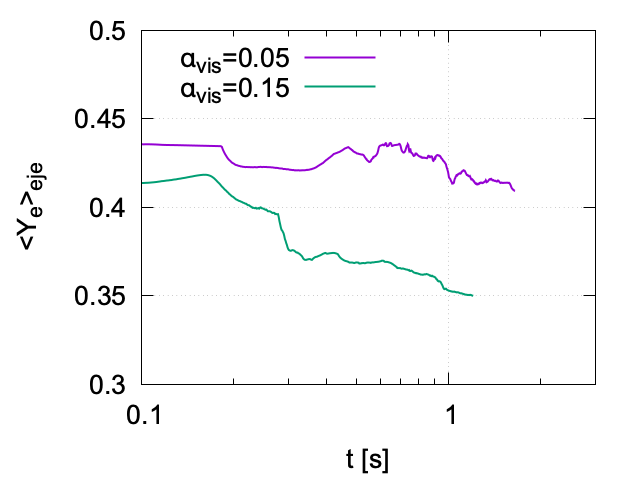}\\
 	 \includegraphics[width=0.32\linewidth]{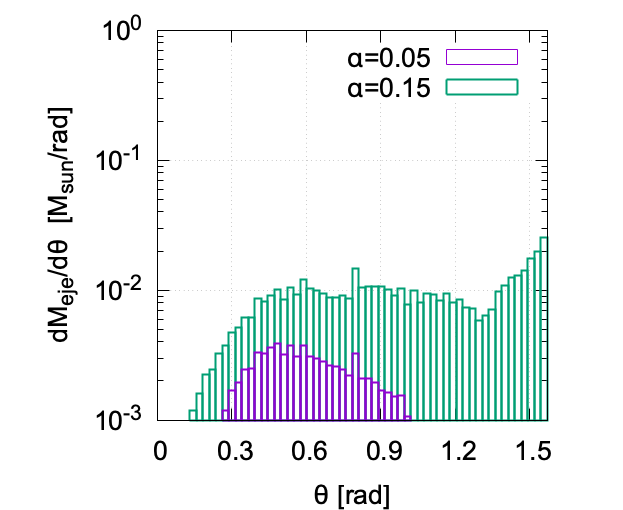}
 	 \includegraphics[width=0.32\linewidth]{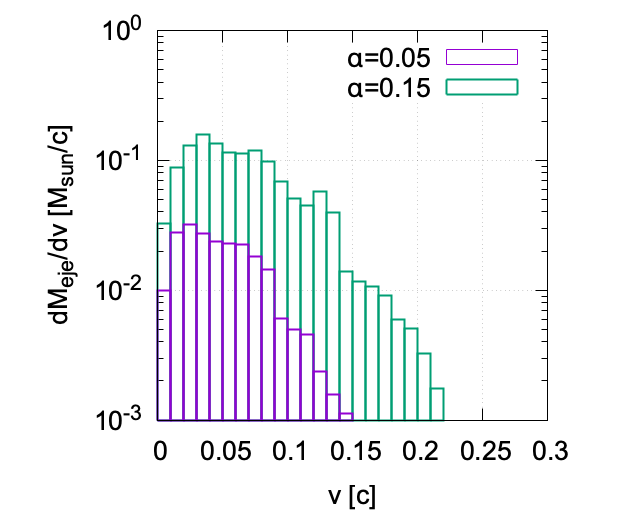}
 	 \includegraphics[width=0.32\linewidth]{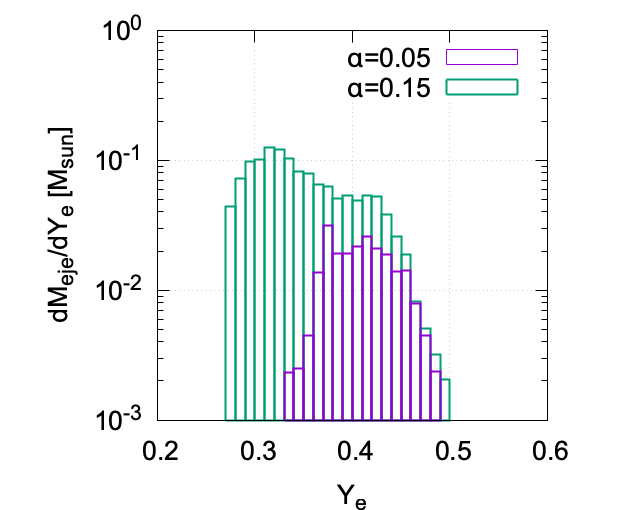}
 	 \caption{{\it Top panels:} Time evolution of the ejecta mass (left), average velocity of ejecta (middle), and mass-averaged $Y_e$ (right). {\it Bottom panels:} Ejecta mass distribution as a function of latitudinal angle (left), velocity (middle), and $Y_e$ (right) at the end of simulations ($1.6\,{\rm s}$ and $1.2\,{\rm s}$ for the models with $\alpha_{\rm vis}=0.05$ and 0.15, respectively).}
	 \label{fig:eje}
\end{figure*}

\subsection{Ejecta}

Figure~\ref{fig:eje} shows the various quantities for the ejecta matter. The top left panel of Fig.~\ref{fig:eje} shows the time evolution of the ejecta mass. The ejecta mass increases for $\sim10^{-3}\,M_\odot$ in the first few $10\,{\rm ms}$. We interpret this component as a consequence of the relaxation of the system from the initial condition and hence not physical. After the initial transition, the ejecta mass is kept constant until $t\approx1\,{\rm s}$ and $0.2\,{\rm s}$ for $\alpha_{\rm vis}=0.05$ and $0.15$, respectively, reflecting the fact that the mass ejection is suppressed as the consequence of efficient neutrino cooling. Thereafter, the ejecta mass shows a rapid increase and continues to increase until the end of the simulations. The total ejecta mass at the end of simulations is $0.002\,M_\odot$ and $0.013\,M_\odot$ for $\alpha_{\rm vis}=0.05$ and $0.15$, respectively. These values are smaller than the ejecta mass obtained in the similar models of the previous study (K8 and K8s in~\cite{Fujibayashi:2020qda}) due to the shorter simulation time in the present work. In fact, the previous study shows that it is necessary to follow a few seconds of the evolution until the mass ejection to be saturated. Nevertheless, the comparison shows that the time evolution of the ejecta mass is broadly in agreement with the previous results~\cite{Fujibayashi:2020qda}.

The bottom left panel of Fig.~\ref{fig:eje} shows the ejecta mass distribution as a function of the ejected latitudinal angle. The ejecta matter is mostly ejected towards $\theta\approx \pi/6$--$\pi/4$. For a lager value of $\alpha_{\rm vis}$, the distribution tends to peak in larger latitudinal angle, and for $\alpha_{\rm vis}=0.15$, a substantial ejecta component is present in the equatorial plane. It should be noted that it is difficult to make a fair comparison with the previous results, since the present simulations are not performed until mass ejection is saturated. Nevertheless, this dependence on the $\alpha_{\rm vis}$ parameter is so far broadly consistent with the trend found in the previous study~\cite{Fujibayashi:2020qda}.

The top middle panel of Fig.~\ref{fig:eje} shows the time evolution of the average ejecta velocity. After the onset of mass ejection ($\approx 1\,{\rm s}$ and $0.3\,{\rm s}$ for $\alpha_{\rm vis}=0.05$ and $0.15$, respectively), the average ejecta velocity settles down to an approximately constant value: $0.05\,c$ and $0.08\,c$ for $\alpha_{\rm vis}=0.05$ and $0.15$, respectively. The model with higher value of $\alpha_{\rm vis}$ has larger ejecta velocity, which reflects an efficient acceleration of the matter in the outer part of the torus. The average ejecta velocity is in quantitative agreement with the previous results (see Fig.10 of~\cite{Fujibayashi:2020qda}). The bottom middle panel of Fig.~\ref{fig:eje} shows the ejecta mass distribution as a function of the ejecta velocity. For both cases of $\alpha_{\rm vis}$, the ejecta matter has its velocity peak around $0.05\,c$, and the ejecta matter with the velocity larger than $0.15\,c$ is minor (an order of magnitude smaller than the peak). These results also agree broadly with those in the previous study~\cite{Fujibayashi:2020qda}.

The top right panel of Fig.~\ref{fig:eje} shows the time evolution of the average ejecta $Y_e$, which is 0.41 and 0.35 for $\alpha_{\rm vis}=0.05$ and $0.15$, respectively, at the end of simulations. They decrease continuously because the matter ejected from the inner part of ejecta, which typically has relatively high velocity and $Y_e$ value, is counted as ejecta first, and the matter ejected from the edge of the torus at the onset of mass ejection, which typically has relatively low velocity and $Y_e$ value, is counted as ejecta later. The bottom left panel of Fig.~\ref{fig:eje} shows the ejecta mass distribution as a function of $Y_e$. The ejecta $Y_e$ distribution shows a peak at $\approx0.4$ and 0.3 and spreads with a range of more than 0.1 and 0.2, respectively. For the higher value of $\alpha_\mathrm{vis}$, the ejecta $Y_e$ value tends to be smaller. This result is in agreement with the previous study~\cite{Fujibayashi:2020qda}.

However, the ejecta $Y_e$ values of the present simulations are higher by more than 0.05 than the similar models of the previous study (see Fig.10 of~\cite{Fujibayashi:2020qda}). One reason for this is the difference in the $Y_e$ evolution of the torus. The average $Y_e$ evolution of the matter is more rapid for the models in the present work (see Fig.~\ref{fig:disk}). In addition, the time at which the mass ejection sets in is delayed by a factor of $\approx2$ in the present models compared to the previous results. Hence, the $Y_e$ values at the onset of mass ejection is higher for the models in the present work than for those in the previous study with the similar setups. However, we note again that a fair comparison with the previous results is not easy because the present simulations are not performed until mass ejection is saturated.

\begin{figure*}
 	 \includegraphics[width=0.48\linewidth]{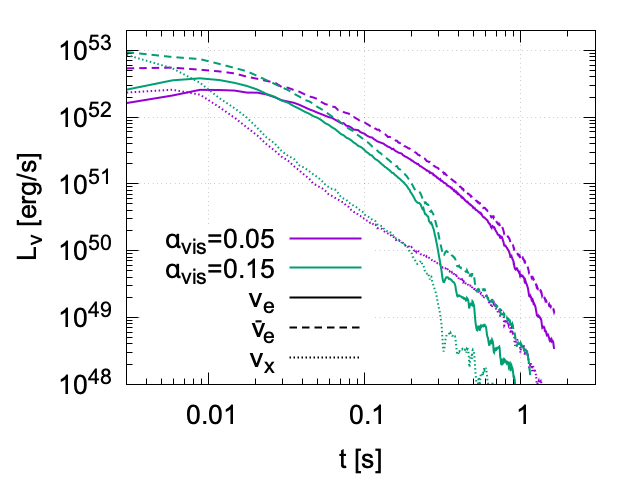}
 	 \includegraphics[width=0.48\linewidth]{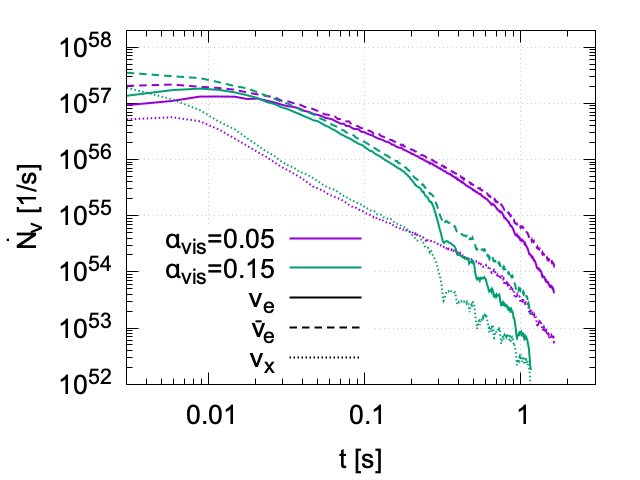}\\
 	 \includegraphics[width=0.48\linewidth]{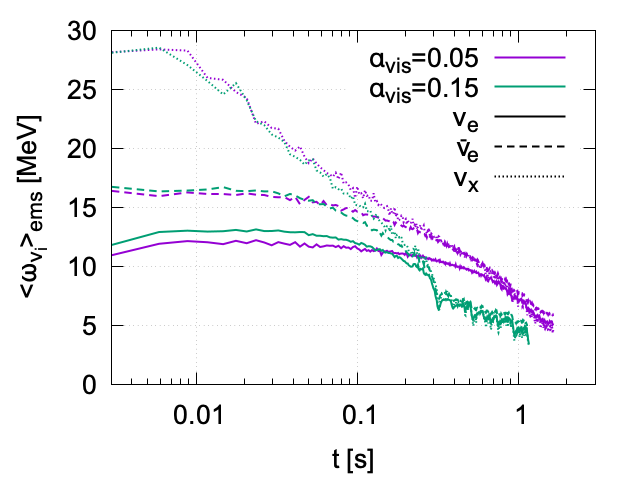}
 	 \includegraphics[width=0.48\linewidth]{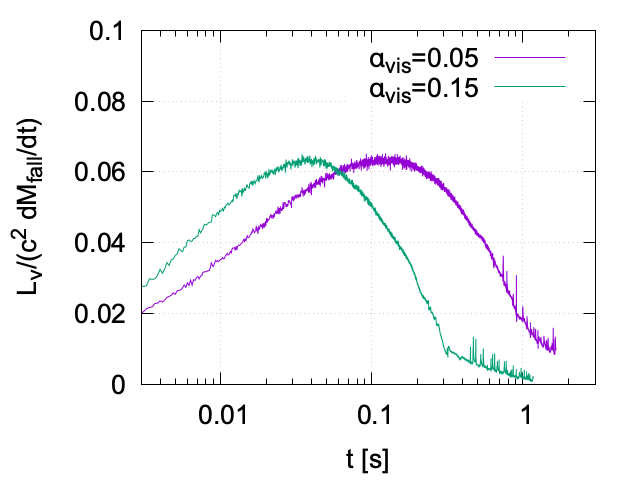}\\
 	 \includegraphics[width=0.48\linewidth]{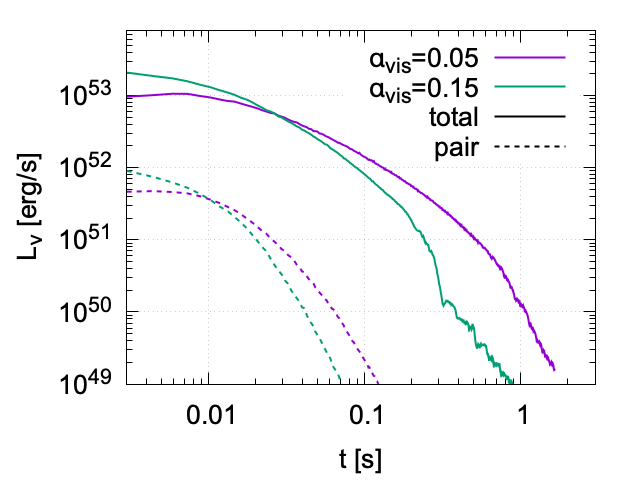}
 	 \includegraphics[width=0.48\linewidth]{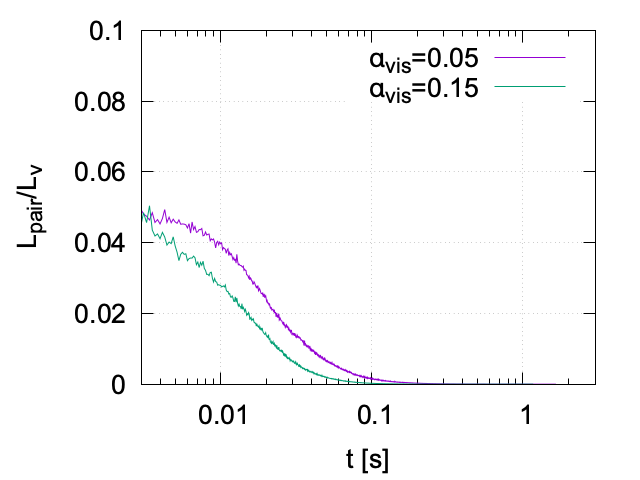}
 	 \caption{Time evolution of various quantities related to neutrino emission: the neutrino luminosity (top left), neutrino number emission rate (top right), average energy of emitted neutrino (middle left), neutrino emission efficiency with respect to the mass accretion rate (middle right), total neutrino luminosity and pair annihilation rate (bottom left), and pair annihilation efficiency with respect to the total neutrino luminosity (bottom right).}
	 \label{fig:nu_detail}
\end{figure*}

\subsection{Neutrinos}
%\ms{I changed number luminosity to number emission rate. I also think that "energy luminosity" is strange term.}

We next focus on the property of neutrino emission from the system. Figure~\ref{fig:nu_detail} shows the time evolution of various quantities related to neutrino emission. The neutrino luminosity of electron antineutrinos is always higher than those of other species by more than a factor of 2 at $t=0.01\,{\rm s}$ and becomes dominant for $t>0.3\,{\rm s}$ thereafter. This is also the case for the number emission rate. This indicates that the matter is kept leptonized during the evolution. Heavy-lepton type neutrinos always have the smallest contribution to the total luminosity, and it is typically more than an order of magnitude smaller than that of electron neutrinos. 

For the first $20\,{\rm ms}$, the model with $\alpha_{\rm vis}=0.15$ shows higher luminosity than the model with $\alpha_{\rm vis}=0.05$ due to the higher mass accretion rate (see Fig.~\ref{fig:t-acc}). However, the model with $\alpha_{\rm vis}=0.15$ shows more rapid decline in the luminosity, which becomes fainter than for $\alpha_{\rm vis}=0.05$ for $t>0.03\,{\rm s}$ because the torus expands more rapidly. The neutrino luminosity at the onset of mass ejection is larger for the model with the larger value of $\alpha_{\rm vis}$. We can understand this from the fact that mass ejection sets in approximately at the time that the neutrino cooling rate drops below the total viscous heating rate ($t\approx1\,{\rm s}$ for $\alpha_{\rm vis}=0.05$ and $t\approx0.3\,{\rm s}$ for $\alpha_{\rm vis}=0.15$), which comes earlier for the model with the larger value of $\alpha_{\rm vis}$ due to the larger heating rate. We also note that the total viscous heating rate at the onset of mass ejection is lower for $\alpha_{\rm vis}=0.05$ due to the torus expansion and mass decrease, which further reduces the neutrino luminosity at that time. After the mass ejection sets in, the neutrino luminosity decreases even steeper reflecting that the rapid decrease in the density and temperature due to the accelerated matter expansion.

The average energy of neutrinos emitted is largest for heavy-lepton type neutrinos, followed by electron antineutrinos, and finally electron neutrinos. This reflects the fact that the respective last scattering surfaces of the neutrinos are located further inside with higher matter temperature. The average neutrino energy shows approximately the same evolution regardless of the value of $\alpha_{\rm vis}$ until the onset of mass ejection. However, at onset of the mass ejection, which happens earlier for the model with larger values of $\alpha_{\rm vis}$, the neutrino energy rapidly decrease reflecting the rapid decrease in the matter temperature (see Fig.~\ref{fig:disk}).

The neutrino emission efficiency with respect to the mass accretion rate is at most $6\%$ independent of the values of $\alpha_{\rm vis}$. However, the time scale on which the efficiency is maintained is more than 5 times longer for the model with a smaller value of $\alpha_{\rm vis}$, indicating that the smaller value of $\alpha_{\rm vis}$ leads to more efficient neutrino emission in the BH-torus system.

We find that the time evolution of the neutrino luminosity agrees well with the results of similar models in the previous study employing the leakage and moment  schemes~\cite{Fujibayashi:2020qda}. In fact, we found that for both models with $\alpha_{\rm vis}=0.05$ and $0.15$, the total neutrino luminosity of the present simulations agrees with the results of the K8 and K8s models in the previous study within $\approx30\%$ after $t=20\,{\rm ms}$ and up to the onset of mass ejection. Since the radiative transfer schemes in~\cite{Fujibayashi:2020qda} and this paper are developed independently, this approximate agreement indicates a good treatment of neutrino radiative transfer employed in both works.

The total pair annihilation energy deposition rate is higher than $10^{51}\,{\rm erg/s}$ for the first few $10\,{\rm ms}$, but steeply declines thereafter, and it becomes smaller than $10^{49}\,{\rm erg/s}$ after $0.1\,{\rm s}$. We should note that the high deposition rate in the first few $10\,{\rm ms}$ can be the artifact of the initial condition in the relaxation phase. The pair annihilation efficiency with respect to the total neutrino luminosity is smaller than $3\%$ after $20\,{\rm ms}$. This result is a factor of few larger than the findings in the previous study~\cite{Just:2015fda}, which may due to the fact that the inner region in the vicinity of the BH within the radius of $10\,{\rm km}$, in which the pair annihilation energy deposition rate is the highest, is not solved in Ref.~\cite{Just:2015fda}. We note that the our treatment of neglecting the electron mass in the deposition rate could also result in the overestimation the deposition rate, although the effect may be minor considering that the average neutrino energy is much higher than $1\,{\rm MeV}$ (see also ~\cite{Richers:2015lma} for the discussion).

The total energy deposited by pair annihilation is $8.8\times10^{49}\,{\rm erg}$ and $9.8\times10^{49}\,{\rm erg}$ for the models with $\alpha_{\rm vis}=0.05$ and $0.15$, respectively. However, most of the energy is deposited within the first $20\,{\rm ms}$, and as is discussed above, the values may suffer from the artifact of the initial condition. The total energy deposited after $20\,{\rm ms}$ is found to be $2.4\times10^{49}\,{\rm erg}$ and $9.4\times10^{48}\,{\rm erg}$ for $\alpha_{\rm vis}=0.05$ and $0.15$, respectively. These values are also consistent with the findings in the previous study~\cite{Just:2015fda}.

\begin{figure*}
 	 \includegraphics[width=0.48\linewidth]{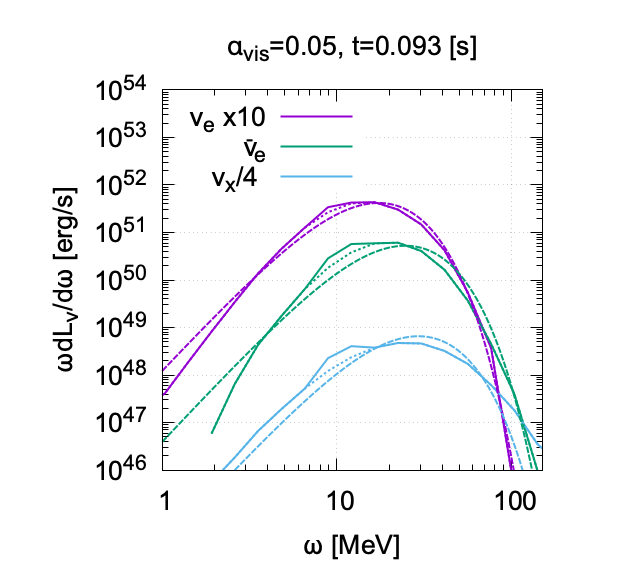}
 	 \includegraphics[width=0.48\linewidth]{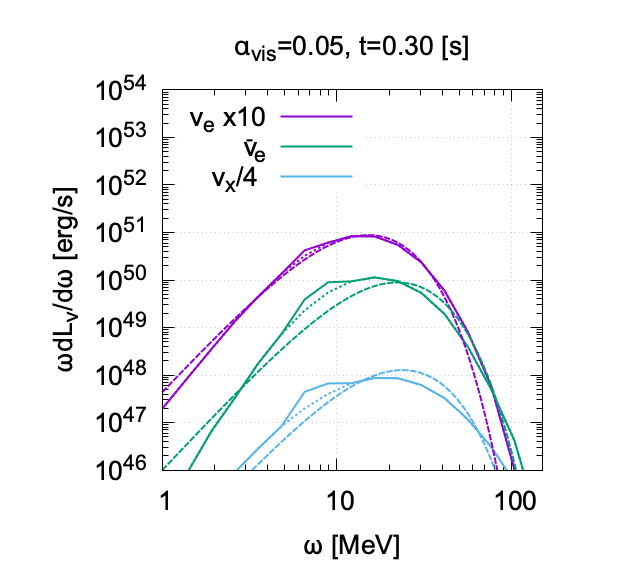}\\
 	 \includegraphics[width=0.48\linewidth]{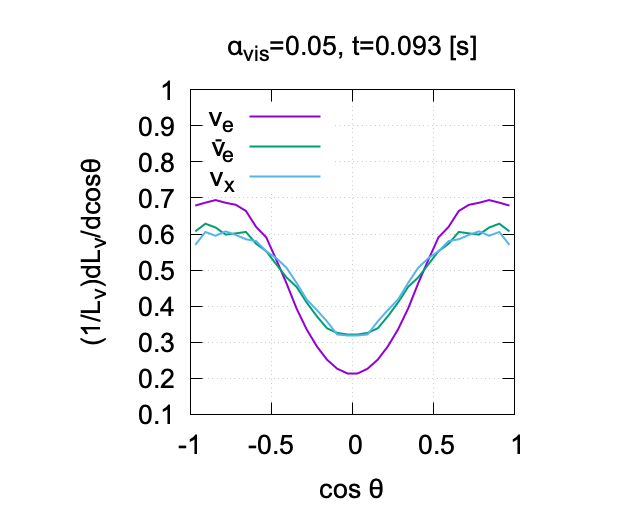}
 	 \includegraphics[width=0.48\linewidth]{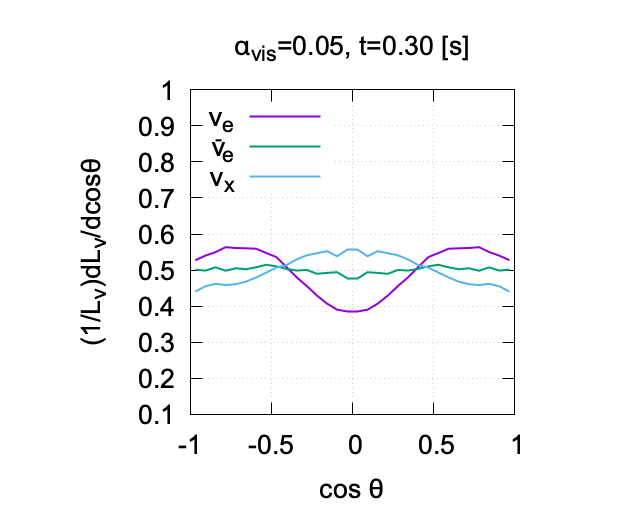}
 	 \caption{{\it Top panels:} energy distributions of emitted neutrinos at $t=0.093\,{\rm ms}$ (left) and $t=0.30\,{\rm ms}$ (right). The dotted curves denote the distribution obtained by only employing MC packets without the residual flag. The dashed curves denote the fitted results employing the zero-chemical potential black-body distribution. {\it Bottom panels:} The angular distributions of emitted neutrinos normalized with the total luminosity at $t=0.093\,{\rm ms}$ (left) and $t=0.30\,{\rm ms}$ (right). Both energy and angular distributions are obtained by correcting MC packets in $r=428$-$450\,{\rm km}$.}
	 \label{fig:nu_dist}
\end{figure*}

The top panels of Fig.~\ref{fig:nu_dist} show the energy distributions of emitted neutrinos at $t=0.093\,{\rm s}$ and $t=0.30\,{\rm s}$ for $\alpha_{\rm vis}=0.05$. The energy distributions are well fitted by the zero chemical potential black-body distributions with the temperature of $4.3\,{\rm MeV}$, $6.0\,{\rm MeV}$, and $7.2\,{\rm MeV}$ at $t=0.093\,{\rm s}$, and $3.7\,{\rm MeV}$, $5.4\,{\rm MeV}$, and $5.8\,{\rm MeV}$ at $t=0.30\,{\rm s}$ for electron neutrinos, electron antineutrinos, and heavy-lepton type neutrinos, respectively (see also~\cite{Richers:2015lma} for the similar results). These values are smaller by a factor of 2--3 than the average energy of emitted neutrinos, which are  $11.3\,{\rm MeV}$, $14.6\,{\rm MeV}$, and $16.3\,{\rm MeV}$ at $t=0.093\,{\rm s}$ and $10.5\,{\rm MeV}$, $12.3\,{\rm MeV}$, and $12.6\,{\rm MeV}$ at $t=0.30\,{\rm s}$ for electron neutrinos, electron antineutrinos, and heavy-lepton type neutrinos, respectively (see Fig.~\ref{fig:nu_detail}), and is approximately consistent with the fact that the average energy of the zero-chemical potential black-body distribution of the temperature of $T_{\rm BB}$ is given by $\approx 3.1\,T_{\rm BB}$.

We note that the bump-like features found around $10\,{\rm MeV}$ in the energy distributions are the artifact of the residual packet prescription (see Sec.~\ref{sec:method}). In fact, these features disappear for the distributions employing MC packets without the residual flag (see the dotted curves in the top panels of Fig.~\ref{fig:nu_dist}). This happens because the MC packets with the residual flag are corrected and recreated with a limited number of MC packets at the beginning of the time step. Since the recreation of the residual packets is done so that the energy-momentum and neutrino numbers are conserved, the neutrino energy of the recreated packets is concentrated in the certain energy. Nevertheless, the total radiation energy of such MC packets with the residual flag is at most $\approx 10\%$ of the total, and the modification of the energy distribution is minor.

The bottom panels of Fig.~\ref{fig:nu_dist} show the angular distribution of emitted neutrinos at $t=0.093\,{\rm s}$ and $t=0.30\,{\rm s}$ for $\alpha_{\rm vis}=0.05$. At $t=0.093\,{\rm s}$, all neutrino species show a similar angular dependence. Neutrino energy is emitted preferentially in the polar direction, reflecting the fact that the projected area of the neutrino last scattering surface becomes the maximum for the face-on view of the torus. Electron neutrinos show a stronger angular dependence than the other species, indicating that electron neutrinos are more optically thick and the last scattering surface has a more oblate shape than that of the other species. At $t=0.3\,{\rm s}$, the angular dependence becomes less significant due to the decrease in the neutrino optical depth of the torus. In particular, electron antineutrinos show approximately isotropic emission, while electron neutrinos still show a mild angular dependence because the optical depth is larger than that for others. Interestingly, heavy-lepton type neutrinos show the opposite angular dependence from the early phase: the more energy is emitted in the equatorial direction than in the polar direction. By checking the neutrino emissivity and energy flux, we find that this is because heavy-lepton type neutrinos are emitted in the vicinity of the BH, and those propagated in the equatorial direction are boosted with the orbital motion.

\begin{figure*} 
 	 \includegraphics[width=0.48\linewidth]{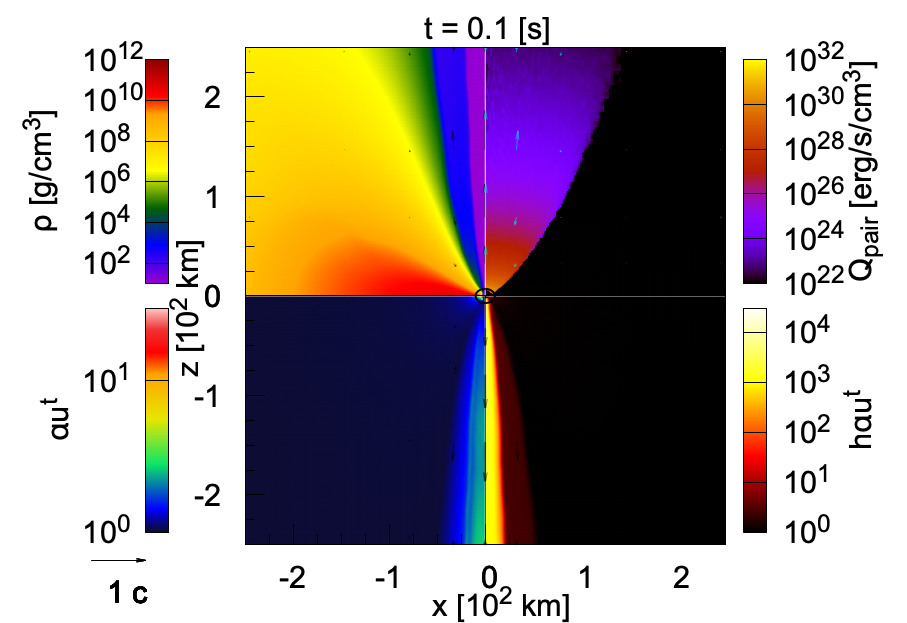}
 	 \includegraphics[width=0.48\linewidth]{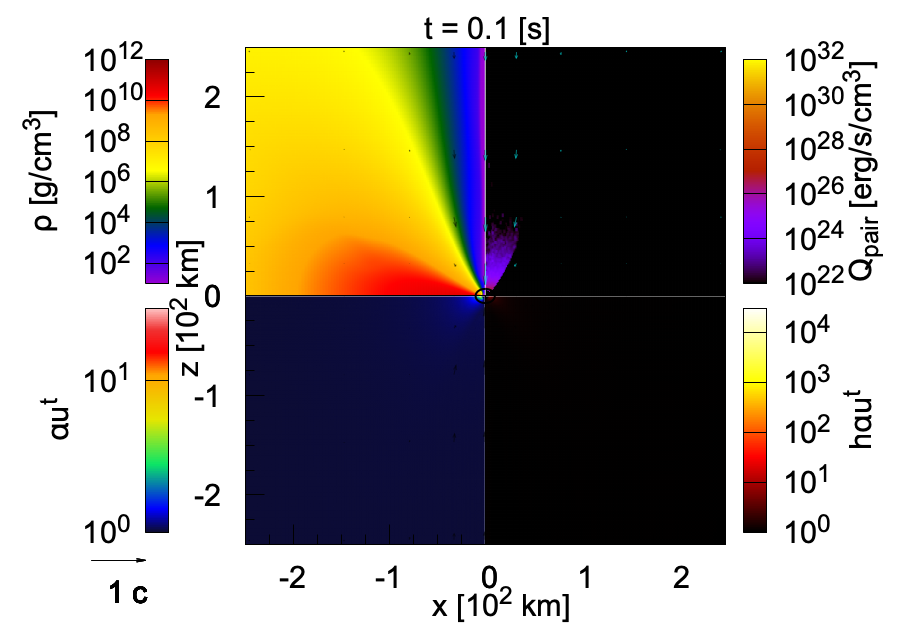}\\
 	 \includegraphics[width=0.48\linewidth]{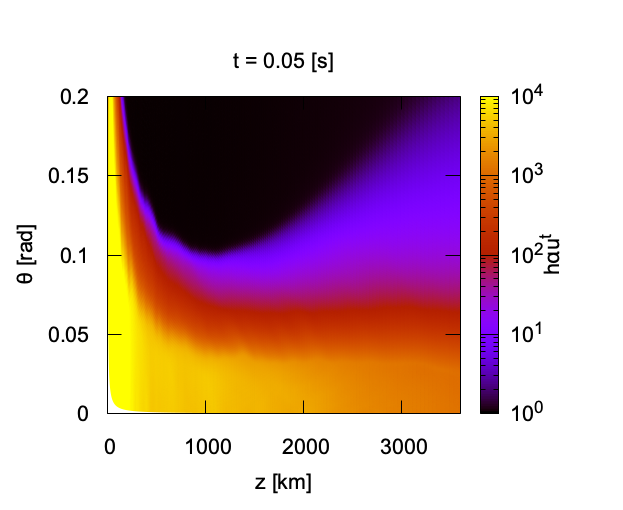}
 	 \includegraphics[width=0.48\linewidth]{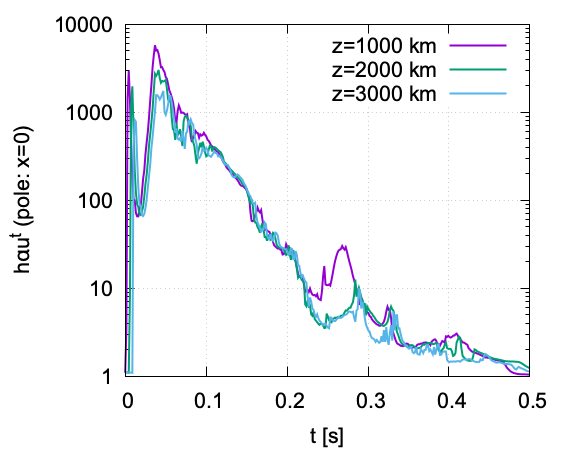}
 	 \caption{{\it Top panels:} snapshots of the baryon mass density, local pair annihilation energy deposition rate, $\alpha u^t$, and $h\alpha u^t$ profiles for the model with $\alpha_{\rm vis}=0.05$ at $t=0.1\,{\rm s}$. The left and right panels denote the results with and without taking $\nu_e{\bar \nu}_e$ pair process, respectively (note that $\nu_x{\bar \nu}_x$ pair process is kept turned on). {\it Bottom left panel:} polar profile of $h\alpha u^t$ for $\alpha_{\rm vis}=0.05$ at $t=0.05\,{\rm s}$ in which $\nu_e{\bar \nu}_e$ pair process is taken into account. {\it Bottom right panel:} Time evolution of $h\alpha u_t$ at the pole ($x=0$) for $z=1000\,{\rm km}$, $2000\,{\rm km}$, and $3000\,{\rm km}$.}
	 \label{fig:jet}
\end{figure*}

To investigate the role of the $\nu_e{\bar \nu}_e$ pair process in the dynamics, we performed a simulation with $\alpha_{\rm vis}=0.05$ but with neglecting $\nu_e{\bar \nu}_e$ pair process (note that $\nu_x{\bar \nu}_x$ pair process is kept turned on). We find that the overall dynamics and evolution of key quantities, such as the mass accretion rate, neutrino luminosity and energy, and mass-averaged thermodynamical valuables, are approximately unchanged regardless of whether $\nu_e{\bar \nu}_e$ pair process is taken into account or not (see App.~\ref{app:comp}). 

%\ms{The term of "pair process of electron-type neutrinos" is a bit confusing. It would be better to simply write $\nu_e \bar \nu_e$ pair process.}

The most and only significant difference which $\nu_e{\bar \nu}_e$ pair process induces is the presence of a relativistic outflow in the early phase. The top panels in Fig.~\ref{fig:jet} show the snapshots of the baryon mass density, local pair annihilation energy deposition rate, $\alpha u^t$ profiles, and $h\alpha u^t$ profiles for the model with $\alpha_{\rm vis}=0.05$ at $t=0.1\,{\rm s}$ with and without taking $\nu_e{\bar \nu}_e$ pair process. Here, $\alpha u^t$ and $h\alpha u^t$ can be regarded as the Lorentz factor taking into account the effect of gravitational potential and terminal Lorentz factor for which case all the thermal energy is converted into the kinetic energy, respectively. We can observe the presence of the relativistic outflow for the case that $\nu_e{\bar \nu}_e$ pair process is taken into account. In fact, the local pair annihilation deposition rate becomes as large as $10^{31}\,{\rm erg/s/cm^3}$ in the vicinity of the event horizon with $\nu_e{\bar \nu}_e$ pair annihilation, which is sufficiently large for the local matter with the density of $\leq 10^{6}\,{\rm g/cm^3}$ to be relativistic within the dynamical time scale (a few $10\,{\rm ms}$). 

The bottom left panel of Fig.~\ref{fig:jet} shows the polar profile of $h\alpha u^t$ at $t=0.05\,{\rm s}$ for the model with $\nu_e{\bar \nu}_e$ pair process being taken into account. We find that the matter with $h\alpha u^t>100$ is collimated within $\theta\approx0.05$-$0.1\,{\rm rad}$ with $\theta$ being the angle measured from the $z$-axis in the presence of the dense torus matter. The right panel of Fig.~\ref{fig:jet} shows the time evolution of $h\alpha u^t$ at the pole ($x=0$) for $z=1000\,{\rm km}$, $2000\,{\rm km}$, and $3000\,{\rm km}$. A relativistic outflow of which $h\alpha u^t$ exceeds 100 is formed around the pole and sustained up to $0.15\,{\rm s}$. However, the value of $h\alpha u^t$ steeply decreases with time and is larger than 10 only up to $\approx 0.2\,{\rm s}$. 

We find that the outflow luminosity in the polar region for which $h\alpha u^t$ is larger than 100 varies from $\sim 10^{49}\,{\rm erg/s}$ to $\sim10^{48}\,{\rm erg/s}$ for $t=0.02$--$0.1\,{\rm s}$, which is more than an order of magnitude lower than the total neutrino pair annihilation deposition rate (see Fig.~\ref{fig:nu_detail}). This happens because the region with the highest neutrino pair annihilation deposition rate is located in the vicinity of the BH's horizon, and the matter there accretes to the BH without being ejected. In fact, we find that the outflow luminosity with $h\alpha u^t>100$ approximately agrees with the total neutrino pair annihilation deposition rate of the region in which the baryon mass density is less than $10^{6}\,{\rm g/cm^3}$ and the fluid velocity is directed outward. By integrating in time, the total energy of the relativistic outflow with $h\alpha u^t>100$ is found to be $3\times10^{47}\,{\rm erg}$ for $t\geq20\,{\rm ms}$, and this corresponds to the isotropic-equivalent outflow energy of $\approx 3\times 10^{50}\,{\rm erg}$ for the collimation angle of $0.05\,{\rm rad}$. This suggests that the relativistic outflow found in our model is energetic enough to explain some of short-hard gamma-ray bursts and the precursors (e.g.,~\cite{Nakar:2007yr} and~\cite{2010ApJ...723.1711T,Xiao:2022quv}; see also Fig.~\ref{fig:jet}).%\ms{Again, isotropic energy seems to be high enough.}

There are several caveats to our results of the relativistic outflow. In the presence of the relativistic outflow, the baryon mass density in the polar region reaches the floor value, and hence, the values of the Lorentz factor and outflow energy are not very reliable. The presence of the high Lorentz factor outflow in the early time ($t\leq50\,{\rm ms}$) could also be the artifact of the initial condition. We also note that, while the matter density outside the torus is set to be a very low value ($10\,{\rm g/cm^3}$) in our initial condition, the matter density in the polar region may have higher density in the realistic situation~(see e.g.,~\cite{Hotokezaka:2012ze}). In such a situation, the formation of a relativistic outflow would be suppressed due to the heavy baryon loading~\cite{Just:2015fda}. In fact, we do not observe the formation of a relativistic outflow for $\alpha_{\rm vis}=0.15$. We interpret this as the consequence that the higher density in the polar region is realized for the model with larger value of $\alpha_{\rm vis}$ due to more rapid expansion of the torus. However, the matter distribution in the polar region and at the torus limb, which is also essential for the collimation of the relativistic outflow, may also suffer from the initial transient behavior at the beginning of the simulations. Hence, the presence of the relativistic outflow can be dependent on the property of the remnant system, and the further investigation is necessary to understand the systematic errors and quantitative dependence.

\begin{figure*} 
 	 \includegraphics[width=0.48\linewidth]{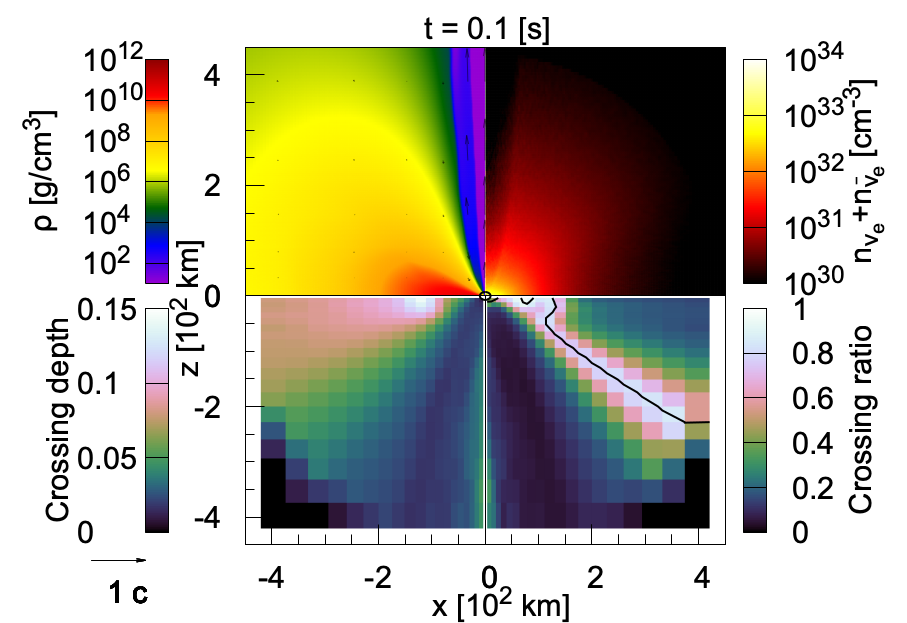}
 	 \includegraphics[width=0.48\linewidth]{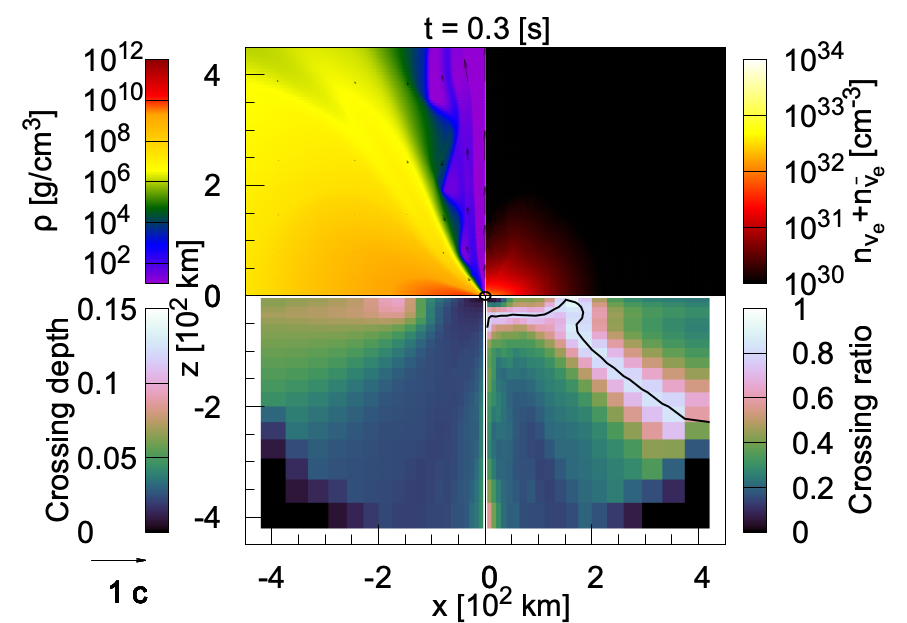}
 	 \caption{Snapshots of the baryon mass density, sum of electron neutrino and electron antineutrino number density, crossing depth, and crossing ratio for the model with $\alpha_{\rm vis}=0.05$ at $t=0.1\,{\rm s}$ (left) and at $t=0.3\,{\rm s}$ (right). The black solid curves denote the location where electron neutrino and electron antineutrino number density are identical. To suppress the statistical error, we take sum of the MC packets in the neighbor $10\times10$ hydrodynamics cells and take $1.5\,{\rm ms}$ average to obtain the neutrino distribution function. We note that the profiles of the crossing depth and ratio are truncated above the radius of $r=300\,M_\odot\approx450\,{\rm km}$ because neutrino radiation fields are solved only within that radius. We also note that the value of the crossing depth around the polar region is not reliable due to the MC shot noise.}
	 \label{fig:ncross}
\end{figure*}

Neutrino distribution functions are directly obtained in our simulation. Such information is useful to indicate the possible occurrence of the fast flavor instability (FFI). Following a previous study~\cite{Richers:2022dqa}, we compute the crossing depth, ${\cal D}$, and crossing ratio, ${\cal R}$, defined by
\begin{align}
    {\cal D}&=\frac{\sqrt{I_+ I_-}}{n_{\nu_e}+n_{{\bar \nu}_e}},\nonumber\\
    {\cal R}&={\rm min}\left(\frac{I_+}{I_-},\frac{I_-}{I_+}\right),
\end{align}
where
\begin{align}
    I_\pm&=\int d\Omega G\Theta\left(\pm G\right),\nonumber\\
    G&=\int d \omega \omega^2 f_{\nu_e}-\int d {\bar \omega} {\bar \omega}^2 f_{{\bar \nu}_e},
\end{align}
with $f_{\nu_e}$ and $f_{{\bar \nu}_e}$ being the distribution functions of electron neutrinos and electron antineutrinos. Here, the crossing depth is related to the growth rate of the instability by $\sim \sqrt{2} G_{\rm F}{\cal D}(n_{\nu_e}+n_{{\bar \nu}_e})$, and the crossing ratio indicates the relative amount of electron neutrino and electron antineutrino excess which is related to the total amount of flavor change~\cite{Richers:2022dqa}. 

Figure~\ref{fig:ncross} shows the snapshots of the crossing depth and crossing ratio for the model with $\alpha_{\rm vis}=0.05$ at $t=0.1\,{\rm s}$ and at $t=0.3\,{\rm s}$. We note that the value of the crossing depth around the polar region is not reliable due to the MC shot noise. We find that the crossing depth is always larger than 0.05 for the latitudinal angle larger than $2\pi/3$ at both $t=0.1\,{\rm s}$ and $0.3\,{\rm s}$. The large value of the crossing depth is found particularly around the equatorial plane, and the position of the peak is shifted slightly outward from the point of the maximum baryon mass density, as found in the previous study~\cite{Richers:2022dqa}. The crossing ratio is the highest around the location where electron neutrino and electron antineutrino number density equates, and the location approximately matches to the edge of the region where the crossing depth is larger than 0.05. The crossing ratio is high ($>0.5$) around the peak of crossing depth while it is relatively low ($<0.2$) around the equatorial plane. This suggests that the neutrino flavor conversion may efficiently take place around the peak of the crossing depth.

The latitudinal extent of the region of which the crossing depth value is larger than 0.05 is slightly shrinking and the peak position shifts outward with its peak value decreasing as the time evolves from $t=0.1\,{\rm s}$ to $0.3\,{\rm s}$. This indicates that the system tends to evolve in a direction where the FFI is less likely to occur. This trend in the time evolution is also broadly consistent with the previous study~\cite{Mukhopadhyay:2024zzl}.

\section{Summary}\label{sec:sum}
In this paper, we presented our new general relativistic MC-based neutrino radiation hydrodynamics code designed to solve axisymmetric systems, for which we made the improvement for several implementation from the previous studies~\cite{Miller:2019dpt,2020ApJ...902...66M,Foucart:2022kon,Sprouse:2023cdm,Mukhopadhyay:2024zzl}. The major improvements are as follows: first, we derived and implemented an extended version of the implicit MC method for multi-species radiation fields. Second, a new numerically efficient and asymptotically correct fitting function for the neutrino pair process kernel function is derived and employed for absorption rates and emissivity. Finally, we introduce new numerical limiters for radiation-matter interaction to ensure stable and physically correct evolution of the system. A higher-order MC scheme for matter-radiation interaction introduced in our previous study~\cite{Kawaguchi:2022tae} is also employed in the code. We demonstrate that the thermalization of one-zone systems is correctly solved by our code for various situations.
%\ms{This sentence is not very clear. You introduced an idea in the previous paper but did not implement it, right?}

We applied our code to a BH-torus system with the BH mass of $3\,M_\odot$, dimmensionless spin of 0.8, and torus mass of $0.1\,M_\odot$, which mimics a post-merger remnant of a binary NS merger for the case that the massive NS collapses to a BH within a short time scale ($\sim10\,{\rm ms}$). We find that the evolution of the system and the various key quantities, such as neutrino luminosity, ejecta mass, torus $Y_e$, and pair annihilation luminosity are in broad agreement with the results of the previous studies employing the leakage or moment schemes~\cite[e.g.,][]{Fujibayashi:2020qda,Just:2021cls} (and also the studies based on MC schemes~\cite{Richers:2015lma,Miller:2019dpt,2020ApJ...902...66M,Foucart:2022kon,Sprouse:2023cdm,Mukhopadhyay:2024zzl}). Quantitatively, we found some differences from the results of the previous studies with similar setup: For example, the average torus $Y_e$ values of the present simulation results are found to be larger ($\approx 0.05$) than in the previous study~\cite{Fujibayashi:2020qda}. While our simplification in the microphysics may be responsible for the difference, the difference in the radiation transfer solver may also be responsible for the quantitative difference between the present and previous results. To clarify the reason for the discrepancy, a detailed comparison under the same initial condition and microphysical setups is crucial.

We found that a mildly relativistic outflow is formed in the polar region for the model with $\alpha_{\rm vis}=0.05$. We confirmed that the relativistic outflow is launched $\nu_e{\bar \nu}_e$ by pair annihilation. In fact, we found that the relativistic outflow is absent for the case that $\nu_e{\bar \nu}_e$ pair process is switched off in the simulation. Except for the presence of the relativistic outflow, the other dynamics and evolution of the system are found to be approximately the same (see App.~\ref{app:comp}).

For the model with $\alpha_{\rm vis}=0.05$, the outflow with the terminal Lorentz factor larger than 100 is sustained up to $\approx0.1\,{\rm s}$, and the region is collimated within $\approx 0.05$-$0.1\,{\rm rad}$ due to the funnel-like matter distribution at the torus limb. The total energy of the relativistic outflow is found to be $\approx 3\times 10^{\rm 47}\,{\rm erg}$. Our result is also broadly in agreement with the result of the model in~\cite{Just:2015fda} in which a relativistic outflow is launched (their model TM1-1415) except for the smaller outflow energy and narrower opening angle. For the collimation angle of $\approx 0.05\,{\rm rad}$, the isotropic-equivalent energy of the relativistic outflow is $\approx 3\times 10^{50}\,{\rm erg}$. The relativistic outflow found in our model is  energetic enough to explain some of short-hard gamma-ray bursts (\cite{Nakar:2007yr}, see also Fig.~\ref{fig:jet}). The time scale and energy of the relativistic outflow are also consistent with the precursors of gamma-ray bursts~\cite{2010ApJ...723.1711T,Xiao:2022quv}. Our result suggests that neutrino pair annihilation can contribute to launch a relativistic outflow in the early phase ($\sim0.01$-$0.1\,{\rm s}$) of the remnant BH-torus evolution, which may be comparable to the time scale of the magnetic field amplification~\cite{Kiuchi:2023obe}. Therefore, investigating the potential role of a pair annihilation driven outflow for explaining the central engine of short gamma-ray bursts in combination with magnetic field dynamics is an important future task.%\ms{It is not clear whether this statement is right, because if the small opening angle is taken into account, the isotropic energy could be $10^{51}$\,erg.}

We should caution that the quantitative properties of the relativistic outflow found in our result may not be very reliable since the baryon mass density in the polar region touches the floor value and the high energy deposition rate by pair annihilation in the very early phase ($\lesssim 10\,{\rm ms}$) of the evolution can be the artifact of the initial condition. We also found that whether or not a relativistic outflow is launched can depend on the effective viscous parameter and the initial profile of the remnant system. The matter distribution in the polar region and at the torus limb, which is also essential for the collimation of the relativistic outflow, may also suffer from the initial transient behavior at the beginning of the simulations. Hence, more systematic studies are necessary to understand the systematic errors and quantitative dependence on the property of the remnant system.

The direct determination of the full distribution function is one of the advantages of employing the MC-based method. To demonstrate its usefulness, we calculated the indicators of the neutrino FFI introduced in~\cite{Richers:2022dqa} directly employing the obtained distribution functions. Our analysis of the FFI showed that neutrino crossing is strongly indicated particularly around the equatorial plane. We found that the profile and strength of the indicator are broadly in agreement with the previous studies employing MC radiative transfer codes~\cite{Richers:2022dqa,Mukhopadhyay:2024zzl}. 

There are several tasks remaining for the development of our code. For example, other than the implementation of more realistic microphysics, numerical techniques to reduce the computational cost in the high absorption/scattering opacity regions will be important to study optically thicker systems such as a remnant massive NS by MC schemes. For this purpose, the implementation of the discrete diffusion technique~\citep{2015ApJS..217....9R,2015ApJ...807...31R,2018PhRvD..98f3007F} or the hybrid method~\citep{Izquierdo:2023fub,Foucart:2017mbt} with moment schemes could significantly help to reduce the computational costs.

%\begin{acknowledgments}
\acknowledgments KK thanks Alexander Alekseenko, Aurore Betranhandy, Francois Foucart, Kenta Kiuchi, Takami Kuroda, Jonah Miller, David Radice, and Sherwood Richers for valuable discussions. Numerical computation was performed on Yukawa21 at Yukawa Institute for Theoretical Physics, Kyoto University and the Sakura and Momiji clusters at Max Planck Computing and Data Facility. This work was supported by Grant-in-Aid for Scientific Research (23H04900) of JSPS/MEXT.
%\end{acknowledgments}

\appendix
\section{Implicit Monte Carlo method for multiple species}\label{app:imc-msp}
In this section, we describe a generalization of the implicit MC method introduced in~\cite{1971JCoPh...8..313F} to multi-species radiation fields.

The time evolution of the monochromatic intensity, $I_\nu^a$, in the fluid rest-frame for the radiation field $a$ is given by
\begin{align}
\frac{d I_\nu^a}{dt}=-\alpha_\nu^a\left(I_\nu^a-\frac{1}{4\pi}u_{\nu,{\rm th}}^a\right),\label{eq:IMC1}
\end{align}
where $d/dt$ is the time derivative in the fluid rest-frame, $\nu$ is the energy (wavelength) of the radiation, and $\alpha_\nu^a$ is the absorption rate. Note that the curvature of the space time and spatial dependence of the fluid velocity are neglected in Eq.~\eqref{eq:IMC1} considering the case that the mean free path of the radiation is small. By integrating out the angular dependence of $I_\nu^a$, we obtain the evolution equation for the monochromatic energy density, $u_\nu^a$, as
\begin{align}
\frac{d u_\nu^a}{dt}=-\alpha_\nu^a\left(u_\nu^a-u_{\nu,{\rm th}}^a\right),
\label{eq:IMC2}
\end{align}
where $u_{\nu,{\rm th}}$ is the monochromatic energy density of the radiation field in the thermal equilibrium, which is determined by the local thermodynamics quantities of the matter.

% \ms{Question: in relativity, is the description here valid? $t$ seems to be replaced by the affine parameter. }

For the case that the change in the matter internal energy is dominated by the interaction with the radiation fields, the evolution of the matter internal energy density $u_{\rm fl}$ is given by
\begin{align}
\frac{d u_{\rm fl}}{dt}&=\sum_{a}\int d\nu \alpha_\nu^a\left(u_\nu^a-u_{\nu,{\rm th}}^a\right).
\end{align}

Assuming that the change in $u_{\nu,{\rm th}}$ is the primary factor for changing $u_{\rm fl}$ during the evolution, the time derivative of $u_{\rm fl}$ can be written as
\begin{align}
\frac{d u_{\nu,{\rm th}}^a}{dt}&=\frac{d u_{\nu,{\rm th}}^a}{d u_{\rm fl}}\frac{d u_{\rm fl}}{dt}\\
&=\beta_\nu^a \sum_{b}\int d{\nu'} \alpha_{\nu'}^b\left(u_{\nu'}^b-u_{{\nu'},{\rm th}}^b\right),\label{eq:IMC3}
\end{align}
where $\beta_nu^a$ is given by $\beta_\nu^a=\displaystyle \frac{d u_{\nu,{\rm th}}^a}{d u_{\rm fl}}$. By integrating Eq.~\eqref{eq:IMC3} in time from $t=t^n$ to $t^{\rm n+1}=t^n+\Delta t$ gives
\begin{align}
\frac{u_{\nu,{\rm th}}^{a,n+1}-u_{\nu,{\rm th}}^{a,n}}{\Delta t}&&=\beta_\nu^a \sum_{b}\int d{\nu'} \alpha_{\nu'}^b\left({\bar u}_{\nu'}^{b}-{\bar u}_{{\nu'},{\rm th}}^{b}\right).\label{eq:IMC4}
\end{align}
Here, the quantities with the superscription of $n$ and $n+1$ denote the values at $t=t^n$ and $t^{n+1}$, respectively, and the quantities with overlines are defined by the time average
\begin{align}
    {\bar X}_\nu^{b}=\frac{1}{\Delta t} \int ^{t^{n+1}}_{t^n}dt X_\nu^{b}.
\end{align}
Note that in derive Eq.~\eqref{eq:IMC4} it was assumed that the change in $\alpha_\nu^a$ and $\beta_\nu^a$ are not significant during the time evolution from $t=t^n$ to $t^{n+1}$.

By employing a parameter, $\lambda_\nu^a$, we can express ${\bar u}_{{\nu},{\rm th}}^{a}$ by
\begin{align}
{\bar u}_{{\nu},{\rm th}}^{a}=u_{{\nu},{\rm th}}^{a,\lambda}=(1-\lambda_\nu^a)u_{{\nu},{\rm th}}^{a,n}+\lambda_\nu^a u_{{\nu},{\rm th}}^{a,n+1}.\label{eq:IMC5}
\end{align}
We note that, while $\lambda_\nu^a$ is yet unspecified, regardless of the choice of $\lambda_\nu^a$, the evolution of $u_{{\nu},{\rm th}}^{a}$ given by substituting ${\bar u}_{{\nu'},{\rm th}}^{b}$ in Eq.~\eqref{eq:IMC4} by Eq.~\eqref{eq:IMC5} is accurate in the linear order of $\Delta t$. In particular, $\lambda_\nu^a=0$ and $1$ correspond to the fully explicit and implicit scheme with respect to $u_{{\nu},{\rm th}}^{a}$. Hence, $\lambda_\nu^a$ can be regarded as the parameter to control the numerical stability of the scheme.

By substituting ${\bar u}_{{\nu'},{\rm th}}^{b}$ in Eq.~\eqref{eq:IMC4} with Eq.~\eqref{eq:IMC5}, multiplying $\alpha_\nu^a\lambda_\nu^a$ to whole the equation, integrating over $\nu$, and taking the summation for the species, we obtain
\begin{align}
\sum_{a}\int d{\nu} &\alpha_{\nu}^a \lambda_{\nu}^a u_{{\nu},{\rm th}  }^{a,n+1}=\sum_{a}\int d{\nu} \alpha_{\nu}^a \lambda_{\nu}^a u_{{\nu},{\rm th}  }^{a,n}\nonumber \\
&+\frac{\sum_b \int d{\nu'} \alpha_{\nu'}^{b}\beta_{\nu'}^b \lambda_{\nu'}^b \Delta t}{1+\sum_b \int d{\nu'} \alpha_{\nu'}^{b}\beta_{\nu'}^b \lambda_{\nu'}^b \Delta t} \nonumber \\
&\times\left[\sum_{b}\int d{\nu'} \alpha_{\nu'}^b {\bar u}_{\nu'}^b-\sum_{b}\int d{\nu'} \alpha_{\nu'}^b u_{{\nu'},{\rm th}  }^{b,n}\right].\label{eq:IMC5-2}
\end{align}

We can use Eq.~\eqref{eq:IMC5-2} to remove $u_{{\nu'},{\rm th}  }^{b,n+1}$ in the right hand side of Eq.~\eqref{eq:IMC4} which appears after substituting ${\bar u}_{{\nu'},{\rm th}}^{b}$ employing Eq.~\eqref{eq:IMC5}. Then, again employing Eq.~\eqref{eq:IMC5}, we get%\ms{ [The first sentence is something wrong.]}
\begin{align}
\alpha_\nu^a u_{\nu,{\rm th}}^{a,\lambda}&=\alpha_\nu^a u_{\nu,{\rm th}}^{a,n}+\frac{\alpha^{a}_\nu\beta_\nu^a \lambda_\nu^a \Delta t}{1+\sum_b \int d{\nu'} \alpha_{\nu'}^{b}\beta_{\nu'}^b \lambda_{\nu'}^b \Delta t} \nonumber \\
&\times\left[\sum_{b}\int d{\nu'} \alpha_{\nu'}^b {\bar u}_{\nu'}^b-\sum_{b}\int d{\nu'} \alpha_{\nu'}^b u_{{\nu'},{\rm th}  }^{b,n}\right].\label{eq:IMC6}
\end{align}
By introducing a parameter $g_\nu^a$ which satisfies
\begin{align}
g_\nu^a \alpha_\nu^a u_{\nu,{\rm th}}^{a,n}&=\alpha_\nu^a u_{\nu,{\rm th}}^{a,n}\nonumber\\
&-\frac{\alpha^{a}_\nu\beta_\nu^a \lambda_\nu^a \Delta t}{1+\sum_b \int d{\nu'} \alpha_{\nu'}^{b}\beta_{\nu'}^b \lambda_{\nu'}^b \Delta t}\sum_{b}\int d{\nu'} \alpha_{\nu'}^b u_{{\nu'},{\rm th}  }^{b,n},\label{eq:gcond}
\end{align}
Eq.~\eqref{eq:IMC6} can be simplified in the following form:
\begin{align}
\alpha_\nu^a u_{\nu,{\rm th}}^{a,\lambda}&=g_\nu^a\alpha_\nu^a u_{\nu,{\rm th}}^{a,n}+f_\nu ^a (1-\left<g\right>)\sum_{b}\int d{\nu'} \alpha_{\nu'}^b {\bar u}_{\nu'}^b.\label{eq:IMC7}
\end{align}
Here, $f_\nu^a$ and $\left<g\right>$ are given by
\begin{align}
    f_\nu^a&=\frac{(1-g_\nu^a)\alpha_\nu^a u_{\nu,{\rm th}}^{a,n}}{\sum_b \int d{\nu'}(1-g_{\nu'}^b)\alpha_{\nu'}^b u_{{\nu'},{\rm th}}^{b,n}},\nonumber\\
    \left<g\right>&=\frac{\sum_b \int d{\nu'} g_{\nu'}^b \alpha_{\nu'}^b u_{{\nu'},{\rm th}}^{b,n}}
    {\sum_b \int d{\nu'} \alpha_{\nu'}^b u_{{\nu'},{\rm th}}^{b,n}}.
\end{align}
Since $g_\nu^a$ provides simpler expression, we actually control $g_\nu^a$ instead of $\lambda_\nu^a$ as the parameter of the scheme to control the numerical stability.

Finally, by integrating Eq.~\eqref{eq:IMC1} in time from $t=t^n$ to $t^{\rm n+1}$ and employing Eqs.~\eqref{eq:IMC4} and ~\eqref{eq:IMC7} we obtain
\begin{align}
\frac{I_\nu^{a,n+1}-I_\nu^{a,n}}{\Delta t}&=-\left<g\right>\alpha_\nu^a{\bar I}_\nu^a+\frac{1}{4\pi}g_\nu^a \alpha_\nu^au_{\nu,{\rm th}}^{a,n}\nonumber\\
&-(1-\left<g\right>)\alpha_\nu^a{\bar I}_\nu^a\nonumber\\
&+\frac{1}{4\pi}f_\nu^a\sum_{b}\int d{\nu'} (1-\left<g\right>)\alpha_{\nu'}^b {\bar u}_{\nu'}^b.\label{eq:IMC8}
\end{align}
Here, we note that ${\bar I}_\nu^a$ and ${\bar u}_\nu^a$ are not replaced by the expression using  $\lambda_\nu^a$ (Eq.~\eqref{eq:IMC5}) because those quantities are directly obtained by solving the MC transport. The first two terms in the right-hand side of this equation can be interpreted as the absorption and emission terms, for which the absorption rate and emissivity are modified by $\alpha_\nu^a\rightarrow\left<g\right>\alpha_\nu^a$ and $\alpha_\nu^a u_{\nu,{\rm th}}^{a,n} \rightarrow g_\nu^a \alpha_\nu^a u_{\nu,{\rm th}}^{a,n}$. The last two terms can be interpreted as
isotropic and inelastic scattering process with the rate of $(1-\left<g\right>)\alpha_\nu^a$, and for which the energy distribution of scattered particles is given by $f_\nu^a$. Indeed, we can easily find that the last two terms cancel out with each other if we integrate the whole equation with respect to the energy and angular dependence and taking the summation for the species. It is worth noting that, for the case that the radiation fields are in thermal equilibrium (i.e., $I_\nu^a={\bar I}_\nu^a=\frac{1}{4\pi}u_{\nu,{\rm th}}^{a,n}$), the right-hand side of the equation vanishes; this guarantees that the system is stationary in the thermal equilibrium state. 

To determine how we control $g_\nu^a$, we focus on the case that the system is very optically thick. For $\alpha_\nu^a \Delta t\gg 1$, the monochromatic intensity should be immediately thermalized, and hence, $I_\nu^{a,n+1}\approx{\bar I}_\nu^{a}\rightarrow\frac{1}{4\pi}u_{\nu,{\rm th}}^{a}$ with $u_{\nu,{\rm th}}^{a}$ being the energy density in thermal equilibrium. Then, by integrating out the energy and angular dependence and taking the summation for the species, $\alpha_\nu^a \Delta t\gg 1$ we obtain
\begin{align}
u_{\rm th}-u^{n}&\approx-\left<g\right>\left<\alpha\right>\Delta t\left(u_{\rm th}-u_{\rm th}^{n}\right)\nonumber\\
&=\left<g\right>\left<\alpha\right>_{\rm tot}\beta\Delta t\left(u_{\rm th}-u^{n}\right).\label{eq:gcond2}
\end{align}
Here, $u_{\rm th}$, $u^n$, and $\beta$ are obtained from $u_{\nu,{\rm th}}^{a}$, $u_{\nu}^{a,n}$, and $\beta_\nu^a$ by integrating out the energy and angular dependence and summing all the species of the radiation field, respectively. $\left<\alpha\right>_{\rm tot}$ is the total Planck-mean of the absorption rate defined by
\begin{align}
    \left<\alpha\right>_{\rm tot}=\frac{\sum_b \int d{\nu'} \alpha_{\nu'}^b u_{{\nu'},{\rm th}}^{b,n}}
    {\sum_b \int d{\nu'}  u_{{\nu'},{\rm th}}^{b,n}}. 
\end{align}

Equation~\eqref{eq:gcond2} suggests that $\left<g\right>\rightarrow 1/\left<\alpha\right>_{\rm tot}\beta\Delta t$ for $\alpha_\nu^a \Delta t\gg 1$. This condition can be satisfied by taking $g_\nu^a$ to be
\begin{align}
    g_\nu^a=g^a={\rm min}\left[\frac{1}{\left<\alpha\right>^a(1+\beta)\Delta t},1\right]
\end{align}
where $\left<\alpha\right>^a$ denotes the Planck-mean of the absorption rate for the radiation field of $a$ defined by 
\begin{align}
    \left<\alpha\right>^a=\frac{ \int d{\nu'} \alpha_{\nu'}^a u_{{\nu'},{\rm th}}^{a,n}}
    { \int d{\nu'}  u_{{\nu'},{\rm th}}^{a,n}}.
\end{align}
We note that the factor $\beta$ is slightly modified by $1+\beta$ motivated by the derivation in~\cite{2021ApJ...920...82F}. We also note that $\Delta t$ should be taken as the time interval in the comoving frame.

\section{One-zone Thermalization test}\label{app:th-test}
To validate our radiative transfer method and implementation of microphysics, we perform several test calculations following the thermalization of effectively one-zone systems. 

\begin{table}
\caption{List of the baryon mass density, initial temperature, and initial $Y_e$ values employed in the one-zone thermalization tests. $\Delta \tau_{\nu_i}\,(\nu_i=\nu_e,{\bar \nu}_e,\nu_x)$ denotes the initial Planck-mean optical depth in each hydrodynamics cell for each neutrino species.}
\centering
\begin{tabular}{c|c|c|c|c}\hline
Model	& $\rho\,[{\rm g/cm^3}]$&	$T\,[{\rm MeV}]$&	$Y_{e,{\rm ini}}$&    $(\Delta \tau_{\nu_e},\Delta \tau_{{\bar \nu}_e},\Delta \tau_{\nu_x})$\\\hline\hline
THtest1   &   $10^{11}$ &   $10$ &   $0.5$& $(0.52,0.30,4.0\times10^{-3})$\\
THtest2   &   $10^{13}$ &   $30$ &   $0.5$& $(4.8\times 10^2,1.3\times 10^2,0.91)$\\
THtest3   &   $10^{13}$ &   $10$ &   $0.5$& $(10,1.7,4.0\times10^{-4})$\\
THtest4   &   $10^{2}$  &   $20$ &   $0.1$& $(0.61,0.61,0.13)$\\\hline
\end{tabular}
\label{tb:thmodel}
\end{table}

For all the models, the initial conditions are composed of matter fields with a uniform baryon mass density, temperature, and electron fraction with no neutrino radiation fields at the beginning. A cylinder-shape computational domain of which cylindrical radius and half-height are given by $\approx15\,{\rm km}$ is considered imposing the axisymmetry and equatorial symmetry. Uniform 20 grids are employed to resolve the cylindrical radius and half of height in the vertical direction. The Courant number is set to be 0.5, which gives a time interval to be $\approx 1\,\mu{\rm s}$. $N_{\rm trg}=120$ and $r_{\rm abs}=0.1$ are employed in these tests, and as the result, total packet numbers vary from $10^6$ to $10^8$. %\ms{ [This sentence is something wrong.]}

Table~\ref{tb:thmodel} lists the model parameters for the one-zone systems examined in this test. The baryon mass density and temperature for model THtest1 broadly agree with those at the peak of the baryon mass density in the initial profile employed in this paper (see Figure~\ref{fig:snap_ini}). Those of model THtest2 broadly agree with those at the peak baryon mass density in the inner edge of the merger remnant torus in the presence of a hypermassive NS~\cite{Fujibayashi:2017xsz}. Model THtest3 is set to have the same density and electron fraction as in model THtest2 but with lower temperature to examine the code in a high electron degeneracy case. Finally, model THtest4 is employed to examine our code in the situation of low baryon mass density and large neutrino numbers.

\begin{figure*}
 	 \includegraphics[width=.41\linewidth]{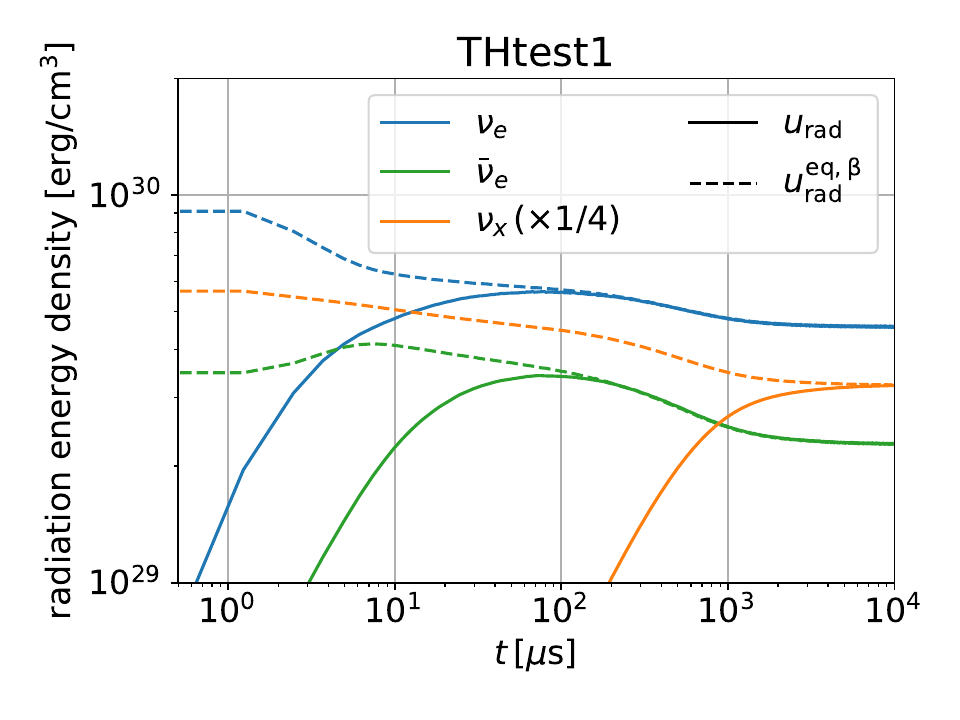}~~
 	 \includegraphics[width=.41\linewidth]{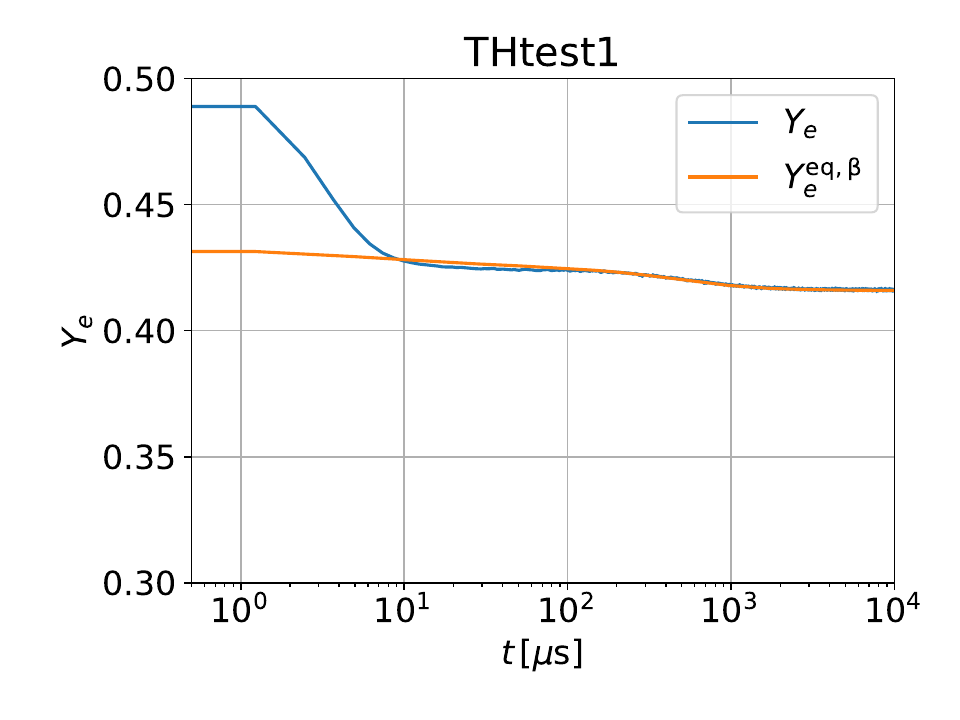}\\
 	 \includegraphics[width=.41\linewidth]{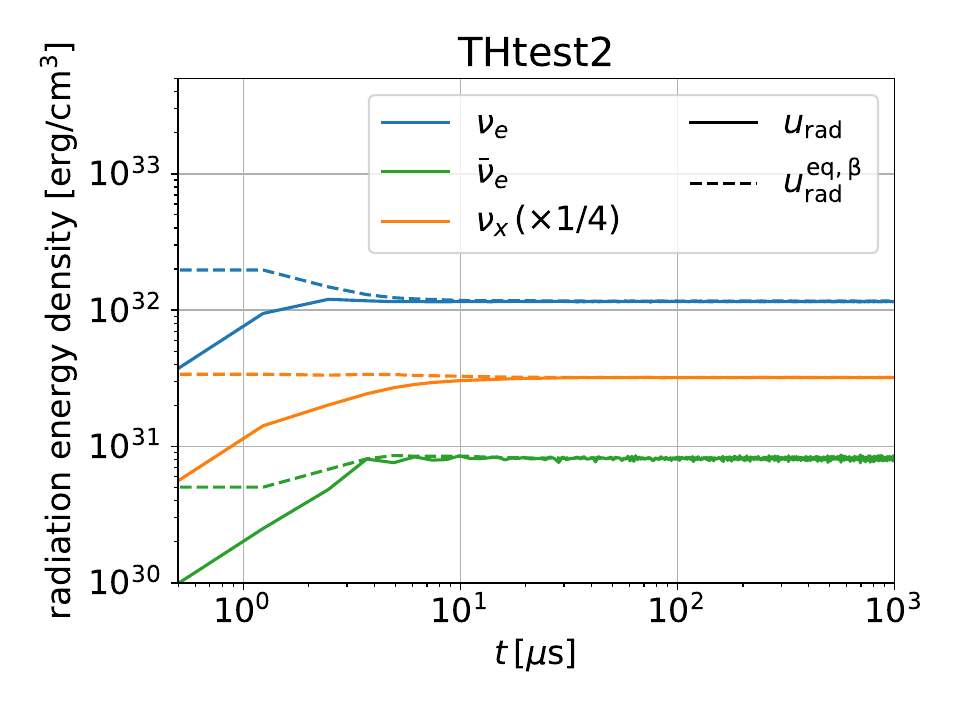}~~
 	 \includegraphics[width=.41\linewidth]{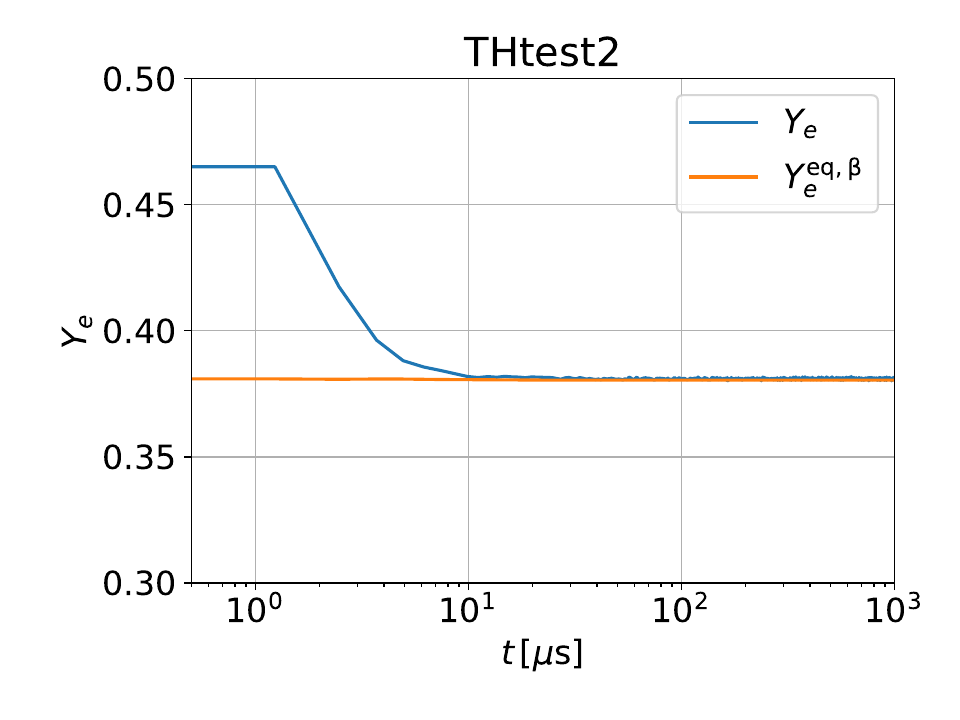}\\
 	 \includegraphics[width=.41\linewidth]{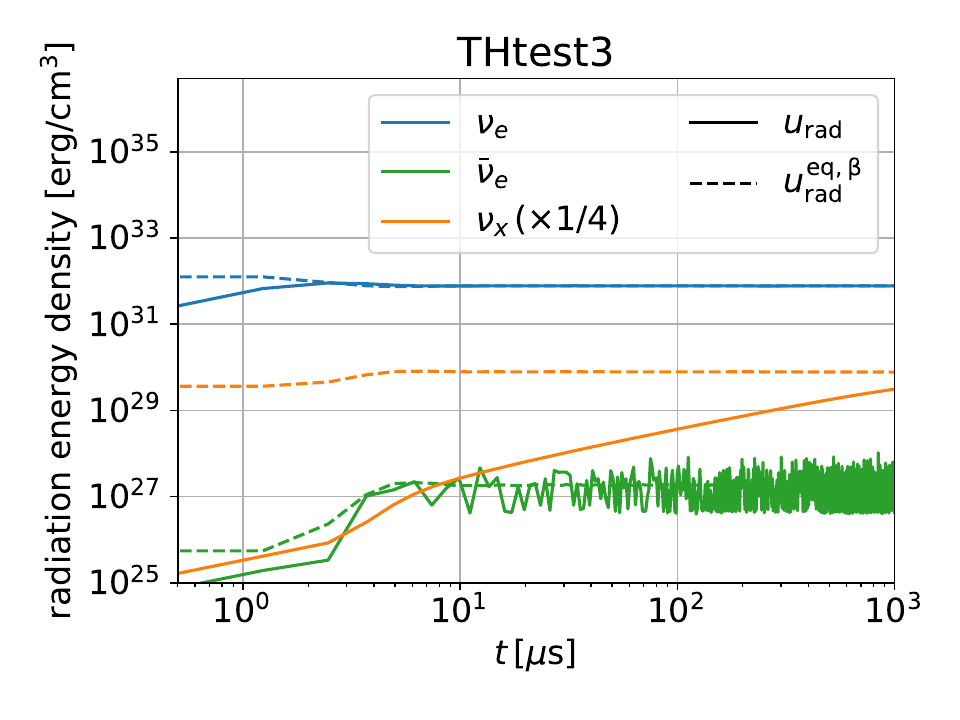}~~
 	 \includegraphics[width=.41\linewidth]{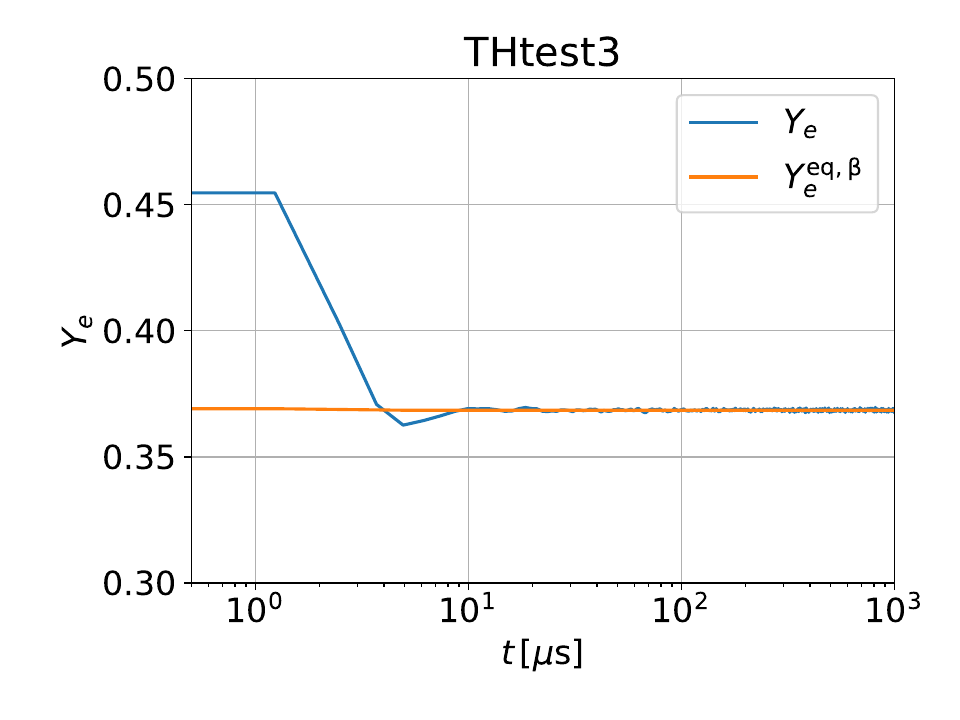}\\
 	 \includegraphics[width=.41\linewidth]{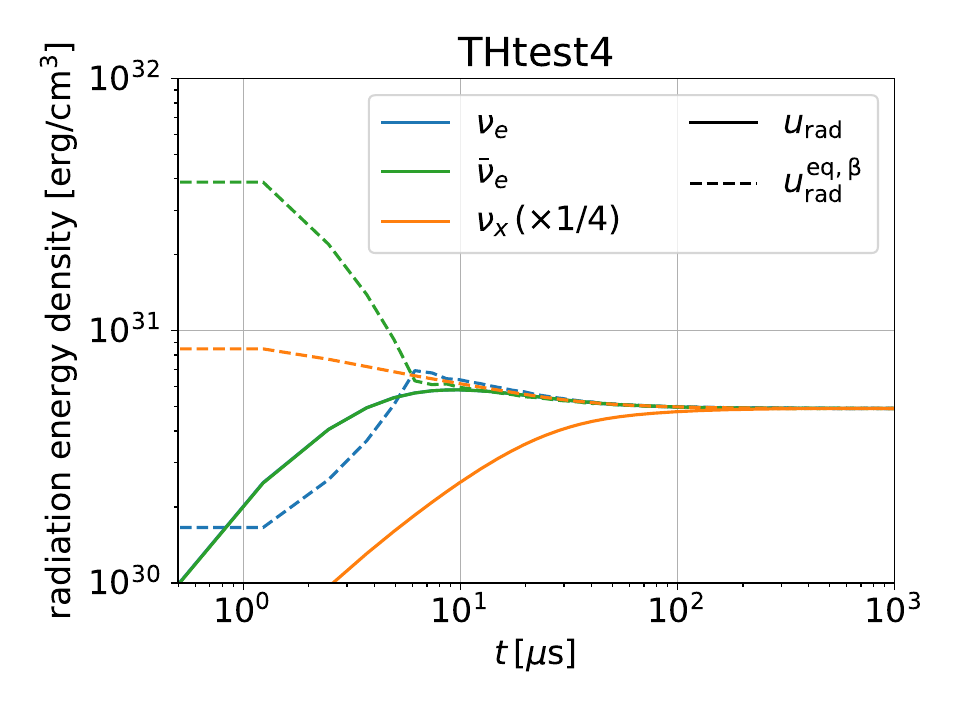}~~
 	 \includegraphics[width=.41\linewidth]{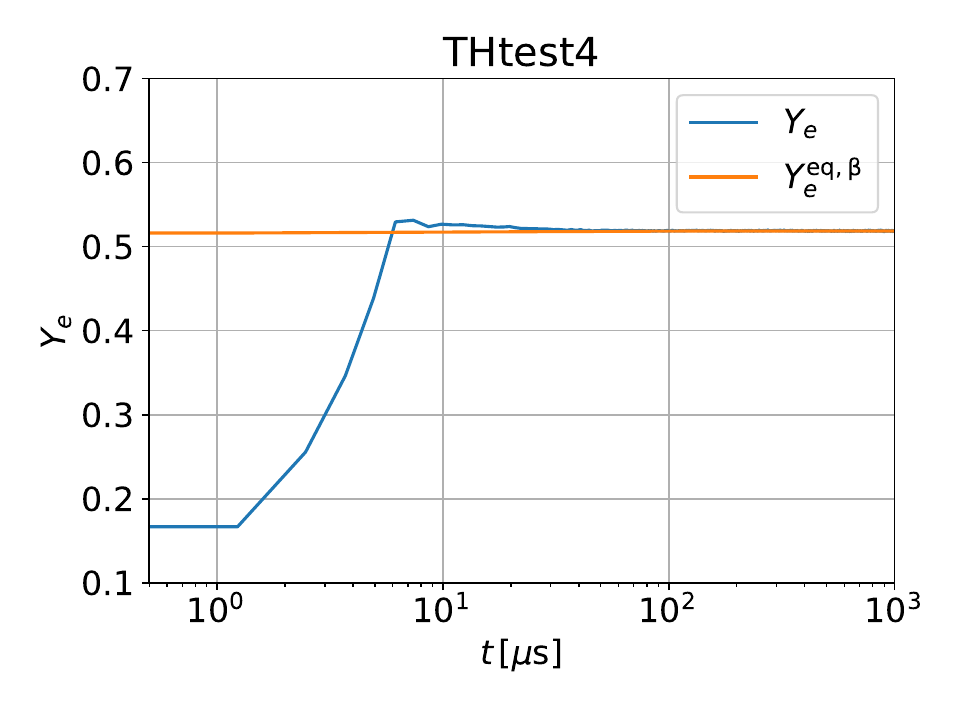}
 	 \caption{Time evolution of the radiation energy density for each neutrino species (left) and electron fraction, $Y_e$ (right) for the one-zone thermalization test problems. The radiation energy density and $Y_e$ values at the $\beta$-equilibrium are also plotted.}
	 \label{fig:thtest}
\end{figure*}

\begin{figure*}
 	 \includegraphics[width=.45\linewidth]{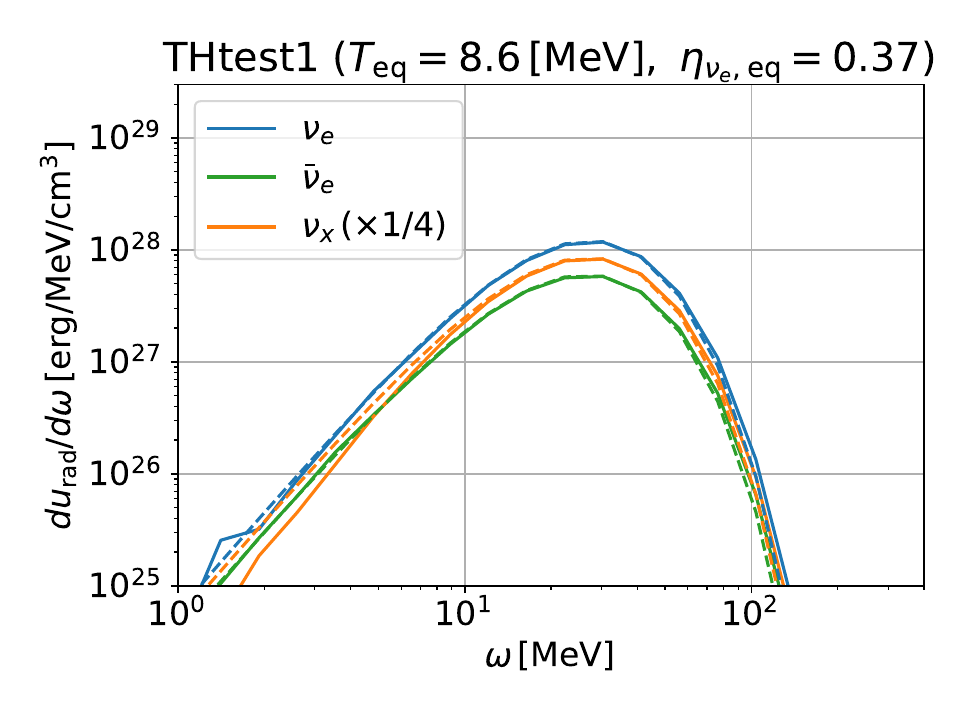}~~~
 	 \includegraphics[width=.45\linewidth]{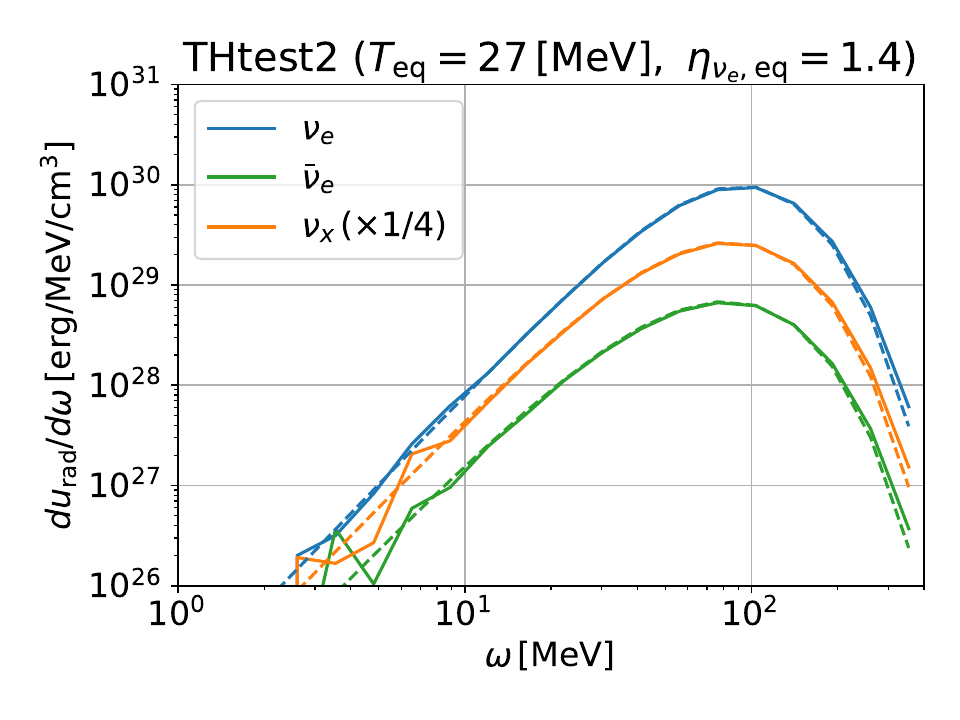}\\
 	 \includegraphics[width=.45\linewidth]{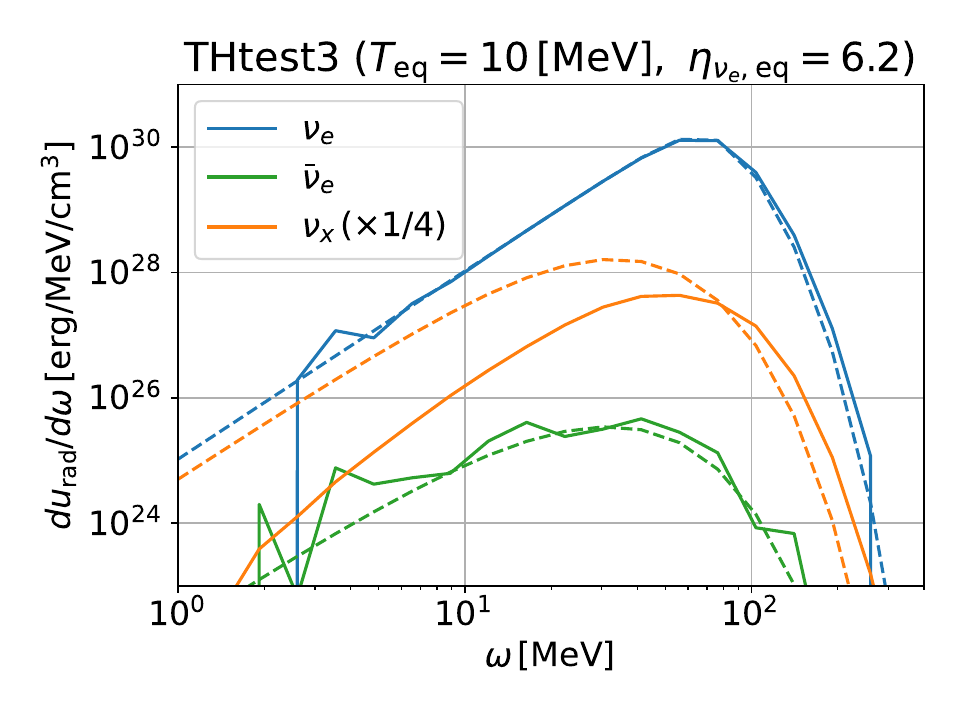}~~~
 	 \includegraphics[width=.45\linewidth]{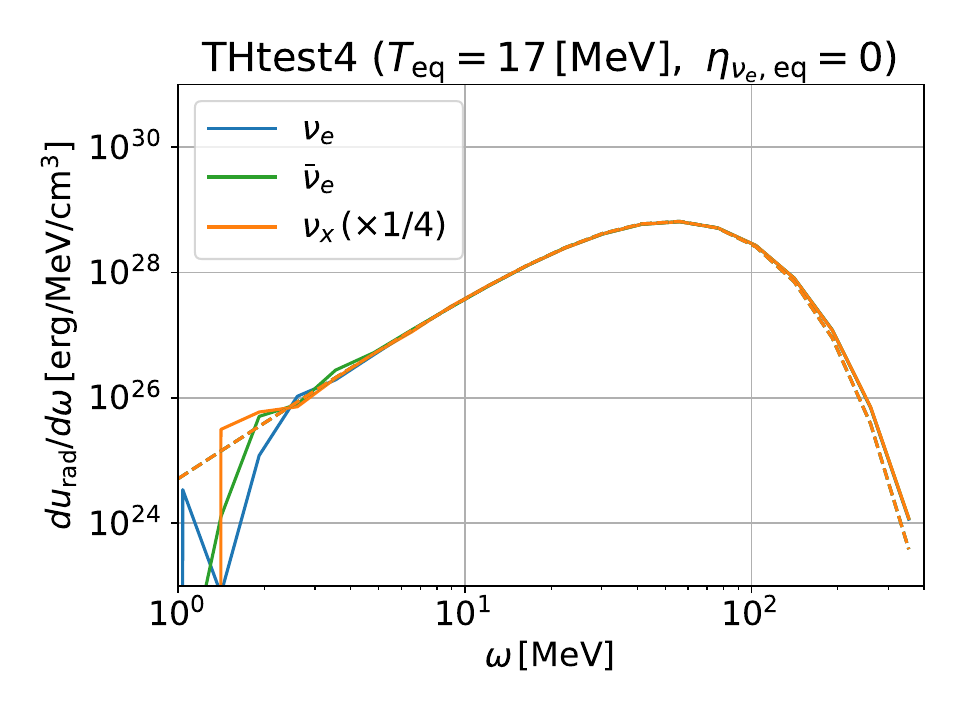}
 	 \caption{The energy distribution of neutrino radiation fields of the one-zone thermalization problems at the end of simulations.}
	 \label{fig:thtest-dist}
\end{figure*}

Figure~\ref{fig:thtest} displays the time evolution of the radiation energy density for each neutrino species, $u_{\rm rad}$, and electron fraction, $Y_e$. The values in the $\beta$-equilibrium state ($u^{\rm eq,\beta}_{\rm rad}$ and $Y_e^{\rm eq,\beta}$, respectively) are also plotted. Figure~\ref{fig:thtest} shows that the radiation energy density for each neutrino species and electron fraction correctly converge to the $\beta$-equilibrium values in the corresponding weak-interaction timescale as the time evolves: We confirm that the time scale for the radiation field settles into the equilibrium state is consistent with that estimated from the absorption rate. Note that the radiation energy density of $\nu_x$ for model THtest3 does not reach the equilibrium value within the simulation time ($\approx1000\,\mu{\rm s}$), as the absorption and emission time scale is quite long ($>2000\,\mu{\rm s}$). We also note that the actual absorption/emission time scales of $\nu_e$ and ${\bar \nu}_e$ for model THtest2 and THtest3 are much shorter than the time interval of the calculation ($1\,{\rm \mu s}$), and hence, those are artificially elongated to be comparable to the time interval under the implicit MC method. The oscillative feature found in the radiation energy density of ${\bar \nu}_e$ for model THtest3 is due to the MC shot noise but it is many orders of magnitude smaller than the energy density of other species.%.\ms{This sentence is not well understood.}

Figure~\ref{fig:thtest-dist} displays the energy distribution of the neutrino radiation field at the end of the one-zone thermalization simulations. The thermal distribution in the $\beta$-equilibrium calculated from the matter temperature and electron degenerate parameter is also shown in the figure. Figure~\ref{fig:thtest-dist} shows that the neutrino radiation fields are settled into equilibrium with correct thermal distributions. Exceptionally, as is found in Fig.~\ref{fig:thtest}, $\nu_x$ of model THtest3 does not follow the thermal distribution, as the radiation field is not yet thermalized due to its long absorption/emission time scale (see Table~\ref{tb:thmodel}). We note that discrepancies between the analytical and simulated distributions are due to the error induced by the finite resolution of the energy binning.

\section{pair annihilation kernel}\label{app:pair}
For given temperature, $T$, and electron degeneracy parameter, $\eta_e$, the kernel function of pair annihilation, $R^{\rm ann}$, in the limit of ignoring the electron mass can be given by Eq.~\eqref{eq:rabs} as a function of the neutrino energy, $\omega$, antineutrino energy, ${\bar \omega}$ , and cosine of the angle between the spatial momentum of neutrino and antineutrino, $\mu$ , respectively. $R^{\rm ann}$ can be rewritten as~\cite{Bruenn:1985en}
\begin{align}
    &R^{\rm ann}\left(\omega,{\bar \omega},\mu\right)=\frac{2G_{\rm F}^2c^4}{3\pi(\hbar c)^4\omega{\bar \omega}}q_\mu{\bar q}^\mu  q_\nu{\bar q}^\nu\nonumber\\
    &\times\left[(C_V^2+C_A^2)I_{\rm sym}\left(x,{\bar x},\mu\right)+2 C_V C_A I_{\rm asym}\left(x,{\bar x},\mu\right) \right]\nonumber\\
    &=\frac{2G_{\rm F}^2}{3\pi(\hbar c)^4}\omega{\bar \omega}\left(1-\mu\right)^2\nonumber\\
    &\times\left[(C_V^2+C_A^2)I_{\rm sym}\left(x,{\bar x},\mu\right)+2 C_V C_A I_{\rm asym}\left(x,{\bar x},\mu\right) \right],
\end{align}
where $x=\omega/k_{\rm B}T$, ${\bar x}={\bar \omega}/k_{\rm B}T$, and $I^{\rm sym}(x,{\bar x},\mu)$ and $I^{\rm asym}(x,{\bar x},\mu)$ are given by
\begin{align}
    I^{\rm sym}(x,{\bar x},\mu)&=\frac{3\pi(\hbar c)^4}{2(C_V^2+C_A^2)G_{\rm F}^2\omega{\bar \omega}\left(1-\mu\right)^2}\\\nonumber
    &\times\left[R^{\rm ann}\left(\omega,{\bar \omega},\mu\right)+R^{\rm ann}\left({\bar \omega},\omega,\mu\right)\right],\\
    I^{\rm asym}(x,{\bar x},\mu)&=\frac{3\pi(\hbar c)^4}{4 C_V C_A G_{\rm F}^2\omega{\bar \omega}\left(1-\mu\right)^2}\\\nonumber
    &\times\left[R^{\rm ann}\left(\omega,{\bar \omega},\mu\right)-R^{\rm ann}\left({\bar \omega},\omega,\mu\right)\right].
\end{align}
Obviously, $I^{\rm sym}(x,{\bar x},\mu)$ and $I^{\rm asym}(x,{\bar x},\mu)$ have the following property:
\begin{align}
    I^{\rm sym}(x,{\bar x},\mu)&=I^{\rm sym}({\bar x},x,\mu)\nonumber\\
    I^{\rm asym}(x,{\bar x},\mu)&=-I^{\rm asym}({\bar x},x,\mu).
\end{align}
We also note that $I^{\rm sym}$ and $I^{\rm asym}$ depend on $\eta_e$ but not on $T$ for fixed values of $x$ and ${\bar x}$. In particular, $I^{\rm sym}$ depends only on the absolute value of $\eta_e$. Interestingly, although we do not use them directly in the present work, the exact analytical expressions for $I^{\rm sym}$ and $I^{\rm asym}$ are found (see App.~\ref{app:ELF}).

We define the angle average, $\left<I\right>$, and the standard deviation around the angle average $\sigma_I$ for both $I=I^{\rm sym}$ and $I=I^{\rm asym}$ by
\begin{align}
\left<I\right>\left(x,{\bar x}\right)=\frac{\int_{-1}^{1}d\mu(1-\mu)^2 I\left(x,{\bar x},\mu\right)}{\int_{-1}^{1}d\mu(1-\mu)^2}
\end{align}
and
\begin{align}
\sigma_I\left(x,{\bar x}\right)=\sqrt{\frac{\int_{-1}^{1}d\mu(1-\mu)^2 \left[I\left(x,{\bar x},\mu\right)-\left<I\right>\left(x,{\bar x}\right)\right]^2}{\int_{-1}^{1}d\mu(1-\mu)^2}},
\end{align}
respectively.

\begin{figure*}
 	 \includegraphics[width=.32\linewidth]{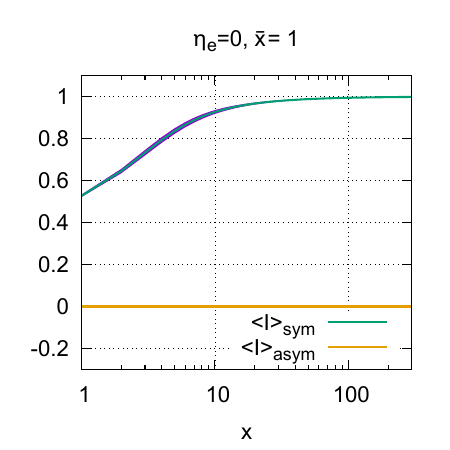}
 	 \includegraphics[width=.32\linewidth]{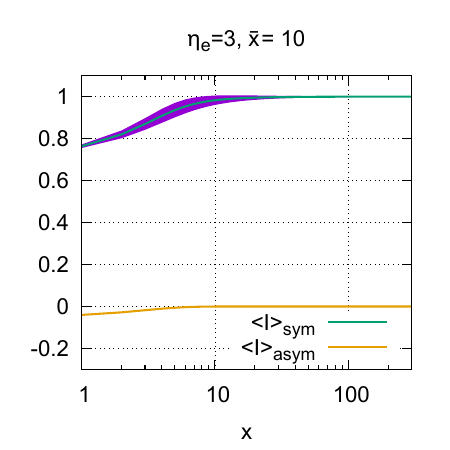}
 	 \includegraphics[width=.32\linewidth]{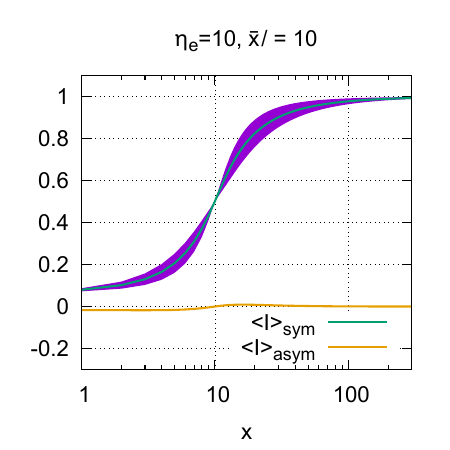}
 	 \caption{$\left<I\right>$ for $I=I^{\rm sym}$ and $I=I^{\rm asym}$ as functions of $x=\omega/k_{\rm B}T$ for several values of ${\bar x}={\bar \omega}/k_{\rm B}T$ and $\eta_e$. The purple filled regions denote the range of the standard deviation, $\sigma_I$, for $I=I^{\rm sym}$.}
	 \label{fig:avcheck}
\end{figure*}

Figure~\ref{fig:avcheck} shows $\left<I\right>$ and $\sigma_I$ for $I=I^{\rm sym}$ and $I=I^{\rm asym}$ as functions of $x=\omega/k_{\rm B}T$ for several values of ${\bar x}={\bar \omega}/k_{\rm B}T$ and $\eta_e$. We find that the angle averaged value of $I=I^{\rm sym}$ is always much larger than that of $I=I^{\rm asym}$. It is also shown that the standard deviation of $I=I^{\rm sym}$ around the angle averaged value is always within $\approx10\%$. Motivated by these facts, we approximate $I^{\rm sym}$ and $I^{\rm asym}$ by
\begin{align}
    I^{\rm sym}(x,{\bar x},\mu)&\approx \left<I^{\rm sym}\right>(x,{\bar x})\nonumber\\
    I^{\rm asym}(x,{\bar x},\mu)&\approx 0,
\end{align}
in our code.

In our code, we further approximate $\left<I^{\rm sym}\right>(x,{\bar x},\mu)$ by employing two fitting functions $\phi_1$ and $\phi_2$ as
\begin{align}
    \left<I^{\rm sym}\right>(x,{\bar x})&\approx\left<I^{\rm sym}_{\rm model}\right>(x,{\bar x})\nonumber\\
    &=\phi_1 \left(x\right) \phi_1\left({\bar x}\right)-\phi_2 \left(x\right) \phi_2\left({\bar x}\right).
\end{align}
Here, we remark that, since $\left<I^{\rm sym}\right>(x,{\bar x},\mu)$ depends on the absolute value of $\eta_e$, $\phi_1$ and $\phi_2$ also depend on the absolute value of $\eta_e$. 
We require $\phi_1$ and $\phi_2$ to satisfy the following conditions:
\begin{enumerate}
    \item 
    \begin{align}
        \int_{x_{\rm min}}^{x_{\rm max}} dx\, \phi_1(x)\phi_2(x)=0.
    \end{align}
    \item 
    For $x \gg 1,\,{\bar x}\gg1$ and $\eta_e\ll 1$
    \begin{align}
        \phi_1(x)\phi_1({\bar x})-\phi_2(x)\phi_2({\bar x})\approx 1.
    \end{align}
    \item
    \begin{align}
        \frac{\partial}{\partial x}\left[\phi_1 \left(x\right) \phi_1\left({\bar x}\right)-\phi_2 \left(x\right) \phi_2\left({\bar x}\right)\right]>0.
    \end{align}
    \item 
    \begin{align}
        \int_{x_{\rm min}}^{x_{\rm max}} dx\int_{x_{\rm min}}^{x_{\rm max}} d{\bar x} K(x,{\bar x})\left[\phi_1 \left(x\right) \phi_1\left({\bar x}\right)-\phi_2 \left(x\right) \phi_2\left({\bar x}\right)\right]\nonumber\\    
        =\int_{x_{\rm min}}^{x_{\rm max}} dx\int_{x_{\rm min}}^{x_{\rm max}} d{\bar x} K(x,{\bar x})\left<I^{\rm sym}\right>(x,{\bar x}),
    \end{align}
    where
    \begin{align}
        K(x,{\bar x})=\frac{x^4}{e^{x}+1}\frac{{\bar x}^3}{e^{{\bar x}}+1}+\frac{x^3}{e^{x}+1}\frac{{\bar x}^4}{e^{{\bar x}}+1}.
    \end{align}
\end{enumerate}
Here, ${x_{\rm min}}$ and ${x_{\rm max}}$ denote the minimum and maximum of the fitting range. The condition 1. is required to uniquely determine $\phi_1$ and $\phi_2$. The condition 2. is required so that the exact expression of the kernel function in the limit of neglecting the phase space blocking for electrons and positrons and the electron mass is reproduced~\cite{Fujibayashi:2017xsz,2021ApJ...920...82F}. The condition 3. is also required so that the kernel function approximated employing $\phi_1$ and $\phi_2$ has the appropriate asymptotic behavior with respect to the neutrino energy. The condition 4. is required so that the total neutrino/antineutrino emissivity is well reproduced for $\eta_\nu=0$ case. 

We tabulate $\phi_1$ and $\phi_2$ for the range of $x\in [{x_{\rm min}},{x_{\rm max}}]=[10^{-2},10^{2}]$ in a logarithmically uniform manner. For $\eta_e=0$ and each value of $\eta_e$ logarithmically selected in the range of $[0.3,14]$, we determine the values of $\phi_1$ and $\phi_2$ by minimizing the following quantity under the conditions given above:
\begin{align}
    \int_{x_{\rm min}}^{x_{\rm max}} dx\int_{x_{\rm min}}^{x_{\rm max}} d{\bar x} \Delta(x,{\bar x})^2\label{eq:diffph}
\end{align}
with   
\begin{align}
\Delta(x,{\bar x})&=W\left[\left<I^{\rm sym}\right>(x,{\bar x})\right]\nonumber\\
&-W\left[\phi_1 \left(x\right) \phi_1\left({\bar x}\right)-\phi_2 \left(x\right) \phi_2\left({\bar x}\right)\right].
\end{align}
Here, $W$ is a function introduced to control the weight of the fitting. After several trial, we decided to use the following form of $W$ to ensure that $\phi_1$ and $\phi_2$ have the desired asymptotic behavior:
\begin{align}
    W(I)=-{\rm ln}\left(\frac{1}{I}-1\right).
\end{align}

\begin{figure*}
 	 \includegraphics[width=.48\linewidth]{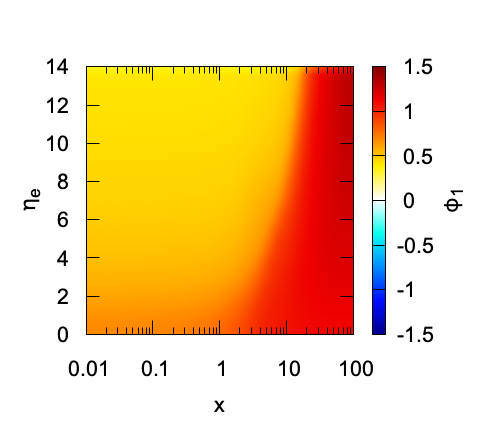}
 	 \includegraphics[width=.48\linewidth]{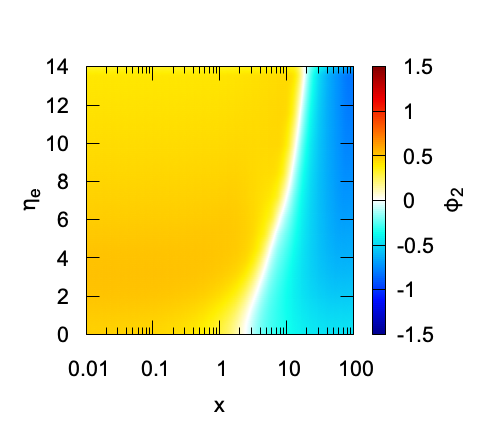}
 	 \caption{$\phi_1$ and $\phi_2$ as functions of $x$ and $\eta_e$ obtained by the numerical fitting.}
	 \label{fig:ph}
\end{figure*}

Figure~\ref{fig:ph} shows the values of $\phi_1$ and $\phi_2$ as functions of $x$ and $\eta_e$ obtained by the fitting. We can see that both $\phi_1$ and $\phi_2$ become small for a large value $\eta_e$. This reflects the fact that the cross-section of pair annihilation is suppressed for the case that electrons are highly degenerate.

\begin{figure*}
 	 \includegraphics[width=.32\linewidth]{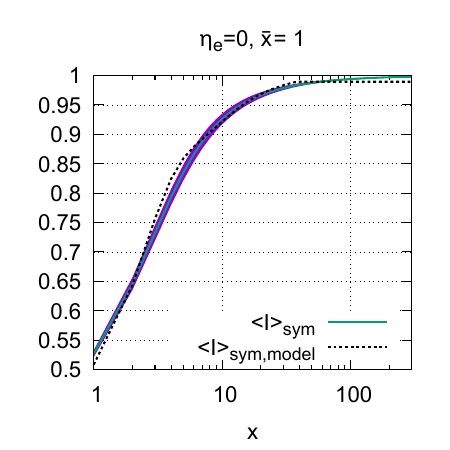}
 	 \includegraphics[width=.32\linewidth]{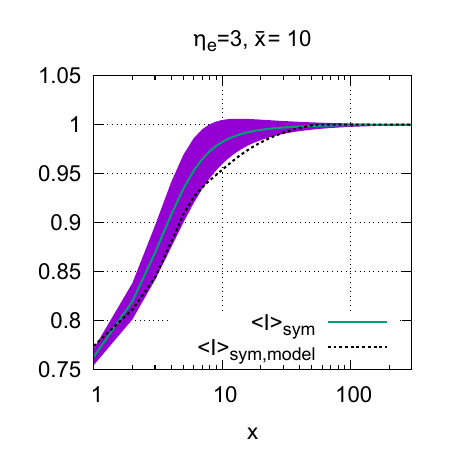}
 	 \includegraphics[width=.32\linewidth]{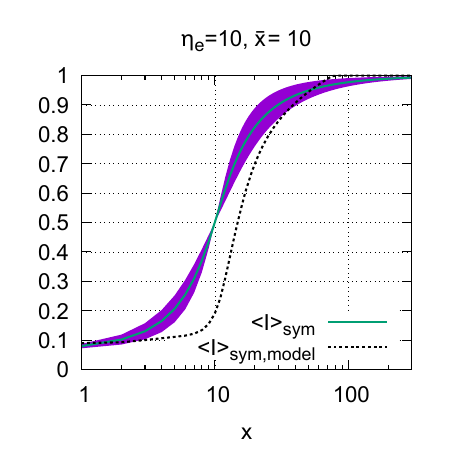}
 	 \caption{Comparison of $\left<I^{\rm sym}\right>$ and $\left<I^{\rm sym}_{\rm model}\right>$ for several values of $\eta_e$ and ${\bar x}$ as functions of $x$. The purple filled regions denote the range of the standard deviation, $\sigma_I$, for $I=I^{\rm sym}$.}
	 \label{fig:mcheck}
\end{figure*}

\begin{figure*}
 \includegraphics[width=.32\linewidth]{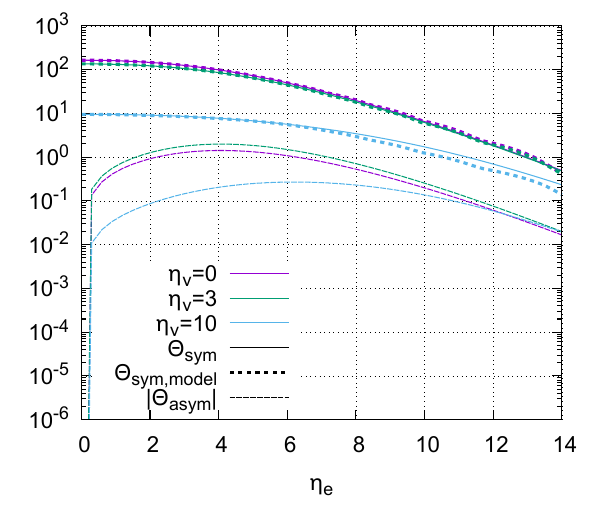}
 \includegraphics[width=.32\linewidth]{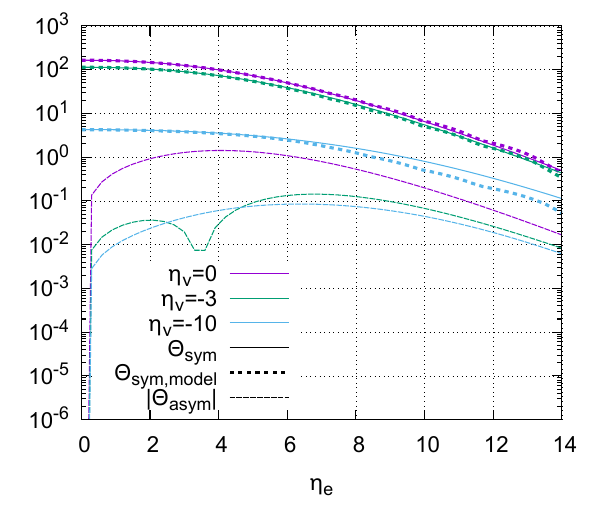}
 \includegraphics[width=.32\linewidth]{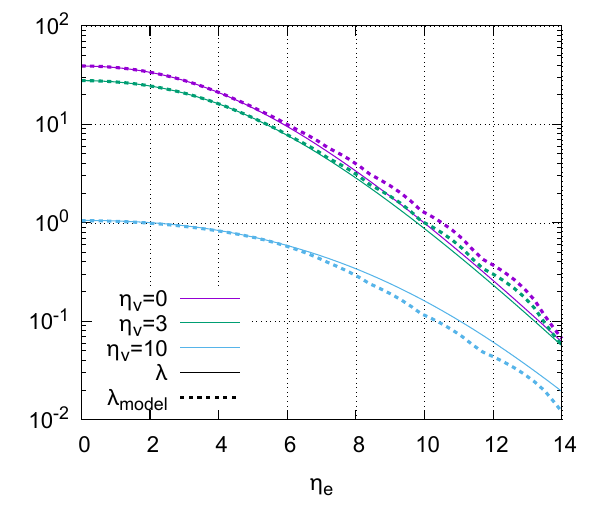}
 \caption{Comparisons of $\Theta_{\rm sym}$ (right and middle) and $\lambda$ (left) calculated from the exact expression of $I^{\rm sym}$ and those calculated from $\left<I^{\rm sym}_{\rm model}\right>$ employing fitting functions $\phi_1$ and $\phi_2$ ($\Theta_{\rm sym, model}$ and $\lambda_{\rm model}$). The horizontal axis denotes the electron degeneracy parameter.%\ms{It would be better to mention what the horizontal axis is.}
 }
 \label{fig:emscheck}
\end{figure*}

Figure~\ref{fig:mcheck} compares $\left<I^{\rm sym}\right>$ and $\left<I^{\rm sym}_{\rm model}\right>$ for several values of $\eta_e$ and ${\bar x}$ as functions of $x$. For $\eta_e\leq3$, we find that $\left<I^{\rm sym}_{\rm model}\right>$ and $\left<I^{\rm sym}\right>$ agree within the standard deviation of $I^{\rm sym}$ around the angle averaged. For $\eta_e\geq10$, the deviation of $\left<I^{\rm sym}_{\rm model}\right>$ from $\left<I^{\rm sym}\right>$ is large particularly when $\left<I^{\rm sym}\right>$ shows the steep change in its value. However, the asymptotic values for $x\ll 1$ and $x\gg1$ are still correctly reproduced.

Neglecting the Fermi blocking effect of neutrinos, the energy-integrated neutrino emissivity and number emissivity of pair process in the thermal equilibrium, $Q_{\rm pair,th}$ and $\Lambda_{\rm pair,th}$, respectively, are given by

\begin{align}
    Q_{\rm pair,th}&(T,\eta_e,\eta_\nu)\nonumber\\&=c\int d^3{\bf Q}\int d^3{\bf {\bar Q}}\,\omega f^{\rm th}(\omega){\bar f}^{\rm th}{\bar \omega})R^{\rm ann}(\omega,{\bar \omega},\mu)\nonumber\\
    &=\frac{32\pi G_{\rm F}^2c/(\hbar c)^4}{3(hc)^6}\left(k_{\rm B}T\right)^9\nonumber\\
    &\times\left[
    (C_V^2+C_A^2)\Theta_{\rm sym}(\eta_e,\eta_\nu)
    +2C_V C_A \Theta_{\rm asym}(\eta_e,\eta_\nu) 
    \right]
\end{align}
and
\begin{align}
    \Lambda_{\rm pair,th}&(T,\eta_e,\eta_\nu)\nonumber\\&=c\int d^3{\bf Q}\int d^3{\bf {\bar Q}}f^{\rm th}(\omega){\bar f}^{\rm th}{\bar \omega})R^{\rm ann}(\omega,{\bar \omega},\mu)\nonumber\\
    &=\frac{32\pi G_{\rm F}^2c/(\hbar c)^4(C_V^2+C_A^2)}{3(hc)^6}\left(k_{\rm B}T\right)^8 \lambda(\eta_e,\eta_\nu).
\end{align}
Here, $\Theta_{\rm sym}$, $\Theta_{\rm asym}$, and $\lambda$ are given by
\begin{align}
    \Theta_{\rm sym/asym}(\eta_e,\eta_\nu)=\int dx\int d{\bar x}\int_{-1}^1 d\mu (1-\mu^2)\nonumber\\
    \times I^{\rm sym/asym}\left(x,{\bar x},\mu\right)\frac{x^4}{e^{x-\eta_\nu}+1}\frac{{\bar x}^3}{e^{{\bar x}+\eta_\nu}+1}
\end{align}
and
\begin{align}
    \lambda(\eta_e,\eta_\nu)=\int dx\int d{\bar x}\int_{-1}^1 d\mu (1-\mu^2)\nonumber\\
    \times I^{\rm sym}\left(x,{\bar x},\mu\right)\frac{x^3}{e^{x-\eta_\nu}+1}\frac{{\bar x}^3}{e^{{\bar x}+\eta_\nu}+1}.
\end{align}

Note the emissivity of antineutrino, ${\bar Q}_{\rm pair,th}$ is given by $Q_{\rm pair,th}$ with $\eta_\nu\rightarrow-\eta_\nu$ and $\Theta_{\rm asym}\rightarrow-\Theta_{\rm asym}$,and the number emissivity of antineutrino, ${\bar \Lambda}_{\rm pair,th}$, agrees with $\Lambda_{\rm pair,th}$ because $\lambda(\eta_e,\eta_\nu)=\lambda(\eta_e,-\eta_\nu)$.

Figure~\ref{fig:emscheck} compares $\Theta_{\rm sym}$ and $\lambda$ calculated from the exact expression of $I^{\rm sym}$ and those calculated from $\left<I^{\rm sym}_{\rm model}\right>$ employing fitting functions $\phi_1$ and $\phi_2$ ($\Theta_{\rm sym, model}$ and $\lambda_{\rm model}$). $\Theta_{\rm asym}$ calculated from the exact expression of $I^{\rm asym}$ is also plotted in Fig.~\ref{fig:emscheck}, which shows that $\Theta_{\rm sym}$ and $\lambda$ employing fitting functions $\phi_1$ and $\phi_2$ reproduce the exact values for $\eta_e\leq10$ and $|\eta_\nu|\leq10$ very well. For $\eta_e\geq10$, $\Theta_{\rm sym, model}$ and $\lambda_{\rm model}$ show discrepancy from the exact values. However, we note that this discrepancy in the emissivity may not affect the dynamics since the emissivity is nevertheless small. Figure~\ref{fig:emscheck} also shows that $\Theta_{\rm asym}$ is always a few order of magnitudes smaller than $\Theta_{\rm sym}$, suggesting that the error neglecting the term of $I^{\rm asym}$ in the kernel has only a minor effect in the results.

\section{Comparison among different simulation setups}\label{app:comp}

\begin{figure}
 \includegraphics[width=0.9\linewidth]{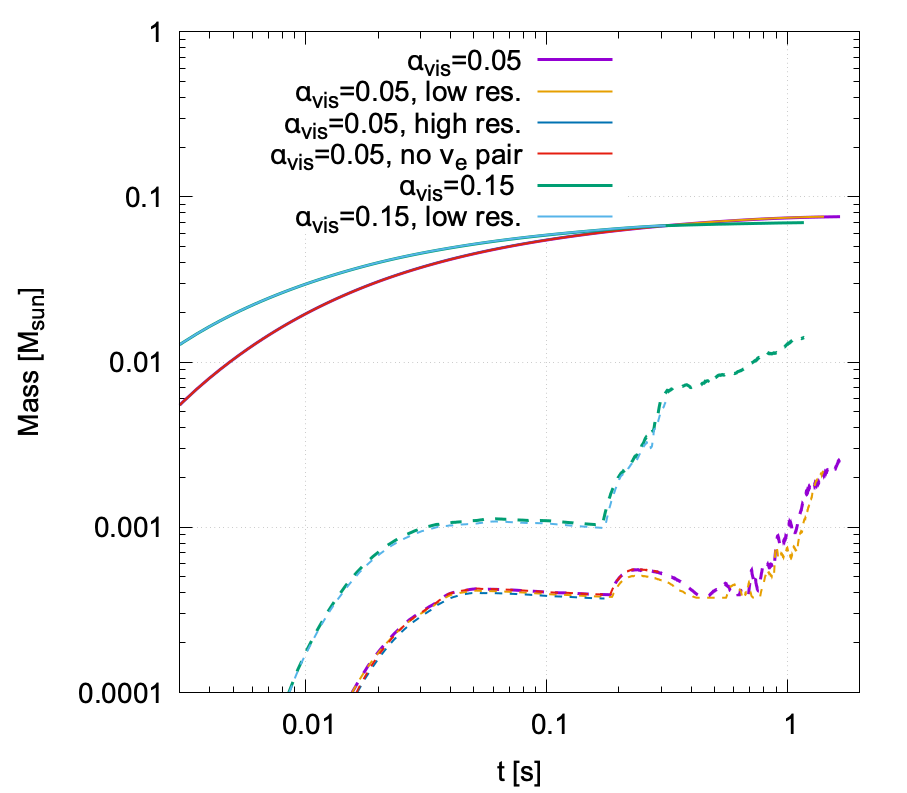}\\
 \includegraphics[width=0.9\linewidth]{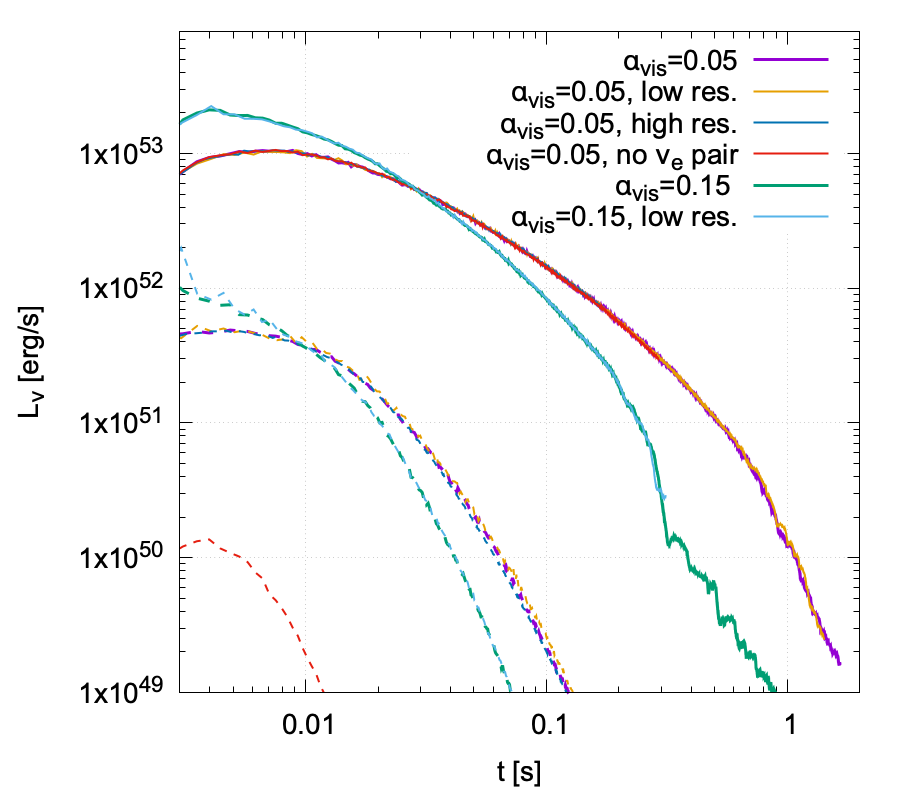}\\
 \includegraphics[width=0.9\linewidth]{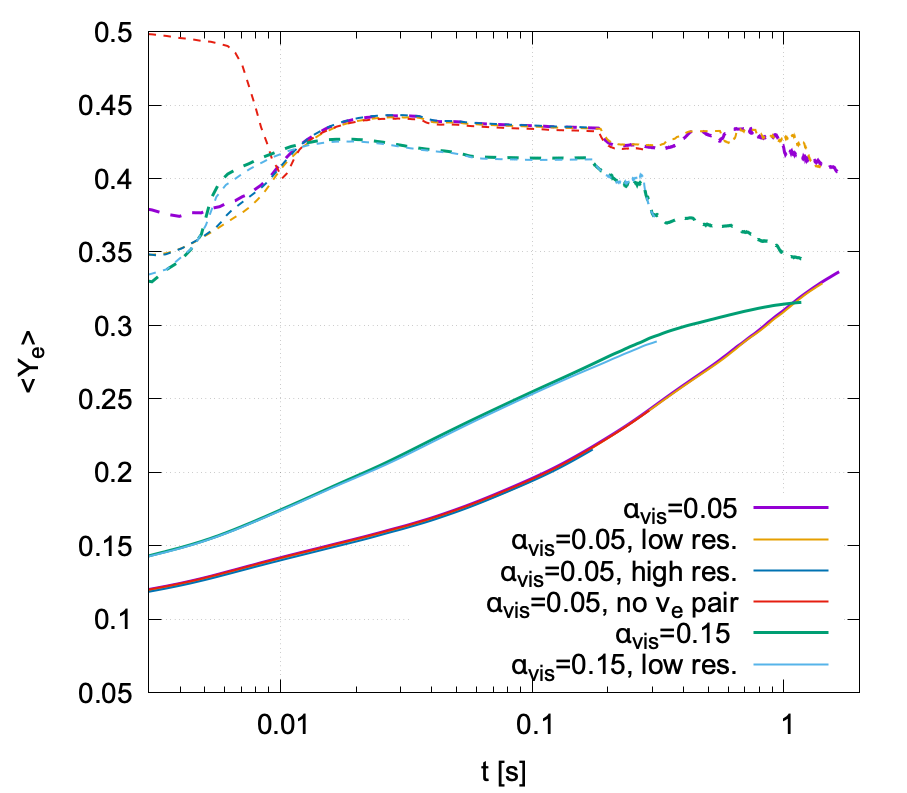}
 \caption{Comparisons of the time evolution of the mass outside the BH (top: solid), ejecta mass (top: dashed), total neutrino luminosity (middle: solid), total pair annihilation rate (middle: dashed), mass-averaged $Y_e$ outside the BH (bottom: solid), and mass-averaged ejecta $Y_e$ (bottom: dashed) for various different simulation setups.}
 \label{fig:comp_setup}
\end{figure}

To clarify how the results depend on the simulation setups, we perform the BH-torus simulations varying the setups. Here, the cases with a smaller grid resolution of 280 cells in each direction and smaller MC packet number of $N_{\rm trg}=32$ for both models with $\alpha_{\rm vis}=0.05$ and $0.15$ (``low res.'' in Fig.~\ref{fig:comp_setup}), the case with larger grid resolution of 480 for the model with $\alpha_{\rm vis}=0.05$ (``high res.''), and the case in which $\nu_e{\bar \nu}_e$ pair process is turned off for the model with $\alpha_{\rm vis}=0.05$ (``no $\nu_e$ pair'') are considered with other setups being the same as described in the main text ($\nu_x{\bar \nu}_x$ pair process is kept turned on). Note that, for the smaller and larger resolution runs, the increasing rates of the grid-spacing are taken to be 1.0143 and 1.0083, respectively.

Figure~\ref{fig:comp_setup} compares the time evolution of the mass outside the BH (top: solid), ejecta mass (top: dashed), total neutrino luminosity (middle: solid), total pair annihilation rate (middle: dashed), mass-averaged $Y_e$ outside the BH (bottom: solid), and mass-averaged ejecta $Y_e$ (bottom: dashed) for various different simulation setups. This shows that all the quantities are approximately in agreement among different setups. Exceptionally, the total pair annihilation rate for the case that $\nu_e{\bar \nu}_e$ pair process is switched off shows significantly small values because pair annihilation of heavy-lepton type neutrinos is only considered.

\section{Exact analytical expression for the pair process kernel function in the zero electron mass limit}\label{app:ELF}

We define a tensor $I^{\mu\nu}$ by
\begin{align}
    I^{\mu\nu}:=\int \frac{d^3p}{2 p^0}\int\frac{d^3{\bar p}}{2 {\bar p}^0}\delta^4\left(Q-p-{\bar p}\right)f(E){\bar f}({\bar E})p^\mu {\bar p}^\nu,
\end{align}
where $E=-u_\mu p^\mu$ and ${\bar E}=-u_\mu{\bar p}^\mu$. We consider the case that $p^\mu$ and ${\bar p}^\mu$ are future directed time-like or null vectors.

We employ the following ansatz for $I^{\mu\nu}$:

\begin{align}
    I^{\mu\nu}&=A \frac{Q^4}{E_Q^2} u^\mu u^\nu+B_1\frac{Q^2}{E_Q} u^\mu Q^\nu +B_2 \frac{Q^2}{E_Q} Q^\mu u^\nu\nonumber\\
    &+C Q^\mu Q^\nu+D Q^2 g^{\mu\nu},
\end{align}
where $E_Q=-u_\mu Q^\mu$ and $Q^2=g_{\mu\nu}Q^\mu Q^\nu$. $A$, $B_1$, $B_2$, $C$, and $D$ are the dimensionless functions of $E_Q$ and $Q^2$. By evaluating  $I^{\mu\nu}u_{\mu}u_{\nu}$, $I^{\mu\nu}u_{\mu}Q_{\nu}$, $I^{\mu\nu}Q_{\mu}u_{\nu}$, $I^{\mu\nu}Q_{\mu}Q_{\nu}$, and $I^{\mu\nu}g_{\mu\nu}$, $A$, $B_1$, $B_2$, $C$, and $D$ are obtained by

\begin{align}
    \left(
    \begin{tabular}{c}
        $A$\\
        $B_1$\\
        $B_2$\\
        $C$\\
        $D$
    \end{tabular}
    \right)
    &=
    \left[
    \begin{tabular}{ccccc}
        $ \frac{Q^4}{E_Q^4}$&$ \frac{Q^2}{E_Q^2}$&$ \frac{Q^2}{E_Q^2}$&                   1&$-\frac{Q^2}{E_Q^2}$\\
        $ \frac{Q^4}{E_Q^4}$&$-\frac{Q^4}{E_Q^4}$&$ \frac{Q^2}{E_Q^2}$&$-\frac{Q^2}{E_Q^2}$&$-\frac{Q^2}{E_Q^2}$\\
        $ \frac{Q^4}{E_Q^4}$&$ \frac{Q^2}{E_Q^2}$&$-\frac{Q^4}{E_Q^4}$&$-\frac{Q^2}{E_Q^2}$&$-\frac{Q^2}{E_Q^2}$\\
        $ \frac{Q^4}{E_Q^4}$&$-\frac{Q^4}{E_Q^4}$&$-\frac{Q^4}{E_Q^4}$&$ \frac{Q^4}{E_Q^4}$&$ \frac{Q^4}{E_Q^4}$\\
        $-\frac{Q^4}{E_Q^4}$&$-\frac{Q^2}{E_Q^2}$&$-\frac{Q^2}{E_Q^2}$&$ \frac{Q^2}{E_Q^2}$&$4\frac{Q^2}{E_Q^2}$
    \end{tabular}
    \right]^{-1}
    \left(
    \begin{tabular}{c}
        $\frac{I^{\mu\nu}u_{\mu}u_{\nu}}{E_Q^2}$\\
        $\frac{I^{\mu\nu}u_{\mu}Q_{\nu}}{E_Q^3}$\\
        $\frac{I^{\mu\nu}Q_{\mu}u_{\nu}}{E_Q^3}$\\
        $\frac{I^{\mu\nu}Q_{\mu}Q_{\nu}}{E_Q^4}$\\
        $\frac{I^{\mu\nu}g_{\mu\nu}}{E_Q^2}$
    \end{tabular} 
    \right).
\end{align}
$I^{\mu\nu}u_{\mu}u_{\nu}$, $I^{\mu\nu}u_{\mu}Q_{\nu}$, $I^{\mu\nu}Q_{\mu}u_{\nu}$, $I^{\mu\nu}Q_{\mu}Q_{\nu}$, and $I^{\mu\nu}g_{\mu\nu}$ are evaluated in the subsequent  paragraphs. In the following we consider the case that $p^\mu$ and ${\bar p}^\mu$ are null vectors (i.e., mass-less particle). For the case that $Q^\mu$ is a space-like vector (i.e., $Q^2 >0$), the arguments of the delta function in $I^{\mu\nu}$ will be 0. Hence, for such a case, $I^{\mu\nu}=0$. In the following, we only consider the case that $Q^\mu$ is a future-directed vector, since otherwise $I^{\mu\nu}$ is always 0.

For the case that $Q^\mu$ is a null or time-like vector (i.e., $Q^2\leq 0$), there always exists a frame in which the spatial components of $Q^\mu$ vanish (i.e., $Q'^\mu=(Q'^0,{\bf 0})$). Taking such a frame, we have
\begin{align}
    I'^{\mu\nu}&=\int \frac{d^3p'}{2 p'{}^{0}} \int\frac{d^3{\bar p'}}{2 {\bar p'}{}^{0}} \delta^3\left({\bf p'}+{\bf{\bar p'}}\right) \nonumber\\
    &\times \delta\left(Q'{}^{0}-p'{}^{0}-{\bar p'}{}^{0}\right)f(E) {\bar f}({\bar E})p'{}^{\mu} {\bar p'}{}^{\nu}.
\end{align}

We focus first on $I^{\mu\nu}g_{\mu\nu}$.
\begin{align}
    I'^{\mu\nu}g_{\mu\nu}&=\int \frac{d^3p'}{2 p'{}^{0}} \int\frac{d^3{\bar p'}}{2 {\bar p'}{}^{0}} \delta^3\left({\bf p'}+{\bf{\bar p'}}\right)\nonumber\\
    &\times \delta\left(Q'{}^{0}-p'{}^{0}-{\bar p'}{}^{0}\right) f(E) {\bar f}({\bar E})p'{}^{\mu} {\bar p'}{}^{\nu}g'_{\mu\nu}\nonumber\\
    &=-\frac{1}{2}\int d^3p' \delta\left(Q'{}^{0}-2p'{}^{0}\right) f(E) {\bar f}({\bar E})\nonumber\\
   &=-\pi\int_0^\infty d p'{}^{0}\int_{-1}^{1} d\mu \left(p'{}^{0}\right)^2\delta\left(Q'{}^{0}-2p'{}^{0}\right) f(E) {\bar f}({\bar E})\nonumber\\
   &=-\frac{\pi}{8}\left(Q'{}^{0}\right)^2\int_{-1}^{1} d\mu f(E) {\bar f}({\bar E}).
\end{align}
%    &~\left(\because p'{}^{0}={\bar p'}{}^{0}\,{\rm and}\,p'{}^{\mu} {\bar p'}{}^{\nu}g'_{\mu\nu}=-2\left(p'{}^{0}\right)^2\,{\rm for}\,{\bf p'}+{\bf{\bar p'}}={\bf 0}\right)\nonumber\\
Here, $\mu$ denotes the cosine of the angle between the spatial components of $\,p'{}^{\mu}$ and $u^\mu$. After integrating the delta functions, $E$ can be written as

\begin{align}
    E&=-u'_\mu p'{}^\mu=u'{}^0p'{}^0\left(1-\beta\mu\right)\nonumber\\
    &=\frac{1}{2}u'{}^0Q'{}^0\left(1-\mu\sqrt{1-\frac{1}{\left(u'{}^0\right)^2}}\right)\nonumber\\
    &=\frac{1}{2}u'{}^0Q'{}^0\left(1-\mu\sqrt{1-\frac{\left(Q'{}^0\right)^2}{\left(u'{}^0\right)^2\left(Q'{}^0\right)^2}}\right),
\end{align}
and similarly, ${\bar E}$ can be written as
\begin{align}
    {\bar E}&=\frac{1}{2}u'{}^0Q'{}^0\left(1+\mu\sqrt{1-\frac{\left(Q'{}^0\right)^2}{\left(u'{}^0\right)^2\left(Q'{}^0\right)^2}}\right).
\end{align}
Here, $\beta$ denotes the norm of the spatial component of $u^\mu$, and we used the normalization condition of $u^\mu$, that is $\left(u'{}^0\right)^2=1+\left(u'{}^0\right)^2\beta^2$. $\left(Q'{}^0\right)^2$ and $u'{}^0Q'{}^0$ can be written in the gauge invariant manner as $-Q^2$ and $E_Q$, respectively, and hence, we have 
\begin{align}
    I^{\mu\nu}g_{\mu\nu}&=-\frac{\pi}{8}(1-\alpha^2)E_Q^2\int_{-1}^{1} d\mu \nonumber\\
    &\times f\left[\frac{1}{2}E_Q(1-\alpha\mu)\right] {\bar f}\left[\frac{1}{2}E_Q(1+\alpha\mu)\right]
\end{align}
with $\alpha=\sqrt{1+\frac{Q^2}{E_Q^2}}$. Note that $\alpha\in [0,1]$.

$I^{\mu\nu}u_{\mu}u_{\nu}$, $I^{\mu\nu}Q_{\mu}u_{\nu}$,$I^{\mu\nu}u_{\mu}Q_{\nu}$, and $I^{\mu\nu}Q_{\mu}Q_{\nu}$ can be also evaluated in the similar manner, and the results are as follows:
\begin{align}
    I^{\mu\nu}u_{\mu}u_{\nu}&=\frac{\pi}{16}E_Q^2\int_{-1}^{1} d\mu \nonumber\\
    &\times(1-\alpha^2\mu^2) f\left[\frac{1}{2}E_Q(1-\alpha\mu)\right] {\bar f}\left[\frac{1}{2}E_Q(1+\alpha\mu)\right],
\end{align}
\begin{align}
    I^{\mu\nu}u_{\mu}Q_{\nu}&=\frac{\pi}{16}(1-\alpha^2) E_Q^3\int_{-1}^{1} d\mu \nonumber\\
    &\times(1-\alpha\mu) f\left[\frac{1}{2}E_Q(1-\alpha\mu)\right] {\bar f}\left[\frac{1}{2}E_Q(1+\alpha\mu)\right],
\end{align}
\begin{align}
    I^{\mu\nu}Q_{\mu}u_{\nu}&=\frac{\pi}{16}(1-\alpha^2) E_Q^3\int_{-1}^{1} d\mu \nonumber\\
    &\times(1+\alpha\mu) f\left[\frac{1}{2}E_Q(1-\alpha\mu)\right] {\bar f}\left[\frac{1}{2}E_Q(1+\alpha\mu)\right],
\end{align}
\begin{align}
    I^{\mu\nu}Q_{\mu}Q_{\nu}&=\frac{\pi}{16}(1-\alpha^2)^2E_Q^4\int_{-1}^{1} d\mu \nonumber\\
    &\times f\left[\frac{1}{2}E_Q(1-\alpha\mu)\right] {\bar f}\left[\frac{1}{2}E_Q(1+\alpha\mu)\right].
\end{align}

Defining $G_s=G_n(E_Q,\alpha)$ by
\begin{align}
    G_n=\int_{-1}^{1} d\mu\,\mu^n f\left[\frac{1}{2}E_Q(1-\alpha\mu)\right] {\bar f}\left[\frac{1}{2}E_Q(1+\alpha\mu)\right],
\end{align}
the results above can be further simplified as
\begin{align}
 \frac{I^{\mu\nu}u_{\mu}u_{\nu}}{E_Q^2}&=\frac{\pi}{16}(G_0-\alpha^2 G_2)\\
 \frac{I^{\mu\nu}u_{\mu}Q_{\nu}}{E_Q^3}&=\frac{\pi}{16}(1-\alpha^2)(G_0-\alpha G_1)\\
 \frac{I^{\mu\nu}Q_{\mu}u_{\nu}}{E_Q^3}&=\frac{\pi}{16}(1-\alpha^2)(G_0+\alpha G_1)\\
 \frac{I^{\mu\nu}Q_{\mu}Q_{\nu}}{E_Q^4}&=\frac{\pi}{16}(1-\alpha^2)^2G_0\\
 \frac{I^{\mu\nu}g_{\mu\nu}}{E_Q^2}&=-\frac{\pi}{8}(1-\alpha^2)G_0
\end{align}

and $A$, $B_1$, $B_2$, $C$, and $D$ can be determined as 
%\begin{align}
%    \left(
%    \begin{tabular}{c}
%        $A$\\
%        $B_1$\\
%        $B_2$\\
%        $C$\\
%        $D$
%    \end{tabular}
%    \right)
%    &=
%    \left[
%    \begin{tabular}{ccccc}
        %$ (1-\alpha^2)^2$&$-(1-\alpha^2)  $&$-(1-\alpha^2)  $&      %       1   &$  (1-\alpha^2)  $\\
        %$ (1-\alpha^2)^2$&$-(1-\alpha^2)^2$&$-(1-\alpha^2)  $&$ (1-%\alpha^2)  $&$  (1-\alpha^2)  $\\
        %$ (1-\alpha^2)^2$&$-(1-\alpha^2)  $&$-(1-\alpha^2)^2$&$ (1-%\alpha^2)  $&$  (1-\alpha^2)  $\\
        %$ (1-\alpha^2)^2$&$-(1-\alpha^2)^2$&$-(1-\alpha^2)^2$&$ (1-%\alpha^2)^2$&$  (1-\alpha^2)^2$\\
        %$-(1-\alpha^2)^2$&$ (1-\alpha^2)  $&$ (1-\alpha^2)  $&$-(1-%\alpha^2)  $&$-4(1-\alpha^2)  $
%    \end{tabular}
%    \right]^{-1}
%    \frac{\pi}{16}\left(
%    \begin{tabular}{c}
%         $G_0-\alpha^2 G_2$\\
%         $(1-\alpha^2)(G_0-\alpha G_1)$\\
%         $(1-\alpha^2)(G_0+\alpha G_1)$\\
%         $(1-\alpha^2)^2G_0$\\
%         $-2(1-\alpha^2)G_0$
%    \end{tabular} 
%    \right)\nonumber\\
%    &=\frac{\pi}{32}\left\{
%    \left(
%    \begin{tabular}{c}
%         $0$\\
%         $0$\\
%         $0$\\
%         $G_0+G_2$\\
%         $G_0-G_2$
%    \end{tabular} 
%    \right)
%    +\frac{2 G_1}{\alpha}
%    \left(
%    \begin{tabular}{c}
%         $0$\\
%         $-1$\\
%         $1$\\
%         $0$\\
%         $0$
%    \end{tabular} 
%    \right)
%    +\frac{G_0-3 G_2}{\alpha^2}
%    \left(
%    \begin{tabular}{c}
%         $1$\\
%         $1$\\
%         $1$\\
%         $1$\\
%         $0$
%    \end{tabular} 
%    \right)
%    \right\}.
%\end{align}
\begin{align}
    &\left(
    \begin{tabular}{c}
        $A$\\
        $B_1$\\
        $B_2$\\
        $C$\\
        $D$
    \end{tabular}
    \right)
    =\nonumber\\
    &\frac{\pi}{32}\left\{
    \left(
    \begin{tabular}{c}
         $0$\\
         $0$\\
         $0$\\
         $G_0+G_2$\\
         $G_0-G_2$
    \end{tabular} 
    \right)
    +\frac{2 G_1}{\alpha}
    \left(
    \begin{tabular}{c}
         $0$\\
         $-1$\\
         $1$\\
         $0$\\
         $0$
    \end{tabular} 
    \right)
    +\frac{G_0-3 G_2}{\alpha^2}
    \left(
    \begin{tabular}{c}
         $1$\\
         $1$\\
         $1$\\
         $1$\\
         $0$
    \end{tabular} 
    \right)
    \right\}.
\end{align}
Note that $G_0-G_2$, $\alpha^{-1}G_1$, and $\alpha^{-2}(G_0-3 G_2)$ have the following asymptotic form for $\alpha\rightarrow0$:
\begin{align}
    G_0-G_2&=\frac{4}{3}f(\frac{1}{2}E_Q){\bar f}(\frac{1}{2}E_Q),\\
    \frac{G_1}{\alpha}=-\frac{2}{3}&E_Q\left[\frac{d f}{dE}(\frac{1}{2}E_Q){\bar f}(\frac{1}{2}E_Q)-f(\frac{1}{2}E_Q)\frac{d{\bar f}}{d {\bar E}}(\frac{1}{2}E_Q)\right]\\
    \frac{G_0-3G_2}{\alpha^2}&=-\frac{1}{15}E_Q^2\left[-2\frac{d f}{dE}(\frac{1}{2}E_Q)\frac{d{\bar f}}{d {\bar E}}(\frac{1}{2}E_Q)\right.\nonumber\\
    &\left.+\frac{d^2f}{dE^2}(\frac{1}{2}E_Q){\bar f}(\frac{1}{2}E_Q)+f(\frac{1}{2}E_Q)\frac{d^2{\bar f}}{d^2 {\bar E}}(\frac{1}{2}E_Q)\right]
\end{align}
\subsubsection{Fermi-Dirac case}
Consider the case that $f(E)$ and ${\bar f}({\bar E})$ are given by the Fermi-Dirac distribution functions with the degenerate parameter of $\eta$ and $-\eta$, respectively, that is
\begin{align}
    f(E)=\frac{1}{e^{\frac{E}{k_{\rm B}T} -\eta}+1}~{\rm and}~{\bar f}({\bar E})=\frac{1}{e^{\frac{\bar E}{k_{\rm B}T} +\eta}+1}.
\end{align}
Then, $G_n$ can be written as $G_n={\cal G}_n(x,\alpha,\eta)$ with
\begin{align}
    {\cal G}_n(x,\alpha,\eta)&=\frac{1}{(\alpha x)^{n+1}({\rm e}^{2 x}-1)}\nonumber\\
    &\times \left[{\cal F}^c_n(x-\eta,\alpha x)-(-1)^n{\cal F}^c_n(-x+\eta,\alpha x)\right.\nonumber\\
    &\left.-{\cal F}^c_n(-x-\eta,\alpha x)+(-1)^n{\cal F}^c_n(x+\eta,\alpha x)\right]
\end{align}
and $x=\frac{E_Q}{2k_{\rm B}T}$. Here, ${\cal F}_n$ denotes the complementary incomplete Fermi-Dirac integral defined by
\begin{align}
    {\cal F}^c_n(x,s)=\int_0^s dt \frac{t^n}{e^{t-x}+1}.
\end{align}
${\cal F}^c_n(x,s)$ can be also written as ${\cal F}^c_n(x,s)=F_n(x)-{\cal F}_n(x,s)$ with the complete Fermi-Dirac integral and incomplete Fermi-Dirac integral define by
\begin{align}
    F_n(x)&=\int_0^\infty dt \frac{t^n}{e^{t-x}+1},\\
    {\cal F}_n(x,s)&=\int_s^\infty dt \frac{t^n}{e^{t-x}+1}\nonumber\\
    &=\sum_{k=0}^n\left(\begin{tabular}{c}$n$\\$k$\end{tabular}\right)s^{n-k}F_k(x-s),
\end{align}
respectively.

${\cal G}_n(x,\alpha,\eta)$ has the following property:
\begin{align}
    {\cal G}_n(-x,\alpha,\eta)&=(-1)^n e^{2x} {\cal G}_n(x,\alpha,\eta)\\
    {\cal G}_n(x,-\alpha,\eta)&=(-1)^n {\cal G}_n(x,\alpha,\eta)\\
    {\cal G}_n(x,\alpha,-\eta)&=(-1)^n {\cal G}_n(x,\alpha,\eta)
\end{align}

%The asymptotic behavior of ${\cal G}_n(x,\alpha,\eta)$ is as follows:

%\begin{align}
%    {\cal G}_n&(x,\alpha,\eta)\rightarrow\frac{\left[1+(-1)^n\right]}{n+1}e^{-2x}~\left(x\gg|\eta|,1\right)\\
%    {\cal G}_n&(x,\alpha,\eta)\rightarrow\nonumber\\ & \frac{\left[1+(-1)^n\right]}{n+1}\frac{1}{(e^{-\eta}+1)(e^{\eta}+1)}~\left(x \ll |\eta|,1\right)\\
%    {\cal G}_n&(x,0,\eta)=\nonumber\\ &\frac{\left[1+(-1)^n\right]}{n+1}\frac{1}{(e^{x-\eta}+1)(e^{x+\eta}+1)}\\
%    {\cal G}_n&(x,\alpha,0)=\nonumber\\ &\frac{
%    \left[1+(-1)^n\right]}{(\alpha x)^{n+1}({\rm e}^{2 x}-1)}\left[{\cal F}^c_n(x,\alpha x)-{\cal F}^c_n(-x,\alpha x)\right]\\
%    {\cal G}_n&(x,\alpha,\eta)\rightarrow\frac{\left({\rm sgn}(\eta)\right)^n}{(\alpha x)^{n+1}}e^{-x-|\eta|}\nonumber\\ &\times\left[\Gamma(n+1,-\alpha x)-\Gamma(n+1,\alpha x)\right]~\left(|\eta|\gg x,1\right)
%\end{align}

It is useful also to derive the form of $G_n$ for the case that $f(E)$ and ${\bar f}({\bar E})$ are the Pauli blocking factors:
\begin{align}
    f(E)&=1-\frac{1}{e^{\frac{E}{k_{\rm B}T} -\eta}+1}=\frac{1}{e^{-\frac{E}{k_{\rm B}T} +\eta}+1},\\
    {\bar f}({\bar E})&=1-\frac{1}{e^{\frac{\bar E}{k_{\rm B}T} +\eta}+1}=\frac{1}{e^{-\frac{\bar E}{k_{\rm B}T} -\eta}+1}.
\end{align}
For this case, $G_n$ can be obtained by substituting $x\rightarrow -x$ and $\eta\rightarrow -\eta$ in the results for the case of the Fermi-Dirac distribution functions, that is, $G_n={\cal G}_n(-x,\alpha,-\eta)=e^{2x}{\cal G}_n(x,\alpha,\eta)$.
\subsubsection{application to neutrino pair production/annihilation rate}
For the case that electron and positron are in the pair-thermal equilibrium states, the production and absorption kernels of neutrino/anti-neutrino in the limit of zero electron mass are given by 
\begin{align}
    R&=\frac{2 G_{\rm F}^2}{(2\pi)^2 \omega{\bar \omega}} \int \frac{d^3p}{p^0}\int \frac{d^3{\bar p}}{{\bar p}^0}\nonumber\\
    &\times\delta^4(q+{\bar q}-p-{\bar p})f(E){\bar f}({\bar E})\nonumber\\
    &\times \left[(C_V+C_A)^2 {\bar p}^\mu q_\mu p^\nu{\bar q}_\nu+(C_V-C_A)^2 p^\mu q_\mu {\bar p}^\nu{\bar q}_\nu\right]\nonumber\\
    &=\frac{8 G_{\rm F}^2}{(2\pi)^2  \omega{\bar \omega}} \left[(C_V^2+C_A^2)I^{\mu\nu}(q_\mu{\bar q}_\nu+{\bar q}_\mu q_\nu)\right.\nonumber\\
    &\left.-2C_VC_AI^{\mu\nu}(q_\mu{\bar q}_\nu-{\bar q}_\mu q_\nu)\right].
\end{align}
Here, $\omega=-u_\mu q^\mu$,  ${\bar \omega}=-u_\mu {\bar q}^\mu$, and $f(E)$ and ${\bar f}({\bar E})$ are given by the Fermi-Dirac distribution or the Pauli blocking factor with the degeneracy parameter of electron and anti-electron depending on whether production or absorption is considered. 

We can rewrite $I^{\mu\nu}(q_\mu{\bar q}_\nu+{\bar q}_\mu q_\nu)$ and $I^{\mu\nu}(q_\mu{\bar q}_\nu-{\bar q}_\mu q_\nu)$ as
\begin{align}
    I^{\mu\nu}&(q_\mu{\bar q}_\nu+{\bar q}_\mu q_\nu)=A \frac{Q^4}{E_Q^2} u^\mu u^\nu(q_\mu{\bar q}_\nu+{\bar q}_\mu q_\nu)\nonumber\\
    &+(B_1+B_2)\frac{Q^2}{E_Q}u^\mu Q^\nu(q_\mu{\bar q}_\nu+{\bar q}_\mu q_\nu)\nonumber\\
    &+C Q^\mu Q^\nu(q_\mu{\bar q}_\nu+{\bar q}_\mu q_\nu)\nonumber\\
    &+D Q^2 g^{\mu\nu}(q_\mu{\bar q}_\nu+{\bar q}_\mu q_\nu)\nonumber\\
    &=\left[A\frac{2\omega {\bar \omega}}{(\omega+{\bar \omega})^2}
    -\frac{1}{2}(B_1+B_2)
    +\frac{1}{2}C+D
    \right]Q^4\nonumber\\
    &=\frac{\pi}{64}\left[3 G_0-G_2-\frac{G_0-3G_2}{\alpha^2}\delta^2\right]Q^4
\end{align}
and
\begin{align}
    I^{\mu\nu}&(q_\mu{\bar q}_\nu-{\bar q}_\mu q_\nu)=
    B_1\frac{Q^2}{E_Q}u^\mu Q^\nu(q_\mu{\bar q}_\nu-{\bar q}_\mu q_\nu)\nonumber\\
    &+B_2\frac{Q^2}{E_Q}Q^\mu u^\nu(q_\mu{\bar q}_\nu-{\bar q}_\mu q_\nu)\nonumber\\
    &=-\frac{1}{2}(B_1-B_2)\frac{\omega-{\bar \omega}}{\omega+{\bar \omega}}\nonumber\\
    &=\frac{\pi}{16}\frac{G_1}{\alpha}\delta Q^4,
\end{align}
respectively, with $\delta=\frac{\omega-{\bar \omega}}{\omega+{\bar \omega}}$. Then, $R$ is given in the form,
\begin{align}
    R&=\frac{G_{\rm F}^2}{32\pi  \omega{\bar \omega}}\left[(C_V^2+C_A^2)\left(3 G_0-G_2-\frac{G_0-3G_2}{\alpha^2}\delta^2\right)\right.\nonumber\\
    &\left.-8C_VC_A\frac{G_1}{\alpha}\delta\right]Q^4\\
    &=\frac{2G_{\rm F}^2}{3\pi}\left[(C_V^2+C_A^2)\frac{3}{16}\left(3 G_0-G_2-\frac{G_0-3G_2}{\alpha^2}\delta^2\right)\right.\nonumber\\
    &\left.-\frac{3}{2}C_VC_A\frac{G_1}{\alpha}\delta\right]\omega{\bar \omega}\left(1-\mu\right)^2.
\end{align}
Here, $\mu$ is the cosine of the angle between the spatial part of $q^\mu$ and ${\bar q}^\mu$ in the fluid rest-frame, which satisfies $q_\mu {\bar q}^\mu=\omega{\bar \omega}(1-\mu)$. For given $\delta$ and $\mu$, $\alpha$ can be given as $\alpha=\sqrt{1-\frac{1}{2}\left(1-\delta^2\right)\left(1-\mu\right)}$. Hence, also with the fact that $x=\frac{E_Q}{2k_{\rm B}T}=\frac{\omega+{\bar \omega}}{2k_{\rm B}T}$, $R$ is given as a function of $\omega$, ${\bar \omega}$, and $\mu$ as well as the temperature, $T$, and electron degeneracy parameter, $\eta$.

\bibliographystyle{apsrev4-2}

\end{document}